\documentclass[aps,twocolumn,showpacs,amsmath,amssymb,pra,superscriptaddress,floatfix,longbibliography]{revtex4-2}

\usepackage{hyperref}
\usepackage{bm}
\usepackage{physics}
\usepackage{braket}
\usepackage{amsmath}
\usepackage{amssymb}
\usepackage{mathtools}
\usepackage{graphicx}
\usepackage{dcolumn}
\usepackage{bm}
\usepackage{multirow}
\usepackage{comment}
\usepackage{leftidx}
\usepackage{subcaption} 

\usepackage{color}
\usepackage{xcolor}
\usepackage{float}
\usepackage{capt-of}
\usepackage{mwe}
\usepackage{amsfonts}
\usepackage[capitalise]{cleveref}
\usepackage[shortlabels]{enumitem}
\usepackage[german,english]{babel}
\usepackage{gensymb}
\usepackage[thinc]{esdiff}
\usepackage{url}
\graphicspath{ {./Images/Free}, {./Images/bound} }
\setlist[enumerate,1]{label={(\roman*)}}
\begin{document}



\title{Theory of resonant x-ray scattering with ultrafast intense pulses}


\author{Akilesh Venkatesh}
\email[]{avenkatesh[at]anl[dot]gov}
\affiliation{Chemical Sciences and Engineering Division, Argonne National Laboratory, Lemont, Illinois 60439, USA}

\author{Phay J. Ho}
\email[]{pho[at]anl[dot]gov}
\affiliation{Chemical Sciences and Engineering Division, Argonne National Laboratory, Lemont, Illinois 60439, USA}


\date{\today}

\begin{abstract}
We present a time-dependent Schr\"odinger equation (TDSE) approach within a non-relativistic quantum electrodynamics (QED) framework to investigate resonant x-ray scattering under intense x-ray pulses. This method enables us to explore how coherent x-ray electron dynamics affect scattering signals from Ne\textsuperscript{+}.  We account for both resonant fluorescence and elastic scattering channels, while also considering competing photoionization and inner-shell decay processes. By computing the angular distribution and energy spectrum of scattered photons, we uncover the interference effects between elastic scattering and resonant fluorescence pathways. Notably, our results reveal a small asymmetry in the energy spectrum as a result of this interference. We discuss the experimental potential for detecting signatures of interference.  Our findings demonstrate that the x-ray Rabi dynamics can be used to control scattering responses and offer new insights into interference mechanisms and scattering efficiency in high-intensity x-ray regimes.
\end{abstract}


\maketitle


\section{Introduction}

X-ray free-electron lasers (XFELs) provide a unique opportunity to explore quantum control of inner-shell electrons on ultrafast timescales. For example, an intense x-ray pulse can drive a coherent, oscillatory population transfer (Rabi oscillation) between two energy levels through resonant coupling, the strength of which is directly proportional to the x-ray field strength. In this strong-field limit, previous works show that x-ray-driven Rabi oscillation can lead to modifications in the spectral line shape of the resonant Auger electron \cite{Hiddenresonance_kanter, Auger_Rabi_Santra, Cavaletto_ResFluor_PRA} and the resonant fluorescence energy spectrum \cite{ Cavaletto_ResFluor_PRA} in atoms, as well as reshaping of propagating x-ray pulses in resonant media \cite{Kai_phay_Neplus}. Current XFEL facilities can generate attosecond pulses with energies reaching hundreds of microjoules in both soft and hard x-ray regimes \cite{Marinelli-APL-2017, Huang-PRL-2017, attoXLEAP2020,Malyzhenkov-2020-PRR, Trebushinin-Photonics-2023}. These high-intensity and nearly Fourier transform-limited pulses allow exploration of x-ray dynamics where the Rabi oscillation rate exceeds the inner-shell decay rate.  In addition to mitigating radiation-induced damage, several studies have investigated the scattering properties of transient resonances to enhance x-ray scattering efficiency and imaging resolution with these ultrashort pulses ~\cite{Ho-2020-NatComm, Tais_2022_Xenon_preprint, Tais_Ne_apstalk}. However, the effects of Rabi oscillations on resonant x-ray scattering processes in intense x-ray pulses remain underexplored.

Rate equation approaches have been a workhorse technique for investigating nonlinear x-ray physics in regimes where the multiphoton ionization rate is much larger than inner-shell decay rate. These approaches provide mechanistic insights into sequential multiphoton excitation and ionization processes, leading to the production of transient core-excited states and high charge states in atoms, molecules, and nanosystems \cite{Rudek-2012-NatPho, Rudek-2013-PRA, Ho-2014-PRL, Ho-2015-PRA, Ho-JCP-2023, Rudenko2017, Vinko-Nature-2012}. However, the rate equation approaches are not adequate for describing coherent dynamics enabled by high-intensity x-ray pulses. 

In this paper, we present a theoretical treatment based on a non-relativistic time-dependent QED framework for studying resonant x-ray scattering in intense x-ray pulses. Previous studies have employed TDSE methods to examine the effects of Rabi dynamics in inner-shell decay \cite{Auger_Rabi_Santra, Cavaletto_ResFluor_PRA} and propagation processes \cite{Kai_phay_Neplus}.  Here, we develop a TDSE-based approach to characterize the effects of coherent x-ray electron dynamics and broad bandwidth on x-ray scattering signals of atoms. Our approach accounts for both resonant fluorescence and elastic scattering channels, as well as competing photoionization and inner-shell decay processes.  This approach enables us to compute the angular distribution and energy spectrum of the scattered photons, resolve contributions from different scattered photon polarizations, and elucidate the nature of the interference between elastic scattering and resonant fluorescence channels in intense x-ray pulses. We find that the angular distribution and energy spectrum of the scattered photons reveal the fingerprints of the interference between the elastic scattering and resonant fluorescence channels.  Our paper show that the degree of this interference, as well as the total scattering response, can be tuned by the x-ray Rabi dynamics.

Our paper is organized as follows: We first describe the formalism for studying resonant x-ray scattering in Sec. \ref{Methods}.  To validate our formalism, we present benchmark studies in Sec. \ref{sec_benchmarking}.  We apply our approach to investigate the scattering response of a transient resonance of Ne and discuss the experimental prospects for observing the interference effects in Sec. \ref{Sec_complete_response}.  We present a summary and outlook in Sec. \ref{conclusion_summary}.  Unless otherwise stated, atomic units will be used in the equations present in this work.

\section{Methods and modelling} \label{Methods}
In this section, we first derive the equations that describe x-ray scattering from an $N$-electron system subject to an ultrafast intense pulse. These equations are then used to obtain the equations for the resonant scattering probability. 

\subsection{Scattering probability amplitude for an $N$-electron system} \label{pos_space_Nelectron}
To describe the x-ray scattering from an $N$-electron system, we extend the approach as in Ref.\cite{NLCPRA_1}, which has been used to treat up to two electrons. This subsection generalizes that result to $N$ electrons. Detailed discussions of the approach are available in Ref.\cite{NLCPRA_1, PhDThesis2022}.

The total vector potential associated with the x-ray field is written as~\cite{Krebs},
\begin{equation}\label{vectorpot}
 \boldsymbol{ \hat{A} } = \boldsymbol{ { A} }_{C}    + \boldsymbol{ \hat{A}}_{Q}.
\end{equation}
where $\boldsymbol{ { A} }_{C} $ is the incident classical field and $\boldsymbol{ \hat{A}}_{Q}$ is the quantized scattered field.  The classical field can be written as,
\begin{equation}\label{classicalvectorpotential}
\begin{split} 
\boldsymbol{A}_C = A_0(t) \sin\big(  {\boldsymbol{k}_{in} \cdot \boldsymbol{r} } - \omega_{in} t \big)    \boldsymbol{\epsilon}_{in} ,
\end{split}
\end{equation}
where $\boldsymbol{k}_{in}$ and $\boldsymbol{\epsilon}_{in}$ refer to the wavenumber and polarization of the incident field respectively and $\boldsymbol{r}$ is the position vector. The envelope $A_0(t)$ is modelled as,
\begin{equation}\label{A0_gaussian}
A_0(t) = \frac{E}{\omega_{in}} \exp\Bigg[\frac{-(2 \ln{2} )t^2}{t^2_{wid}} \Bigg].
\end{equation}
Here E, $\omega_{in}$, and $t_{wid}$ refer to the peak electric field, the photon energy, and the pulse duration, respectively.
The quantized scattered field is given by~\cite{Loudon},
\begin{equation}\label{APDX_Eqn_quantized vector potential}
    \boldsymbol{\hat{A}}_Q
    = \sqrt{ \frac{2\pi}{  V  } } \sum_{\boldsymbol{k},\boldsymbol{\epsilon}}  \frac{1}{\sqrt{\omega_{k}}} \bigg[  \boldsymbol{\epsilon} e^{i\boldsymbol{k\cdot r}} \hat{a}_{\boldsymbol{k},\boldsymbol{\epsilon}} +  \boldsymbol{\epsilon}^* e^{-i\boldsymbol{k\cdot r} }
    { \hat{a}_{\boldsymbol{k},\boldsymbol{\epsilon}} ^\dagger  }\bigg].
\end{equation}
Here, $V$ and $\boldsymbol{r}$ denote the quantization volume  and the position vector, respectively. The operators $\hat{a}_{\boldsymbol{k},\boldsymbol{\epsilon}}^{\dagger}$ and $\hat{a}_{\boldsymbol{k},\boldsymbol{\epsilon}}$ denote the creation and annihilation operators for a photon with momentum $\boldsymbol{k}$ and polarization $\boldsymbol{\epsilon}$. Here $\boldsymbol{k} \cdot \boldsymbol{\epsilon} = 0$. The outgoing photon energy $\omega_k = ck$, where $c$ is the speed of light in vacuum.

The total Hamiltonian for the interaction of the electromagnetic field with an $N$-electron system is given by,
\begin{equation}\label{full hamiltonian}
\begin{split}
  \hat{H} = & \sum\limits_{ b=1}^{N} \bigg[\frac{(\boldsymbol{\hat{P}}_b + \boldsymbol{\hat{A}}(\boldsymbol{r}_b))^2}{2} + \hat{V_a}(|\boldsymbol{r}_b|) \bigg]  \\
  &+ \sum_{\boldsymbol{k},\boldsymbol{\epsilon}}\omega_{k}  \hat{a}_{\boldsymbol{k},\boldsymbol{\epsilon}}^{\dagger} \hat{a}_{\boldsymbol{k},\boldsymbol{\epsilon}}
  + \sum\limits_{{i=1,b>i}}^{N} \hat{V}_{ee}(|\boldsymbol{r}_b-\boldsymbol{r}_i|)~.
\end{split}
\end{equation}
Here $\boldsymbol{\hat{P}}_b$, $\hat{V_a}(|\boldsymbol{r}_b|)$ correspond to the momentum operator and potential energy operator (due to the nucleus) acting on $b$\textsuperscript{th} electron, respectively, at position $\boldsymbol{r}_b$. $\hat{V}_{ee}$ describes interaction between electron pairs. 

In this approach, the incident classical field is treated non-perturbatively, whereas the outgoing quantized field is treated perturbatively to the first order. This approach is effective for capturing processes that involve multiple incoming photons but only one outgoing photon. The Hamiltonian in Eq.~(\ref{full hamiltonian}) can be separated into contributions from the incident and the outgoing field.
\begin{equation}\label{hamiltonian_separate}
\begin{split}
  \hat{H} = & \sum\limits_{ b=1}^{N} \hat{H}_{C}(\boldsymbol{r}_b) + \sum_{\boldsymbol{k},\boldsymbol{\epsilon}}\omega_{k}  \hat{a}_{\boldsymbol{k},\boldsymbol{\epsilon}}^{\dagger} \hat{a}_{\boldsymbol{k},\boldsymbol{\epsilon}} \\
  + &\sum\limits_{ b=1}^{N} ( \boldsymbol{\hat{P}}_b  + \boldsymbol{A}_{C}(\boldsymbol{r}_b)) \cdot \boldsymbol{\hat{A}}_{Q}(\boldsymbol{r}_b)
\end{split}
\end{equation}
where
\begin{equation} \label{definition_H_C}
    \hat{H}_{C} (\boldsymbol{r}_b) =  \frac{(\boldsymbol{\hat{P}}_b + \boldsymbol{A}_{C} (\boldsymbol{r}_b))^2}{2} + \hat{V}_{a}(|\boldsymbol{r}_b|)   + \sum\limits_{i=1}^{N} \hat{V}_{ee}(|\boldsymbol{r}_b-\boldsymbol{r}_b|) ~.          
\end{equation}
The wavefunction ansatz for the system is,
\begin{equation} \label{wfn_ansatz_Nelectron}
\begin{split}
    \ket{\psi_{total} } = & \psi^{(0)} (\boldsymbol{r_1}, \boldsymbol{r_2},.. \boldsymbol{r_N} ,t) \ket{0}  \\
    &+ \sum_{\boldsymbol{k},\boldsymbol{\epsilon}} \psi_{\boldsymbol{k},\boldsymbol{\epsilon}} ^{(1)}(\boldsymbol{r_1}, \boldsymbol{r_2},.. \boldsymbol{r_N},t) e^{-i \omega_{k} t} {\hat{a}_{\boldsymbol{k},\boldsymbol{\epsilon}} }^{\dagger}  \ket{0},
\end{split}
\end{equation}
where the first term describes the unscattered wave, while the second term accounts for the scattered wave associated with one outgoing photon. In the first term, $\ket{0}$ represents the vacuum state of the scattered photon in the Fock basis. The quantity $\psi^{(0)}$  represents the wave function at time $t$ of the $N$-electron system interacting with a classical electromagnetic field in the absence of scattered photons. The quantity $\psi_{\boldsymbol{k},\boldsymbol{\epsilon}} ^{(1)}(\boldsymbol{r_1}, \boldsymbol{r_2},.. \boldsymbol{r_N},t)$ describes the scattering probability amplitude at time $t$ for a photon to scatter with momentum $\boldsymbol{k}$ and polarization $\boldsymbol{\epsilon}$ and the N-electrons located at positions $\boldsymbol{r_1}$, $\boldsymbol{r_2}$, .. $\boldsymbol{r_N}$ respectively.

Following the approach for one-electron system described in Ref.~\cite{NLCPRA_1}, similar equations for $\psi^{(0)}$ and $\psi_{\boldsymbol{k},\boldsymbol{\epsilon}} ^{(1)}$ a $N$-electron system can be derived. 
\begin{equation} \label{TDSE_psi0}
  i \frac{\partial \psi ^ {(0)} }{\partial t} - \sum\limits_{ b=1}^{N} \hat{H}_{C_b}\psi ^{ (0) } = 0 ,
\end{equation}
\begin{equation} \label{Exact_eqn_psi1}
\begin{split} 
  i \frac{\partial \psi ^ {(1)}_{\boldsymbol{k},\boldsymbol{\epsilon}} }{\partial t} - \sum\limits_{ b=1}^{N} \hat{H}_{C_b} (\boldsymbol{r}_b) \psi ^{ (1) }_{\boldsymbol{k},\boldsymbol{\epsilon}} =  & \sqrt{\frac{2\pi}{ V\omega_{k}} } e^{i\omega_{k} t } \sum\limits_{ b=1}^{N} e^{-i\boldsymbol{k}\cdot \boldsymbol{r}_b}     \\    
  & \times \boldsymbol{\epsilon}^* \cdot (\boldsymbol{\hat{P}}_b + \boldsymbol{A}_{C}(\boldsymbol{r}_b) ) W(t) \psi ^{(0)}.
\end{split}
\end{equation}
Here $W(t)$ is a windowing function that adiabatically turns on and off the perturbative interaction between the electrons and the quantized outgoing field~\cite{NLCPRA_1, Sakurai_adv}. For non-resonant scattering, a windowing function that smoothly reaches a value of 1 for the duration of the pulse is sufficient. 

\subsection{Resonant scattering from an $N$-electron system}\label{Res_scattering_formalism} 

In this section, we derive expressions for resonant scattering by expanding the unscattered wave $\psi ^{(0)}$ and the scattered wave $\psi ^ {(1)}_{\boldsymbol{k},\boldsymbol{\epsilon}}$ in the eigen-basis of the field-free Hamiltonian.  The unscattered wave can be expanded as, 
\begin{equation} \label{Expand_psi0_eigenstates}
\ket{\psi^{(0)} (t)} = \sum\limits_{ j=1}^{n} C^{(0)}_{j}(t)  \ket{\psi_{j} }.
\end{equation}
where $C^{(0)}_j(t)$ is the time-dependent probability amplitude associated with electronic eigenstate $\ket{\psi_{j} }$. For elastic scattering, the final electronic states of interest are bound states. Hence, a finite basis consisting of all the relevant bound states during resonant scattering is used to describe the electronic part of the system. The continuum electronic states, which may become occupied via Auger decay or photoionization pathways, are modelled using decay rates. 
Therefore the norm of $\psi^{(0)}$ which describes the population of the system will not be preserved for non-zero decay rates.  

Consider the action of the classical interaction Hamiltonian [Eq.~(\ref{definition_H_C})] on $\ket{\psi^{(0)}(t)}$,
\begin{equation} \label{H_psi0_equation}
\begin{split}
\sum\limits_{ b=1}^{N} \hat{H}_{C_b} & \ket{\psi^{(0)}(t)}  \\
& = \bigg( H_{sys}
+ \sum\limits_{ b=1}^{N} \boldsymbol{A}_{C}(\boldsymbol{r}_b) \cdot \boldsymbol{P}_b \bigg) \sum\limits_{ j=1}^{n} C^{(0)}_{j}(t)  \ket{\psi_{j}} \\
& = \bigg( H_{sys} + \boldsymbol{A}_C(t) \cdot \sum\limits_{ b=1}^{N} \boldsymbol{P}_b \bigg) \sum\limits_{ j=1}^{n} C^{(0)}_{j}(t)  \ket{\psi_{j}}.
\end{split}
\end{equation}
Here $H_{sys}$ is the sum of the field-free Hamiltonian of the system and the complex decay terms that describe photoionization anf inner-shell decays. To obtain Eq.~(\ref{H_psi0_equation}), two approximations have been employed. Firstly, we neglect the interaction term associated with $A^2_{C_j}$ as it does not play a role in the resonant dynamics. Secondly, we applied the dipole approximation. This is valid for two reasons: first, our x-ray wavelength is much larger than the size of the atomic orbitals concerned. Second, even for relatively short wavelengths, the dipole approximation has been found to be adequate for calculating photo-absorption probabilities of core-shell states as photoabsorption happens close to the nucleus~\cite{fano1985propensity}. 

The term $H_{sys}$ in Eq.~(\ref{H_psi0_equation}) can be written as, 
\begin{equation} \label{definition_H_system}
\begin{split}
H_{sys} & = \sum\limits_{ j=1}^{n} \bigg[ E_j - \frac{i}{2} \bigg( \Gamma_{a,j} + \Gamma_{p, j}(t) \bigg) \bigg] \ket{\psi_{j} }\bra{\psi_{j}} \\
& = \sum\limits_{ j=1}^{n} \xi_j(t) \ket{\psi_{j} }\bra{\psi_{j}} 
\end{split}
\end{equation}
The quantities $E_j$, $\Gamma_{a,j}$ and $\xi_j(t)$ are the eigenenergy, the Auger decay rate, and the complex eigenenergy of $\ket{\psi_{j}}$ respectively. $\Gamma_{p, j}(t)$ are the time-dependent photoionization rate from the state $\ket{\psi_{j}}$. 
\begin{equation} \label{definition_Gamma_j}
\Gamma_{p, j}(t) = \frac{I(t) \sigma_{j}}{\omega_{in}}.
\end{equation}
Here $\sigma_{j}$ is the one-photon photoionization cross section from the state $\ket{\psi_{j} }$ and $I(t)$ is the instantaneous pulse intensity.

Using Eqs.(\ref{Expand_psi0_eigenstates} and (\ref{H_psi0_equation}), Eq.~(\ref{TDSE_psi0}) can be written as, 
\begin{equation} \label{eqn_psi0_nstate}
\begin{split}
i \frac{\partial C^{(0)}_{m} }{\partial t}  = ~ & \xi_m C^{(0)}_{m} \\
   &+ \boldsymbol{A}_C \cdot \sum\limits_{ j=1}^{n} C^{(0)}_{j} \bra{ \psi_{m} }\sum\limits_{ b=1}^{N} \boldsymbol{P}_b \ket{\psi_{j} } .
\end{split}
\end{equation}
Similar to Eq.~(\ref{Expand_psi0_eigenstates}), one can expand $\psi_{\boldsymbol{k},\boldsymbol{\epsilon}} ^{(1)}$ to obtain, 

\begin{equation} \label{Expand_psi1_eigenstates}
\ket{\psi_{\boldsymbol{k},\boldsymbol{\epsilon}} ^{(1)}(t)} = \sum\limits_{ j=1}^{n} C^{(1)}_{j, \boldsymbol{k} \boldsymbol{\epsilon}}(t)  \ket{\psi_{j} }.
\end{equation}
Here, $C^{(1)}_{j, \boldsymbol{k} \boldsymbol{\epsilon}}(t)$ is the scattering probability amplitude for electronic state $\ket{\psi_{j} }$ and for the outgoing photon with momentum $\boldsymbol{k}$ and polarization $\boldsymbol{\epsilon}$.

Before rewriting Eq.~(\ref{Exact_eqn_psi1}) in the above basis [Eq.~(\ref{Expand_psi1_eigenstates})], there is one important difference in the approach compared to the derivation of the equation for $C^{(0)}_{m}$ [Eq.~(\ref{eqn_psi0_nstate})]. Here, the decay terms in the Hamiltonian are excluded from the time-evolution of the scattered states as the decay terms will cause the norm of $\psi_{\boldsymbol{k},\boldsymbol{\epsilon}} ^{(1)}$ the scattering probability density to decrease over time, even after the incident pulse. This is unphysical for the following reasons. First, the scattered photons have already left the system, so their number cannot decrease with time. Second, including decay for the scattered states would make the scattering probability depend on the propagation time, which can be unphysical. Decay terms are included in the time evolution of the electronic wave function $\psi^{(0)}$ to model the population that remains in the electronic states of interest [Eq.~(\ref{Expand_psi0_eigenstates})]. This would effectively limit the maximum amount of resonant fluorescence based on the linewidth of the excited state. However, these are irrelevant in the calculation of $\psi_{\boldsymbol{k},\boldsymbol{\epsilon}} ^{(1)}$ if only the outgoing photons are measured, not the follow-up dynamics of the electrons. Including all continuum states and performing a second-order perturbation calculation in the quantized field could potentially account for the effect of $\psi_{\boldsymbol{k},\boldsymbol{\epsilon}} ^{(1)}$ on $\psi^{(0)}$ without decay rates, but this is beyond the scope of this work. 

Using Eq.~(\ref{Expand_psi1_eigenstates}) and the dipole approximation for the matrix elements involving $\boldsymbol{P}_b$, Eq.~(\ref{Exact_eqn_psi1}) can be expressed as,

\begin{widetext}
\begin{equation} \label{eqn_psi1_step1}
\begin{split}
i \frac{\partial C^{(1)}_{m, \boldsymbol{k} \boldsymbol{\epsilon}} }{\partial t} - \bigg[ E_m C^{(1)}_{m, \boldsymbol{k} \boldsymbol{\epsilon}} + \boldsymbol{A}_C \cdot \sum\limits_{ j=1}^{n} C^{(1)}_{j, \boldsymbol{k} \boldsymbol{\epsilon}} \bra{ \psi_{m} }\sum\limits_{ b=1}^{N} \boldsymbol{P}_b \ket{\psi_{j} } \bigg] \
 = & \sqrt{\frac{2\pi}{ V\omega_{k}} } e^{i\omega_{k} t } \sum\limits_{ j=1}^{n} C^{(0)}_{j}(t) 
 \bigg[ \boldsymbol{\epsilon}^* \cdot \bra{ \psi_{m} }\sum\limits_{ b=1}^{N}  \boldsymbol{P}_b \ket{\psi_{j} }           \\
 &  + \boldsymbol{\epsilon}^* \cdot 
\bra{ \psi_{m} }\sum\limits_{ b=1}^{N} e^{-i\boldsymbol{k}\cdot \boldsymbol{r}_b}  \boldsymbol{A}_{C_b} \ket{\psi_{j} } \bigg]  W(t).
\end{split}
\end{equation}

Further more using Eq.~(\ref{classicalvectorpotential}) and applying the rotating wave approximation, Eq.~(\ref{eqn_psi1_step1}) can be written as,
\begin{equation} \label{eqn_psi1_nstate}
\begin{split}
i \frac{\partial C^{(1)}_{m, \boldsymbol{k} \boldsymbol{\epsilon}} }{\partial t} - \bigg[ E_m C^{(1)}_{m, \boldsymbol{k} \boldsymbol{\epsilon}} + \boldsymbol{A}_C \cdot \sum\limits_{ j=1}^{n} C^{(1)}_{j, \boldsymbol{k} \boldsymbol{\epsilon}} \bra{ \psi_{m} }\sum\limits_{ b=1}^{N} \boldsymbol{P}_b \ket{\psi_{j} } \bigg] 
 = & \sqrt{\frac{2\pi}{ V\omega_{k}} } e^{i\omega_{k} t } \sum\limits_{ j=1}^{n} C^{(0)}_{j}(t) 
 \bigg[ \boldsymbol{\epsilon}^* \cdot \bra{ \psi_{m} }\sum\limits_{ b=1}^{N} \boldsymbol{P}_b \ket{\psi_{j} }           \\
 &  - \frac{i}{2} A_0(t) e^{-i\omega_{in} t } ~\boldsymbol{\epsilon}^* \cdot \boldsymbol{\epsilon}_{in} 
\bra{ \psi_{m} }\sum\limits_{ b=1}^{N} e^{i ( \boldsymbol{k}_{in} - \boldsymbol{k})\cdot \boldsymbol{r}_b}  \ket{\psi_{j} } \bigg] W(t).
\end{split}
\end{equation}
\end{widetext}

The source terms on the right-hand side of Eq.~(\ref{eqn_psi1_nstate}) shows the channels that contribute to the scattering probability amplitude.  If the final scattered state ($m$) is a ground state, the source terms then consist of both resonant fluorescence (first term) and elastic scattering (second term) respectively. The corresponding independent channel amplitudes can be calculated by only including that source term. Also, the time-dependent source terms in Eq.~(\ref{eqn_psi1_nstate}) can be interpreted as the contribution of the occupied intermediate states to the two outgoing photon channels.  Previous works~\cite{buffa1988_amplitude_ResFluor, Robinson_Berman1984_buffa_citation} have employed a similar treatment for calculating resonant fluorescence driven by short pulses. But, these works are not in the x-ray domain and do not include elastic scattering.

In calculating the scattering probability amplitude [Eq.~(\ref{eqn_psi1_nstate})]for the final state to be an excited state, only the source term for elastic scattering should contribute as there can be no spontaneous emission from the ground state to the excited state. This is indeed the case as the contribution from this source term goes to zero, if a windowing function $W(t)$ is used to adiabatically turn on and off the source terms during the pulse~\cite{Sakurai_adv}. Under resonant conditions, one can choose to not use a windowing function, if this term is dropped~\cite{buffa1988_amplitude_ResFluor, Robinson_Berman1984_buffa_citation} as it may introduce a spurious contribution. To illustrate this, consider the trivial case when there is no incident x-ray pulse. Then, in the absence of a windowing function, this term would lead to a non-zero contribution to the scattering probability amplitude, which is unphysical.

In Eq.~(\ref{eqn_psi1_nstate}), if $m = j$ the quantity $\bra{\psi_{m} }\sum\limits_{ b=1}^{N} e^{i ( \boldsymbol{k}_{in} - \boldsymbol{k})\cdot \boldsymbol{r}_b} \ket{\psi_{j} }$ is the form factor of state $\ket{\psi_{m} }$ of the atom or molecule being studied. $\bra{ \psi_{m} }\sum\limits_{ b=1}^{N} \boldsymbol{P}_b \ket{\psi_{j} }$ can be computed from the corresponding transition dipole matrix elements. In this study of Ne\textsuperscript{+}, the form factors for elastic scattering and the transition dipole matrix elements were computed using the Hartree-Fock-Slater electron structure theory~\cite{HFS_Phay_2017}. The differential equations in Eq.~(\ref{eqn_psi0_nstate}) and Eq.~(\ref{eqn_psi1_nstate}) were then solved simultaneously using a previously developed leap-frog approach~\cite{NLCPRA_1}. The results obtained were verified for a few cases with a sixth order Runge-Kutta (RK6) method~\cite{numericalrecipes}. The scattering probability amplitude obtained is used to calculate the scattering probability, differential cross section and total cross section, which are prescribed in the next section (Sec.~\ref{section_dcs}). 

\subsection{Differential cross section} \label{section_dcs}
The probability of scattering a photon into mode $\boldsymbol{k},\boldsymbol{\epsilon}$ with the associated angles ($\theta_s$, $\phi_s$) and the electrons to be in the final state $\ket{\psi_j}$, is given by $ P_ {j, \boldsymbol{k},\boldsymbol{\epsilon}} = \abs{{ C^{(1)}_{j, \boldsymbol{k} \boldsymbol{\epsilon}}} }^2$. If the final scattered electronic state is not measured, the scattering probability is given by
\begin{equation} \label{summed_scatteringprobability}
     P_ {\boldsymbol{k},\boldsymbol{\epsilon}} = \sum\limits_{ j=1}^{n} { C^{(1)}_{j, \boldsymbol{k} \boldsymbol{\epsilon}}}^*  C^{(1)}_{j, \boldsymbol{k} \boldsymbol{\epsilon}} .
 \end{equation}
For a given scattering angle, a set of two outgoing photon polarizations that are orthogonal to $\boldsymbol{k}$ and to each other are,
\begin{equation} 
\begin{split} \label{polarization_choices}
     &\boldsymbol{\epsilon_1} = (-\sin\theta_s,~\cos\theta_s\cos\phi_s,~ \cos\theta_s\sin\phi_s) \\
     &\boldsymbol{\epsilon_2} = (0,~ -\sin\phi_s,~ \cos \phi_s)
\end{split}
\end{equation}
The differential cross section (DCS) for a photon to scatter into an solid angle $d\Omega$ can be obtained by summing over the two outgoing photon polarizations as
\begin{equation} \label{dcs}
        \dv{\sigma}{\Omega}^{(1)} =
        \frac{V \omega_{in} }{(2\pi)^3}\frac{\int \sum\limits_{ \boldsymbol{\epsilon} } P_ {\boldsymbol{k},\boldsymbol{\epsilon}} k^2dk }{\int I(t) dt } .
\end{equation}
Here $I(t)$ is the time-dependent intensity of the incident x-ray pulse. In this work, we restrict to the case of those outgoing photons that are nearly elastically scattered, that is, $\omega_k \approx \omega_{in}$. In Eq.(\ref{dcs}), for a given $\boldsymbol{\epsilon}$, the integration is performed in the neighbourhood of $\abs{\boldsymbol{k}} \approx \abs{ \boldsymbol{k_{in}}  }$ to cover the whole energy spectrum of the outgoing photons from resonant fluorescence and elastic scattering. For this work, the number of $\omega_k$ grid points and the range of $\omega_k$ are selected to ensure that the computed cross sections are converged to about 1\% of the value.

For linear x-ray regime, the scattering probability [Eq.~(\ref{summed_scatteringprobability})] is proportional to the intensity, therefore the differential cross section (DCS) is independent of intensity. However, for the intense field regime, the DCS for resonant scattering will depend on the intensity. The total cross section $\sigma^{(1)}$ is then given by
\begin{equation} \label{totalcs}
        \sigma^{(1)} =\int d\Omega  \dv{\sigma}{\Omega}^{(1)}.
\end{equation}
We note that the outgoing photons can in general come from elastic scattering, resonant fluorescence or stimulated emission. However, the scattering probability $P_ {\boldsymbol{k},\boldsymbol{\epsilon}}$ [Eq.~(\ref{summed_scatteringprobability})] describes resonant fluorescence and elastic scattering channels and not stimulated emission. Stimulated emission is in the direction of the incident driving x-ray field and can be calculated directly from $\ket{\psi^{(0)} (t)}$.

\section{Applications} \label{Sec_applications}

\begin{figure}
\resizebox{90mm}{!}{\includegraphics{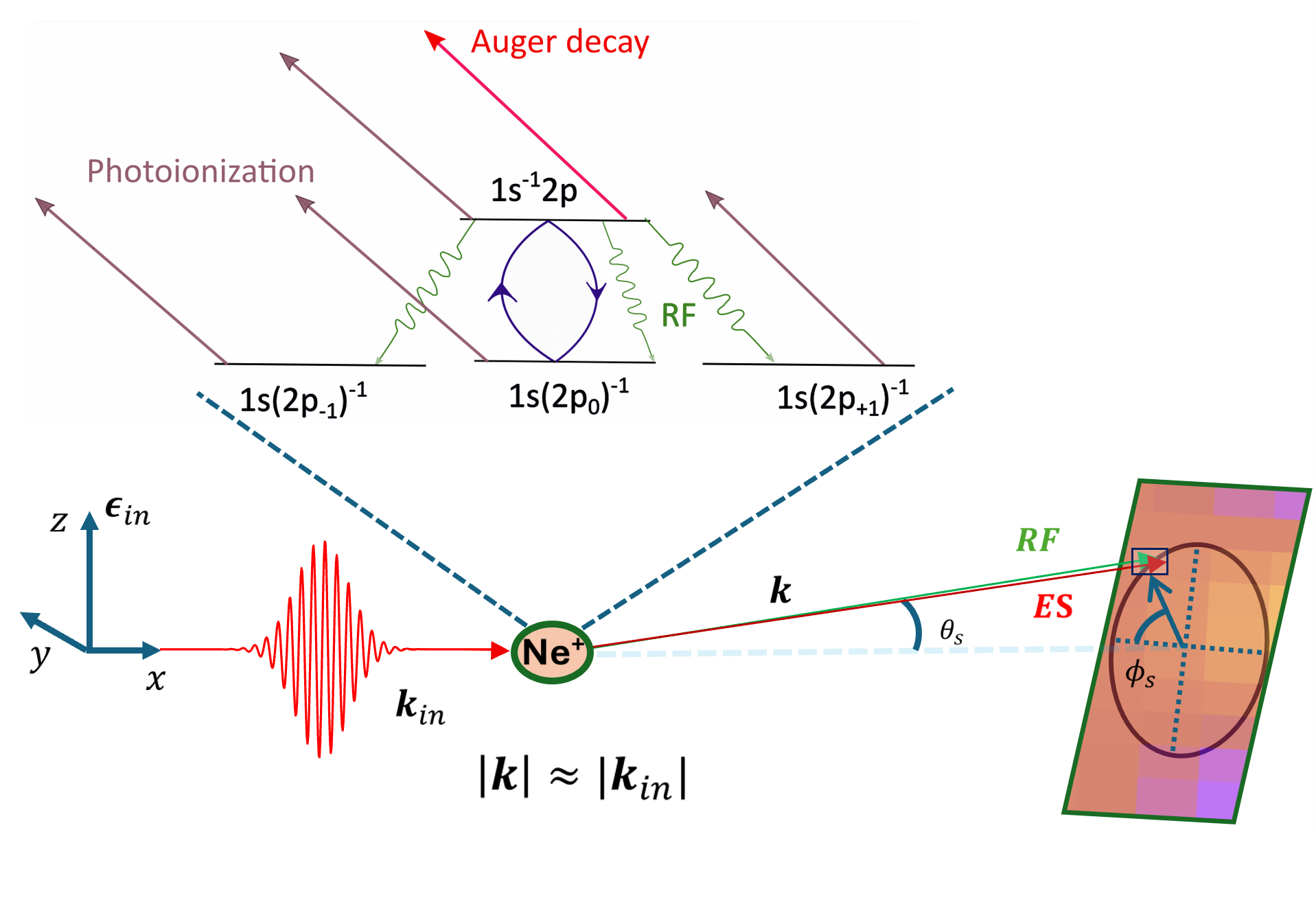}}
\caption{\label{Schematic_diagram}\label{Coordinate_system}
Schematic diagram~\cite{Companion_letter} showing both elastic scattering (ES) and resonant fluorescence (RF) channels for the outgoing photons from a resonantly driven Ne\textsuperscript{+} atom. The electronic states are shown in the inset figure which includes two decay channels, photoionization and Auger decay.
}
\end{figure}

In this section, we use the approach described in Sec.~\ref{Res_scattering_formalism} to study resonant x-ray scattering from an Ne\textsuperscript{+} atom.
This atom in an intense x-ray pulse has been studied theoretically with a two-level description ~\cite{Cavaletto_ResFluor_PRA} for a special case of resonant fluorescence at a specific angle. As shown in Fig.~\ref{Schematic_diagram}~\cite{Companion_letter}, we use a four-level description for the electronic states to fully capture the dependence of elastic scattering and resonant fluorescence on the scattering angle and outgoing photon polarization. The four electronic states are $1s2p_{-1}^{-1}$, $1s2p_0^{-1}$, $1s2p_{+1}^{-1}$, and $1s^{-1}2p$ with the first three levels being degenerate. The spin-orbit interaction and relativistic effects have been ignored as it is not expected to play a significant role here.

The incident x-ray photon energy  $\omega_{in}$ of 849.8 eV is chosen to be resonant with the $1s2p^{-1} \rightarrow 1s^{-1}2p$ transition and the incident photon momentum $\boldsymbol{k}_{in}$ and the polarization $\boldsymbol{\epsilon}_{in}$ are chosen to be in $\hat{x}$ and $\hat{z}$ direction, respectively (Fig.~\ref{Schematic_diagram}~\cite{Companion_letter}). 
In the intense field regime, Rabi oscillations can occur between the core-hole excited state $1s^{-1}2p$ and the ground state $1s2p_0^{-1}$. The other degenerate ground states, $1s2p_{-1}^{-1}$ and $1s2p_{+1}^{-1}$, do not participate in the Rabi oscillations. The pulse can also photoionize any of the valence electrons in the Ne\textsuperscript{+}. The photoionization cross section for the triply degenerate ground and core-excited states are $8.4\times 10^{-4}$ a.u. ($\sim$23.6 kilobarns) and the core-excited state was estimated to be $1.1\times 10^{-3}$ a.u. ($\sim$31.7 kilobarns) respectively~\cite{Cavaletto_ResFluor_PRA, Elettra_database}.  Auger decay from the core-hole excited state is accounted for, where the core-hole lifetime is about 2.4 fs and Auger decay is the dominant decay channel with a rate $\Gamma_{a,3} \sim 0.01$ a.u.

The characteristics of Rabi dynamics during a short pulse is well characterized by the pulse area, $Q$~\cite{Pulsearea_defn_Eberly, Cavaletto_ResFluor_PRA}, defined as, 
\begin{equation} \label{pulsearea_defn}
    Q = \int_{-\infty}^{\infty} \Omega(t) dt  .
\end{equation}
Here $\Omega(t) = \boldsymbol{E}_{in}(t)\cdot \boldsymbol{\mu} $ where $\Omega(t)$, $\boldsymbol{E}_{in}(t)$, and $\boldsymbol{\mu}$ refer to the resonant Rabi frequency, the incident time-dependent electric field and the transition dipole moment between the ground state and core-excited state, respectively. In the absence of decay channels, pulse areas of even-multiples of $\pi$ leave the system mostly in the ground state, whereas pulse areas that correspond to odd-multiples of $\pi$ leave a substantial population in the core-excited state.

\subsection{Benchmarking the approach} \label{sec_benchmarking}

We first benchmark our approach by comparing our results of the resonant fluorescence channel with the work of Cavaletto and coworkers~\cite{Cavaletto_ResFluor_PRA} using the same parameters. The polarization of the outgoing photon is chosen to be in the same direction as the polarization of the incident x-ray field, and $\theta_s=90\degree$ and $\phi_s = 0\degree$ (see Fig.~\ref{Schematic_diagram}). A pulse duration of 2 fs is used.  Also,  we used the values of the transition dipole moments for Ne\textsuperscript{+} in Ref.~\cite{Cavaletto_ResFluor_PRA}.

First, we find excellent agreement between our calculated results and Ref.~\cite{Cavaletto_ResFluor_PRA} for the time-dependent probabilities of the states during Rabi dynamics for both $\pi$-type and 2$\pi$-type pulses [Eq.~(\ref{pulsearea_defn})]. 

Next, we compare energy spectrum, $S$, given as,
\begin{equation} \label{energy_spectrum}
        S(\omega_k, \Omega) = \frac{V\omega_k^3}{(2\pi)^3 c^3}  \sum\limits_{ j=1}^{n}  \abs{{ C^{(1)}_{j, \boldsymbol{k} \boldsymbol{\epsilon}}}}^2.
\end{equation}
This quantity when integrated over all emitted frequency can be interpreted as the total emitted energy into a solid angle $d\Omega$ with the chosen outgoing photon polarization $\boldsymbol{\epsilon}$ from the interaction of the atom with the incident pulse.

\begin{figure}
\resizebox{80mm}{!}{\includegraphics{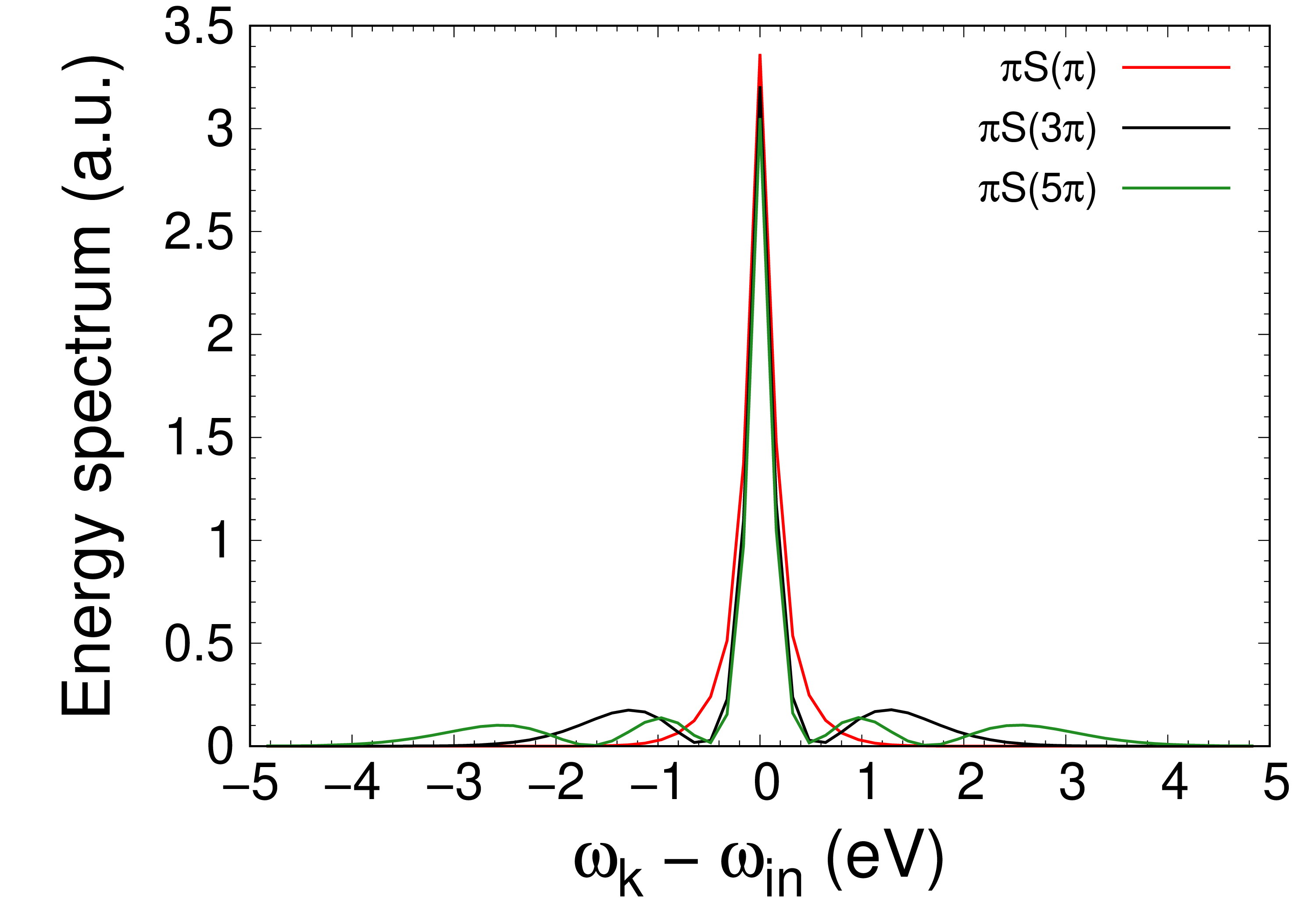}}
\resizebox{80mm}{!}{\includegraphics{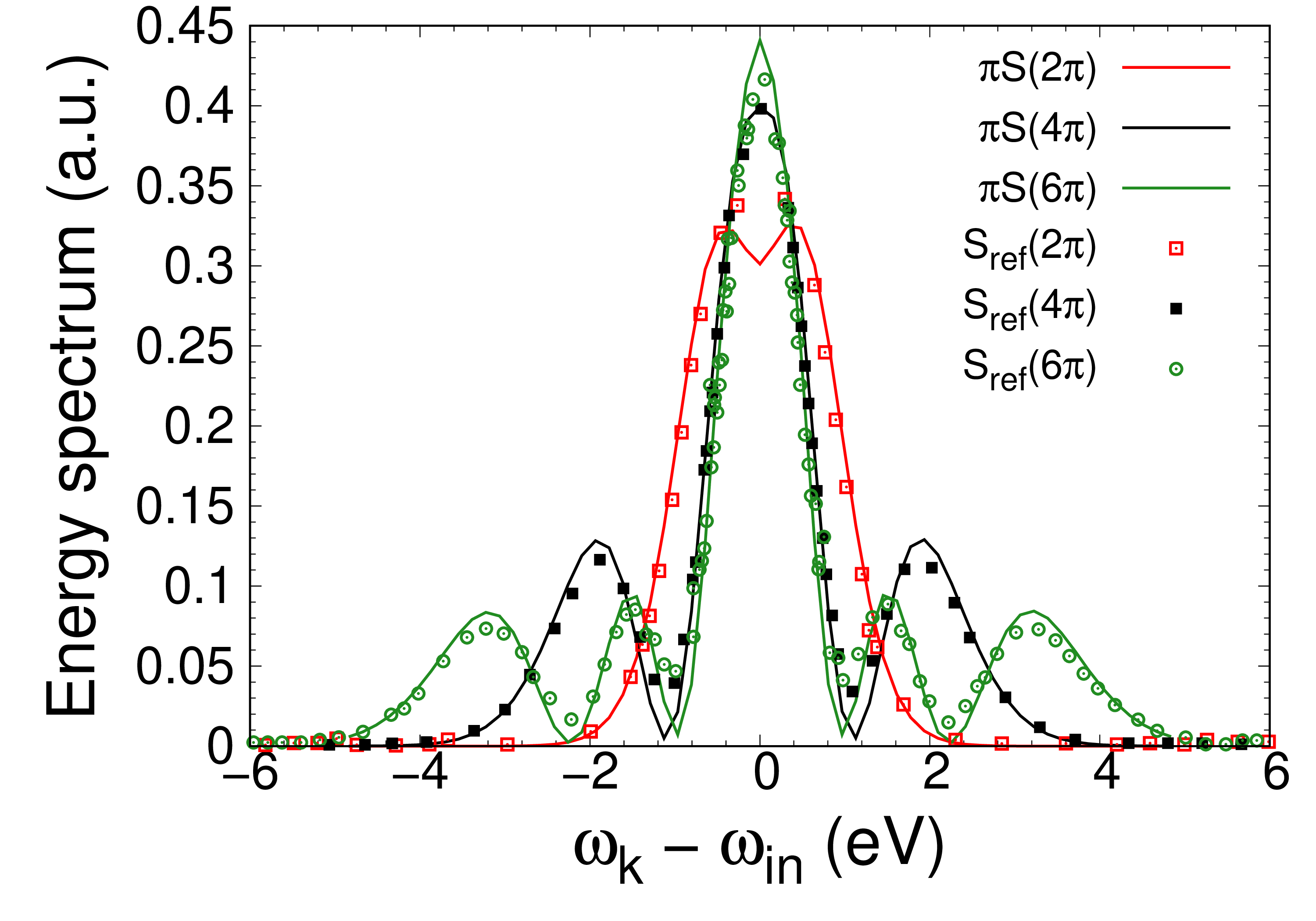}}
\caption{\label{Fig_Cavaletto_comparison}
The results for the energy spectrum vs energy transferred for resonant fluorescence from an Ne\textsuperscript{+} atom. The solid curves are the results of the calculation shown in this work and the points (S\textsubscript{ref}) were extracted from the plots shown in Cavaletto et al.~\cite{Cavaletto_ResFluor_PRA}. The approach described in this work shows good agreement with the results from Ref.~\cite{Cavaletto_ResFluor_PRA}.
}
\end{figure}

Comparing the calculated energy spectrum (Fig.~\ref{Fig_Cavaletto_comparison}) from resonant fluorescence with that of Ref.~\cite{Cavaletto_ResFluor_PRA} first, for the case of $\pi$-type pulses, we find good agreement provided we multiply the calculated energy spectrum by a factor of $\pi$. This is due to a difference in our definitions of the energy spectrum as the definition in Ref.~\cite{Cavaletto_ResFluor_PRA} is missing a factor of $\pi$~\cite{Cavaletto_discussion}. To verify our calculation is correct, we compute the number of fluorescence photon using the resonant fluorescence channel cross section $\sigma_F^{(1)}$,
\begin{equation} \label{NF}
        N_F = \frac{\int I(t) dt }{\omega_{in}} \times \sigma_F^{(1)}
\end{equation}
Our computed $N_F$ agrees with the result obtained from a rate equation to about 1 \% accuracy. The rate equation used is given by, 
\begin{equation}
     \dv{N_{F}}{t} =  \Gamma_{SE}   N_{1s^{-1}} (t) 
\end{equation}
where $\Gamma_{SE}$ is the total spontaneous emission decay rate, which is $1.4\times 10^{-4}$ a.u. taken from Ref.~\cite{Cavaletto_ResFluor_PRA}, and $N_{1s^{-1}} (t)$ is the time-dependent population of the excited state ($1s^{-1}2p$), which is obtained by solving the Eq.~\ref{eqn_psi0_nstate}.

For the 2$\pi$-type pulses, using the same scaling factor of $\pi$, our result shows a reasonable agreement with the results from Cavaletto et al.~\cite{Cavaletto_ResFluor_PRA}. However, we find two small differences, which appears to stem from the different approaches. One difference is that there is a small minimum at $\omega-\omega_{in} = 0$ in the calculated result for the 2$\pi$ pulse, which appears to have been predicted previously \cite{rzazewski1984_multipeakresfluorescence,robinson1986_temporaldiffraction}. Another small difference is in the regions around the sidebands ($\omega-\omega_{in}\approx \pm 1.2$ eV,  $\pm 2.4$ eV), where Ref.~\cite{Cavaletto_ResFluor_PRA} shows a non-zero minima, whereas our calculation shows a near-zero minima. Previous calculations \cite{rzazewski1984_multipeakresfluorescence, lewenstein1986theory_santraref, robinson1986_temporaldiffraction} show that for a given pulse area, the characteristics of the minima depend on the ratio of Fourier energy bandwidth of the pulse to the energy bandwidth associated with the lifetime of the state. Calculations with a longer pulse duration indicate that the zeroes of the curve become minimas as the Fourier bandwidth of the incident pulse decreases.

Another independent benchmark for the approach (Sec.~\ref{Res_scattering_formalism}) was performed by comparing the results obtained by solving Eq.~(\ref{eqn_psi0_nstate}) and Eq.~(\ref{eqn_psi1_nstate}) for a few angles with the results from Kramers-Heisenberg differential cross section (KHDCS) in the limit of weak fields and a long pulse duration. The KHDCS for the initial and final electronic state to be the same is given by~\cite{Sakurai_adv},
\begin{equation} \label{eqn_KH_dcs}
\begin{split}
\bigg(\dv{\sigma}{\Omega} \bigg)_{KH}= ~ & r^2_0 \bigg| \epsilon \cdot \epsilon_{in}
             - \bigg( \frac{\bra{A} \hat{\boldsymbol{P}} \cdot \epsilon \ket{I} \bra{I} \hat{\boldsymbol{P}} \cdot \epsilon_{in} \ket{A} }{E_I - E_A - \omega_{in} - i\Gamma_I/2}  \\
            & + \frac{\bra{A} \hat{\boldsymbol{P}} \cdot \epsilon_{in} \ket{I} \bra{I} \hat{\boldsymbol{P}} \cdot \epsilon \ket{A}}{E_I - E_A + \omega_{in}} \bigg) \bigg|^2
\end{split}
\end{equation}
Here $r_0$ is the classical electron radius, $\hat{\boldsymbol{P}}$ is the total momentum operator, and the states $\ket{A}$ and $\ket{I}$ refer to the ground state and the core-hole state respectively. The quantity $E_I$ and $E_A$ are the binding energies of these states and $\Gamma_I$ is the decay rate of the core-hole state. In the weak-field regime, since there is no significant population transfer from the ground state to the excited state, for the elastic case the final electronic state can be assumed to be the same as the initial state. 
To make comparison with the monochromatic case assumed in the KHDCS, a pulse duration much larger than the core-hole lifetime is used in our approach. We find excellent agreement between our calculated results and the KHDCS. Since Eq.~\ref{eqn_KH_dcs} includes both the elastic scattering and resonant fluorescent amplitudes, it serves as an additional check for our calculations obtained using Eq.~(\ref{eqn_psi1_nstate}).
\\

\subsection{Complete single-atom response for Ne\textsuperscript{+}} \label{Sec_complete_response}
In this subsection, we compute single-atom scattering response spanning the entire range of scattering angles, outgoing photon polarization directions and for different pulse conditions. We study how the resonant fluorescence channel and non-resonant elastic scattering channel contributions together shape the total single-atom response. The initial state of the Ne\textsuperscript{+} atom is assumed to be an equal superposition of the three degenerate ground states that is, $\ket{\psi_i} = \frac{1}{\sqrt{3}} \big[ \ket{ 1s2p_{-1}^{-1} } + \ket{ 1s2p_0^{-1}} + \ket{ 1s2p_{+1}^{-1} } \big]$. The motivation for this choice is to allow for multiple interfering pathways between the two channels and to increase the fraction of elastic scattering in the total response, thereby illustrating the interference.  We explore the scattering response for different initial states in the companion letter~\cite{Companion_letter}.

\begin{figure}
\begin{minipage}[t][3ex][t]{0.001\textwidth}
(a)
\end{minipage}
\begin{minipage}[t]{0.23\textwidth}
        \vspace{0.2cm}
        \includegraphics[width=\linewidth]{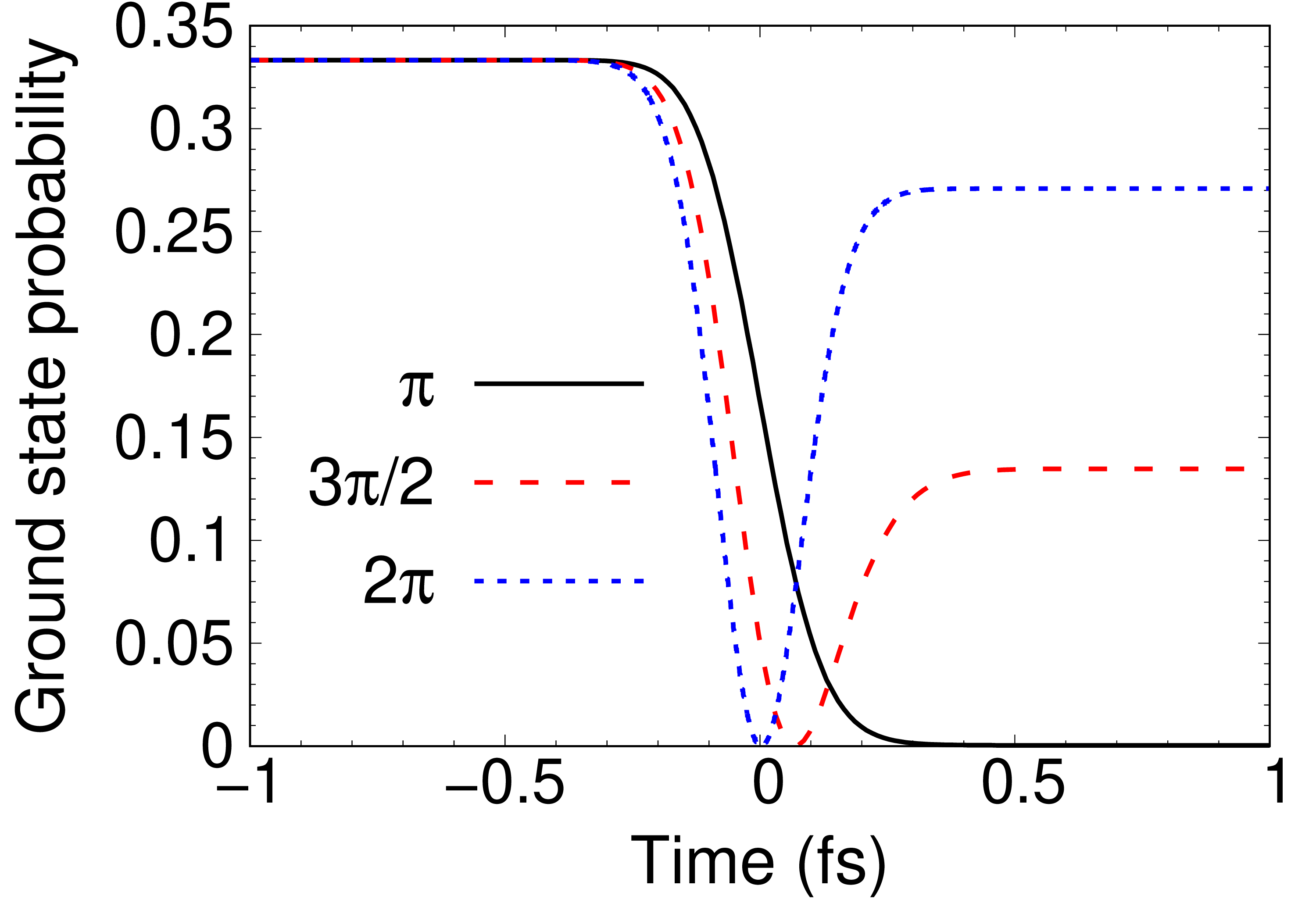}
\end{minipage}
\begin{minipage}[t][3ex][t]{0.001\textwidth}
          (c)
\end{minipage}
\begin{minipage}[t]{0.23\textwidth}
        \vspace{0.2cm}
        \includegraphics[width=\linewidth]{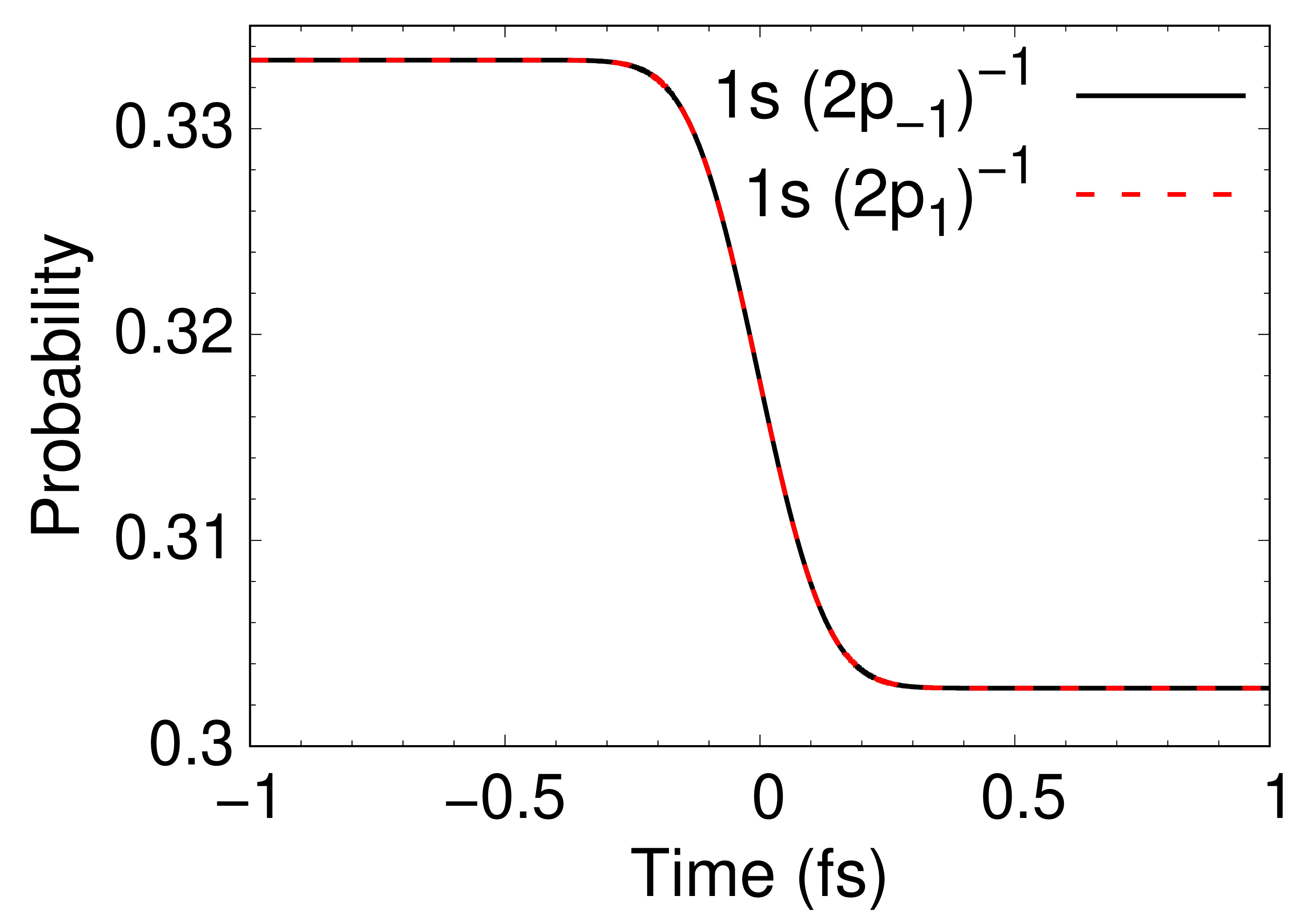}
\end{minipage}
\vspace{0.1cm}
\begin{minipage}[t][3ex][t]{0.001\textwidth}
        (b)
\end{minipage}
\begin{minipage}[t]{0.23\textwidth}
        \vspace{0.2cm}
        \includegraphics[width=\linewidth]{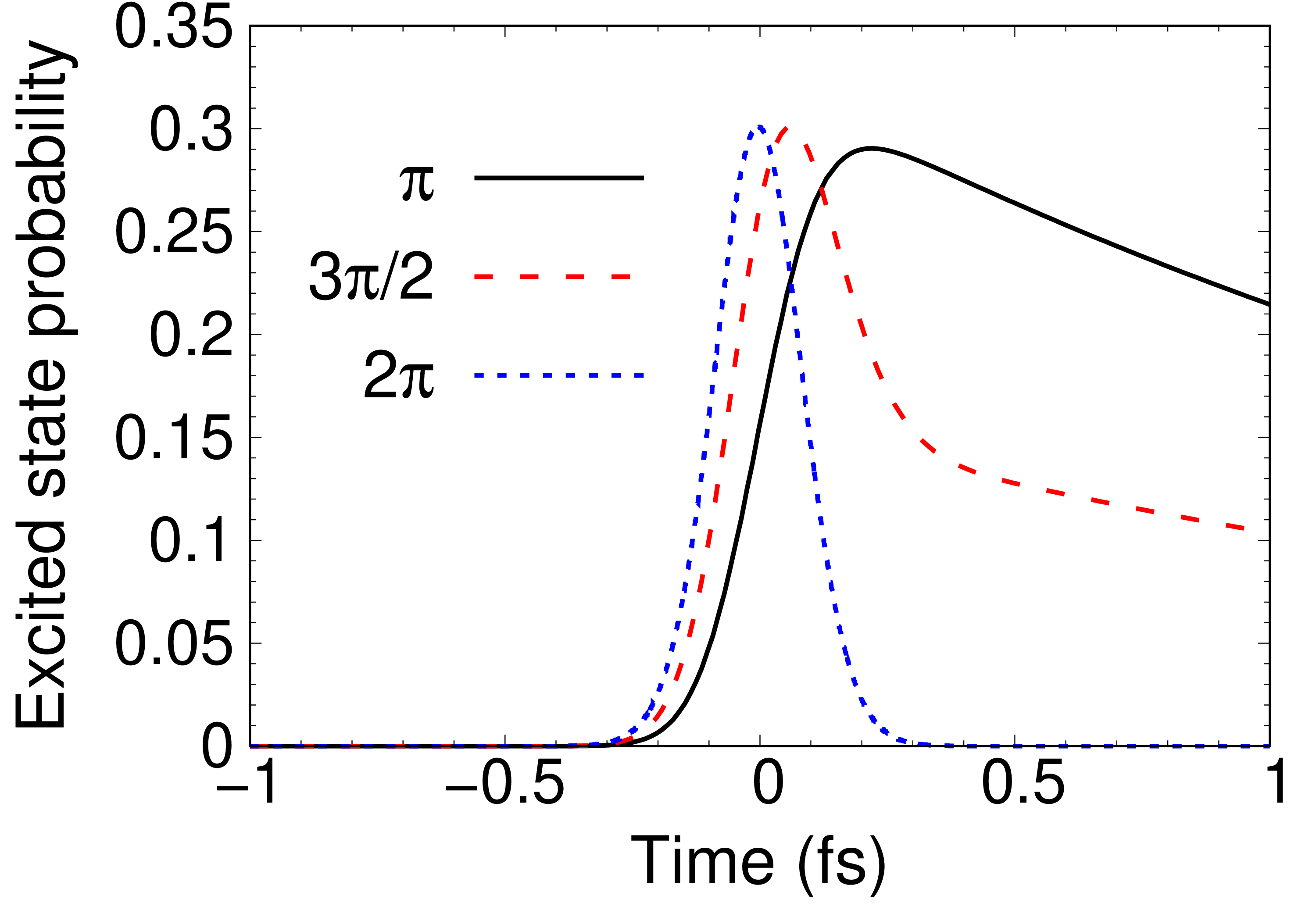}
\end{minipage}
\begin{minipage}[t][3ex][t]{0.001\textwidth}
       (d)
\end{minipage}
\begin{minipage}[t]{0.23\textwidth}
        \vspace{0.2cm}
        \includegraphics[width=\linewidth]{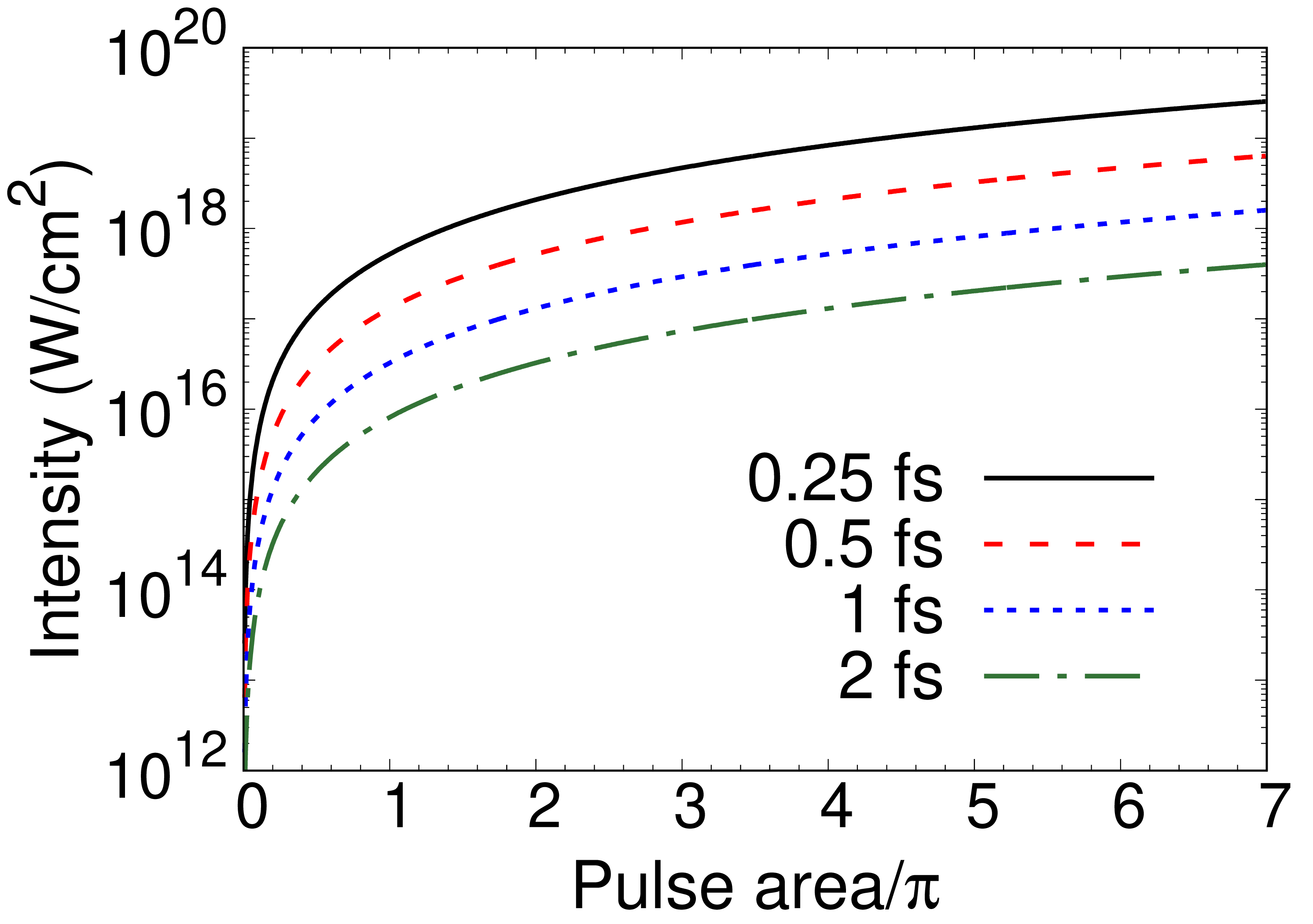}
\end{minipage}
\caption{ Rabi dynamics in an Ne\textsuperscript{+} atom driven by resonant x-ray pulses with 849.8 eV. Time-dependent population of the (a) ground state $1s2p_0^{-1}$ and (b) core-excited state $1s^{-1}2p$, respectively for 3 pulse areas. (c) shows the population of the two degenerate states undergoing photoionization decay for a pulse area of $2\pi$.  (d) The intensities required for a Gaussian pulse to generate different pulse areas for different pulse durations. The pulse duration is 0.25 fs for (a) to (c).
}
\label{Fig_Rabi_dynamics_p25fs}
\end{figure}

Fig.\ref{Fig_Rabi_dynamics_p25fs} illustrates the population dynamics of the different states of a single Ne\textsuperscript{+} atom when exposed to an intense, 0.25 fs x-ray pulse polarized in the z-direction. Figs.~\ref{Fig_Rabi_dynamics_p25fs}a and \ref{Fig_Rabi_dynamics_p25fs}b show that Rabi oscillations drive population between the states $1s2p_0^{-1}$ and $1s^{-1}2p$.  The other electronic states $1s2p_{-1}^{-1}$ and $1s2p_{+1}^{-1}$ do not participate in the Rabi dynamics, but they are subjected to photoionization(Fig.~\ref{Fig_Rabi_dynamics_p25fs}c). 

It is evident from Figs.\ref{Fig_Rabi_dynamics_p25fs}a and \ref{Fig_Rabi_dynamics_p25fs}b that a $2\pi$ pulse leaves the system mostly in the ground state, while a $\pi$ pulse leaves it primarily in the core-excited state at the end of the pulse. A $3\pi/2$-pulse results in intermediate values of population in both the ground state and core-excited state. Adding $n$ multiples of $2\pi$ to the pulse area corresponds to $n$ Rabi oscillations during the pulse. The intensity of the incident pulse required to generate a pulse area for different pulse durations is shown in Fig.~\ref{Fig_Rabi_dynamics_p25fs}d. This range of intensities and pulse duration is presently achievable at XFEL facilities~\cite{LCLS_specswebsite, Euxfel_specswebsite}.


\begin{figure*} [hbt!] 
\begin{minipage}[t][][t]{\textwidth}
       \hspace{0cm}
       \textbf{Elastic}
       \hspace{2.7cm}
       \textbf{Res. Fluor.}
       \hspace{2cm}
       \textbf{Incoherent sum}
       \hspace{1.9cm}
       \textbf{Coherent sum}
\end{minipage}
\centering
\begin{minipage}[b][][b]{\textwidth}
\vspace{0.3cm}
\hspace{-0.3cm}
\begin{subfigure}{0.23\linewidth}
  \includegraphics[width=\linewidth]{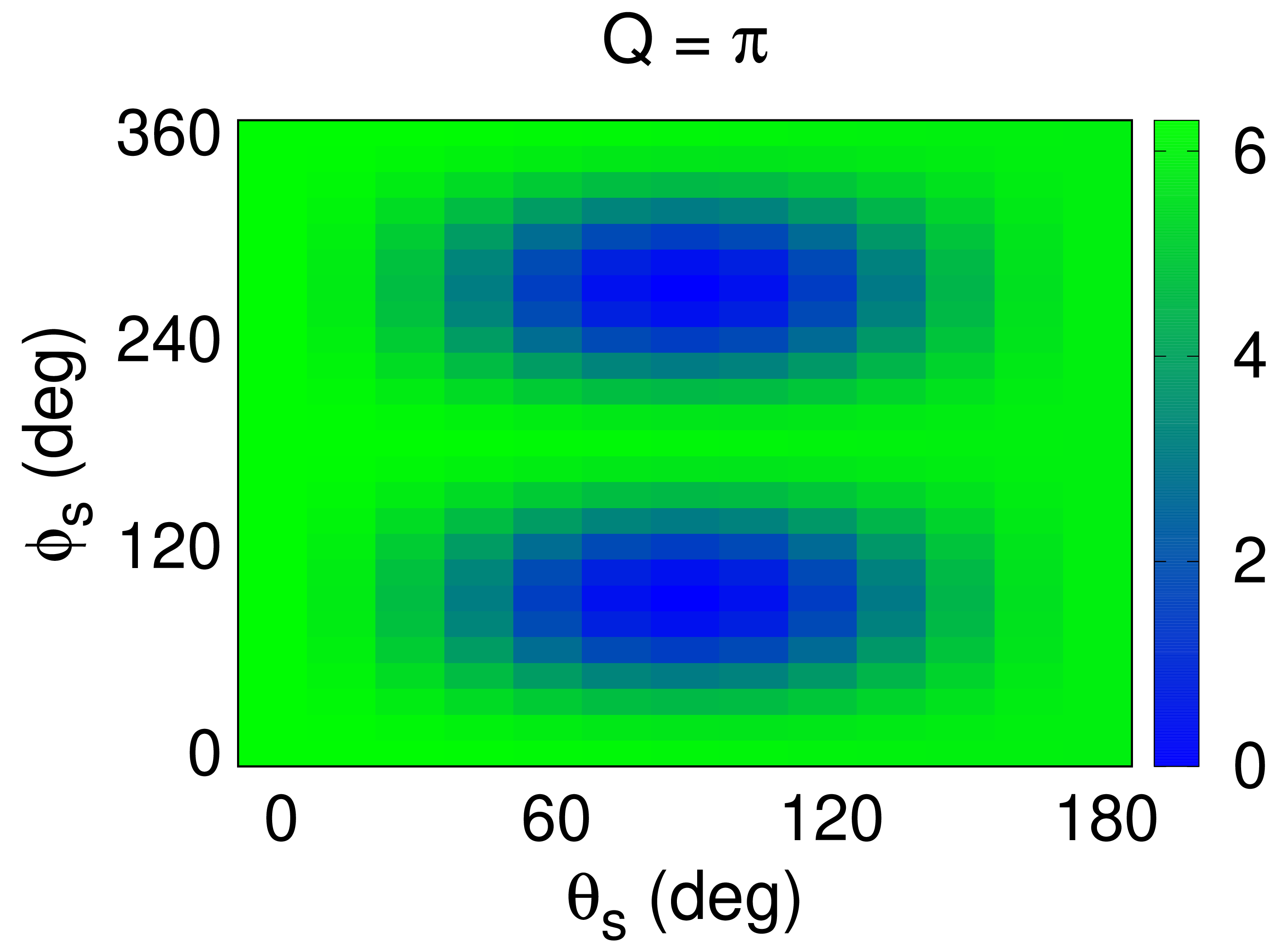}
  \label{Fig_DCS_Aconly_pi}
\end{subfigure}
\hspace{0.1cm}
\begin{subfigure}{0.23\linewidth}
  \includegraphics[width=\linewidth]{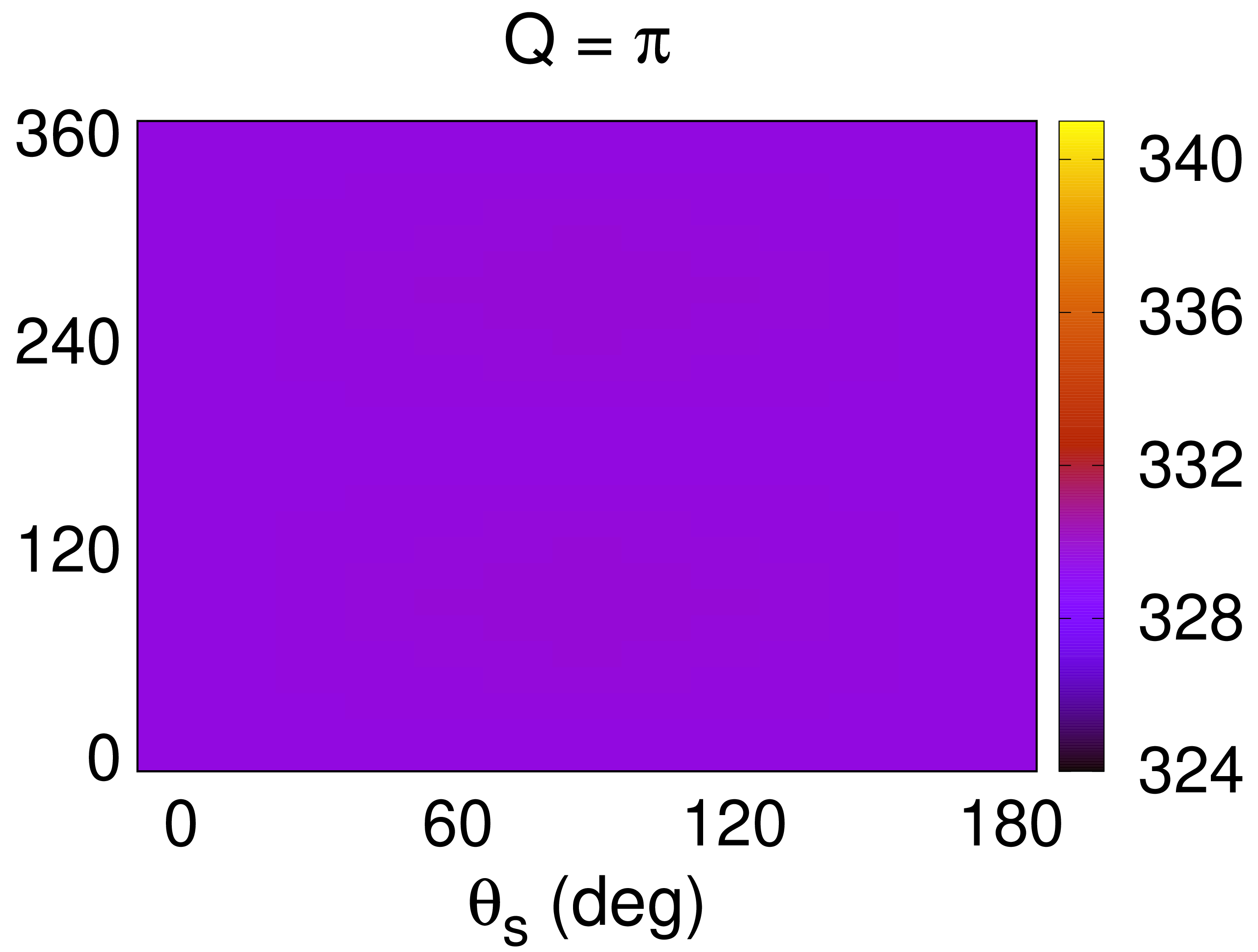}
  \label{Fig_DCS_AcOFF_pi}
\end{subfigure} 
\hspace{0.1cm}
\begin{subfigure}{0.23\linewidth}
  \includegraphics[width=\linewidth]{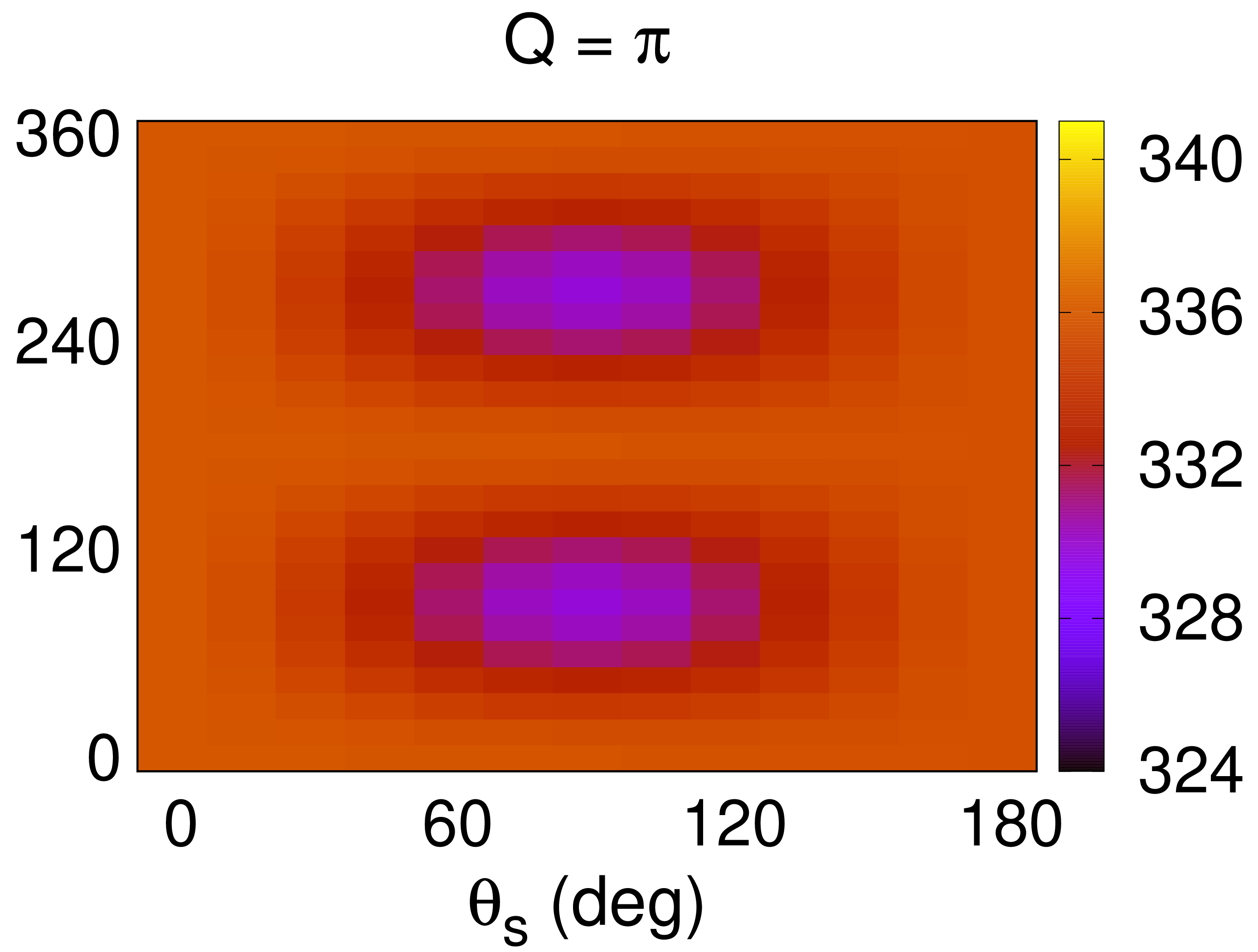}
  \label{Fig_DCS_incoherentsum_pi}
\end{subfigure} 
\hspace{0.1cm}
\begin{subfigure}{0.25\linewidth}
  \includegraphics[width=\linewidth]{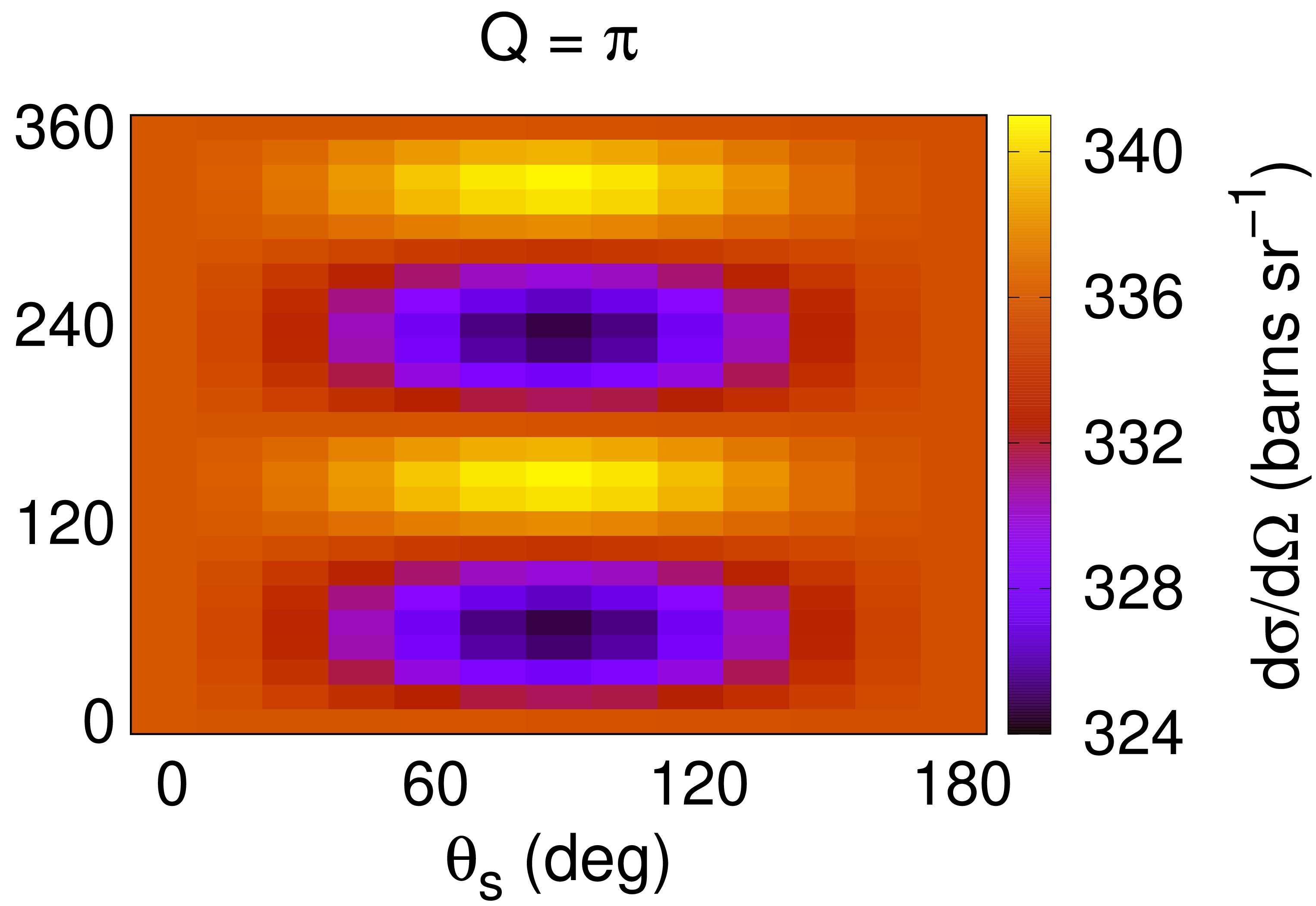}
  \label{Fig_dcs_pi}
\end{subfigure} 

\medskip 
\hspace{-0.3cm}
\begin{subfigure}{.23\linewidth}
  \includegraphics[width=\linewidth]{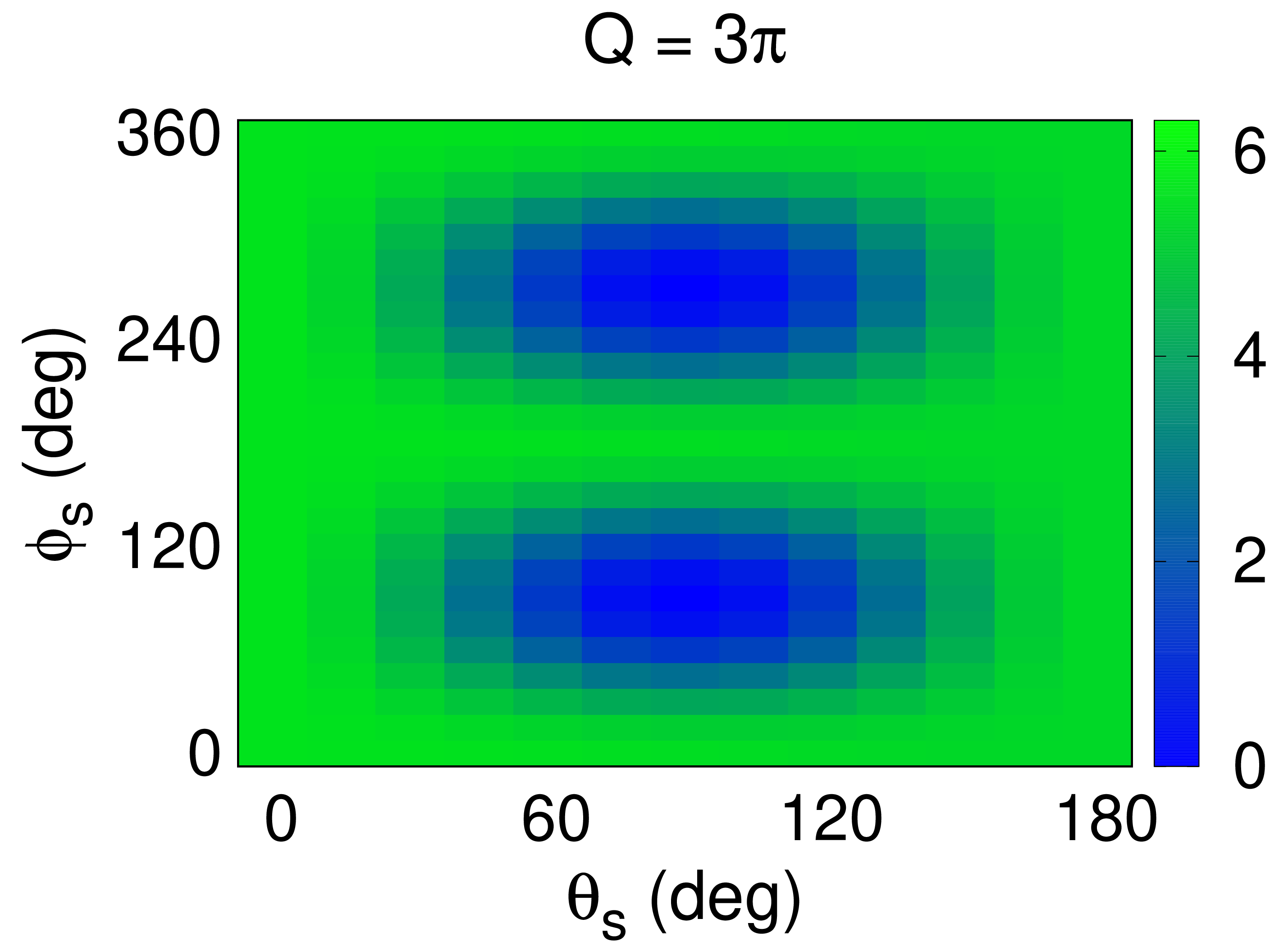}
  \label{Fig_DCS_Aconly_3pi}
\end{subfigure}
\hspace{0.1cm}
\begin{subfigure}{.23\linewidth}
  \includegraphics[width=\linewidth]{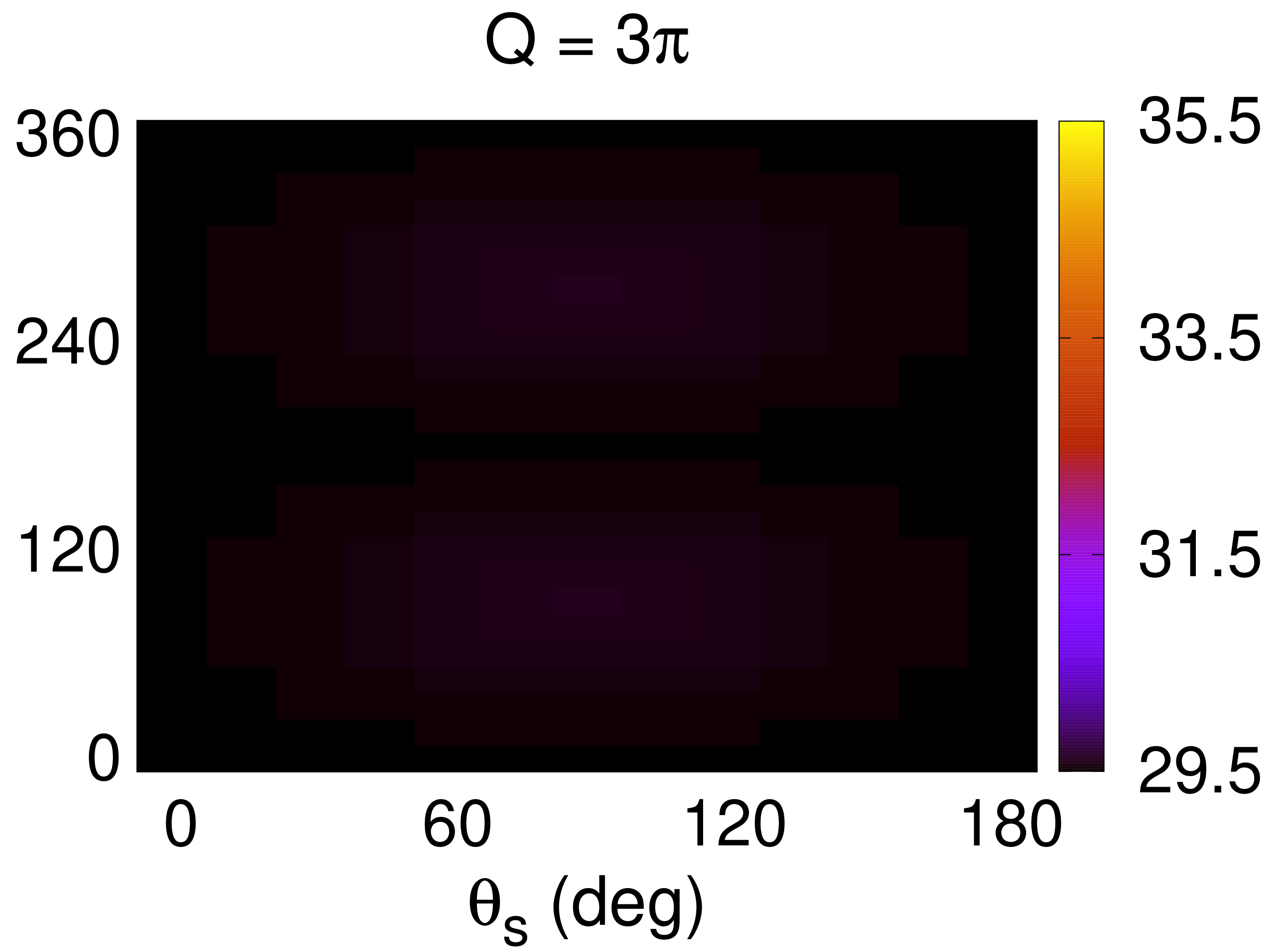}
  \label{Fig_DCS_AcOFF_3pi}
\end{subfigure} 
\hspace{0.1cm}
\begin{subfigure}{.23\linewidth}
  \includegraphics[width=\linewidth]{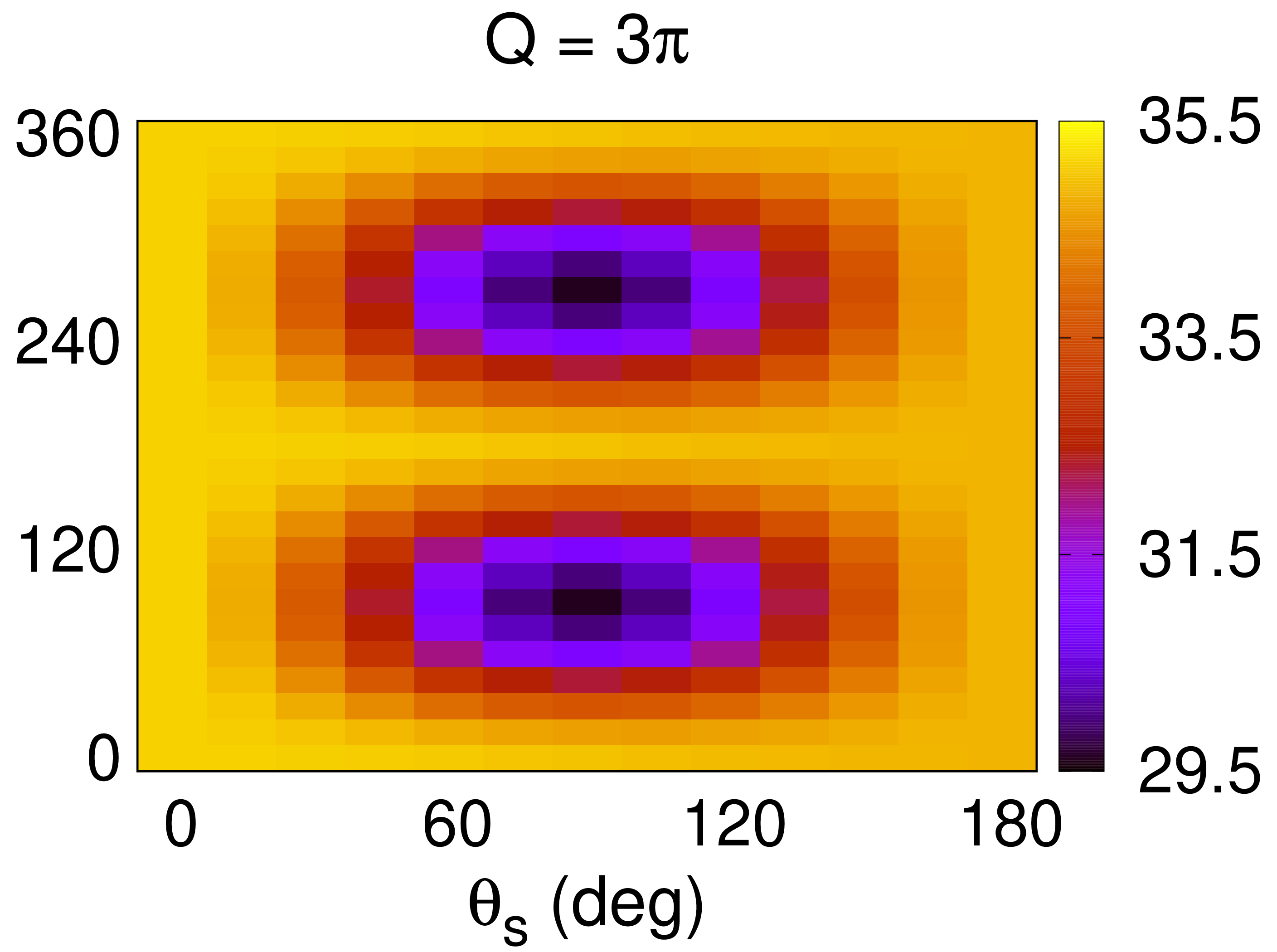}
  \label{Fig_DCS_incoherentsum_3pi}
\end{subfigure} 
\hspace{0.1cm}
\begin{subfigure}{.255\linewidth}
  \includegraphics[width=\linewidth]{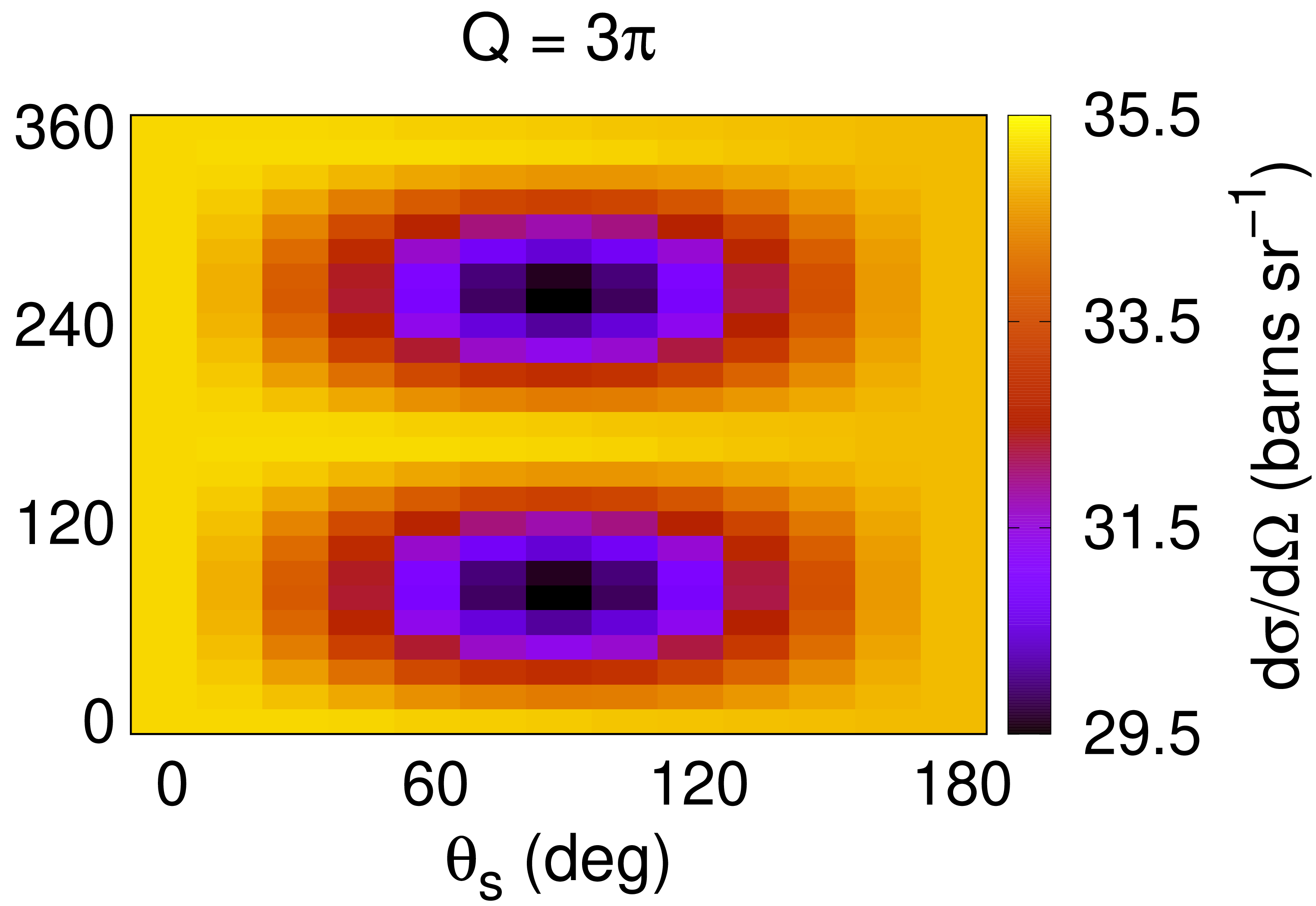}
  \label{Fig_dcs_3pi}
\end{subfigure} 

\medskip 
\hspace{-0.3cm}
\begin{subfigure}{.235\linewidth}
  \includegraphics[width=\linewidth]{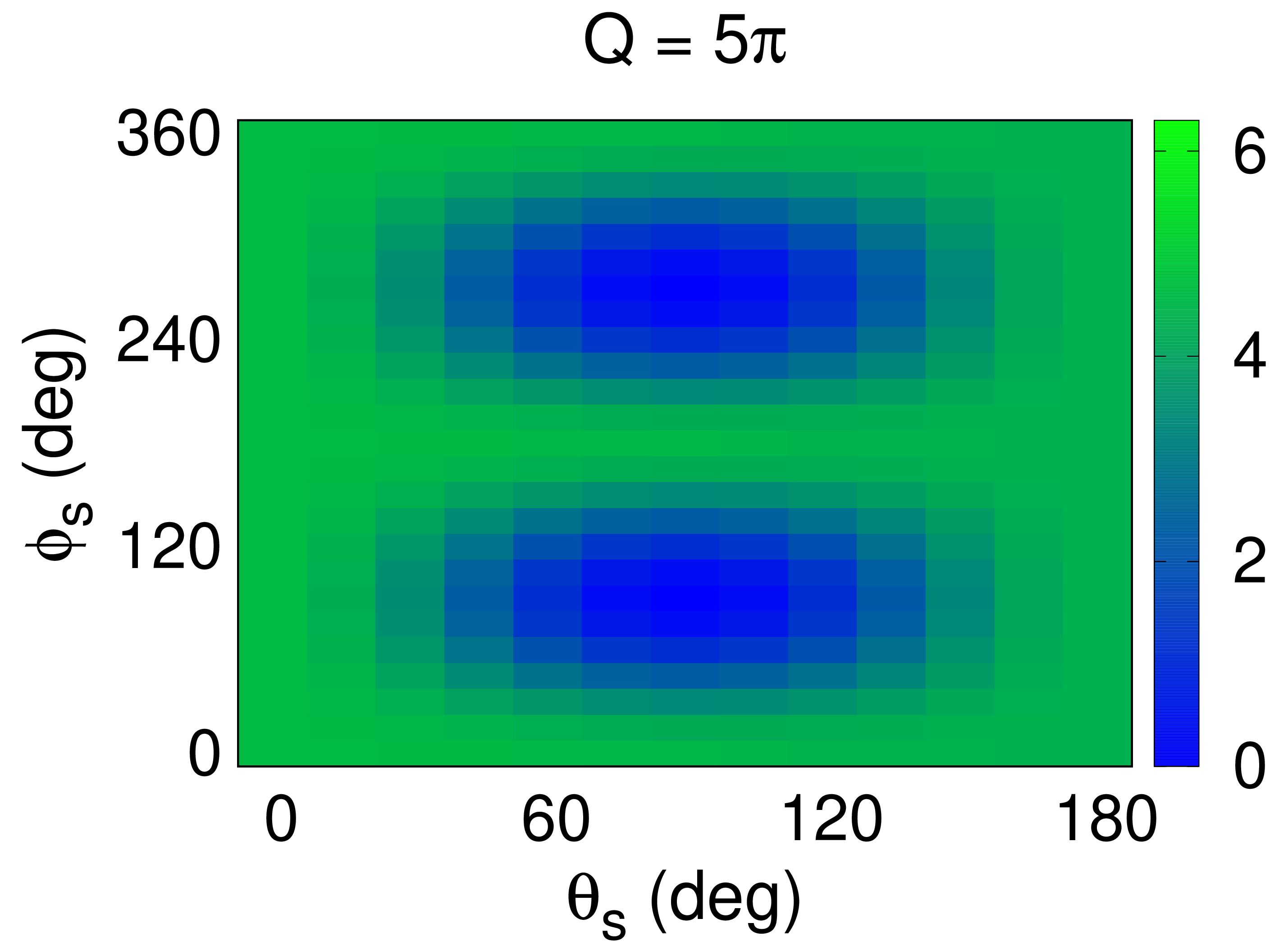}
  \caption{}
  \label{Fig_DCS_Aconly_5pi}
\end{subfigure}
\hspace{0.1cm}
\begin{subfigure}{.23\linewidth}
    \includegraphics[width=\linewidth]{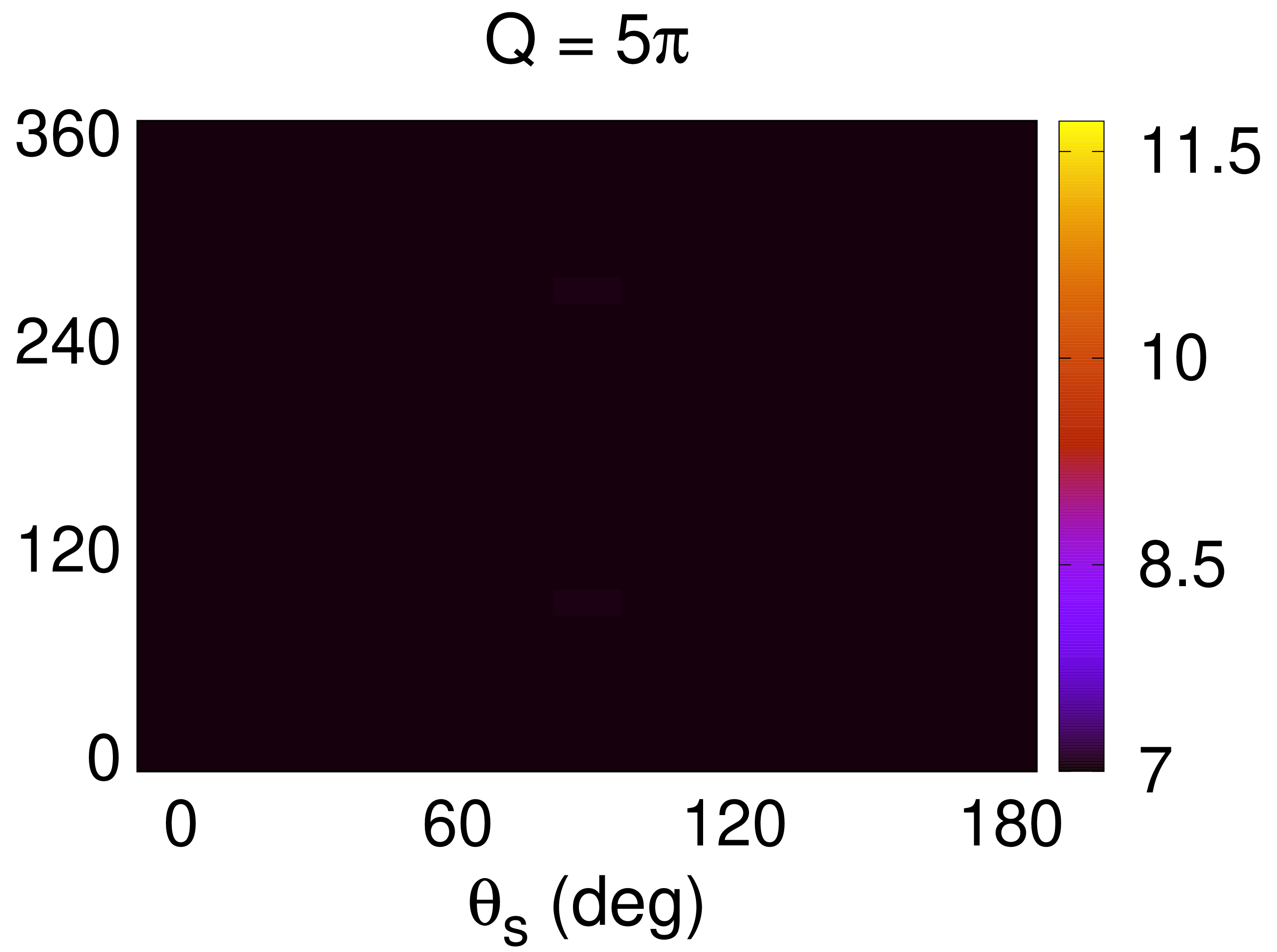}
  \caption{}
  \label{Fig_DCS_AcOFF_5pi}
\end{subfigure} 
\hspace{0.1cm}
\begin{subfigure}{.23\linewidth}
    \includegraphics[width=\linewidth]{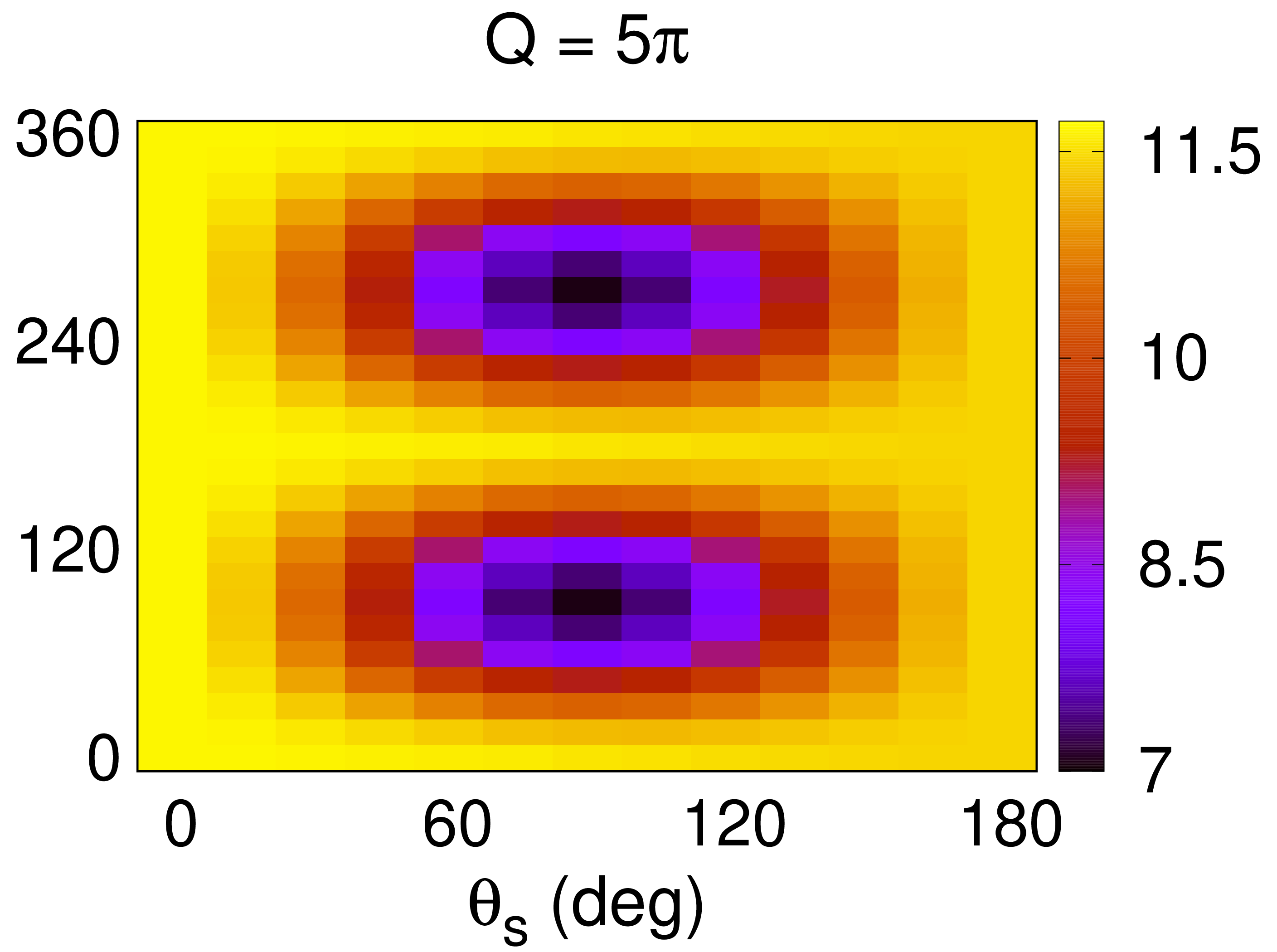}
  \caption{}
  \label{Fig_DCS_incoherentsum_5pi}
\end{subfigure} 
\hspace{0.1cm}
\begin{subfigure}{.255\linewidth}
    \includegraphics[width=\linewidth]{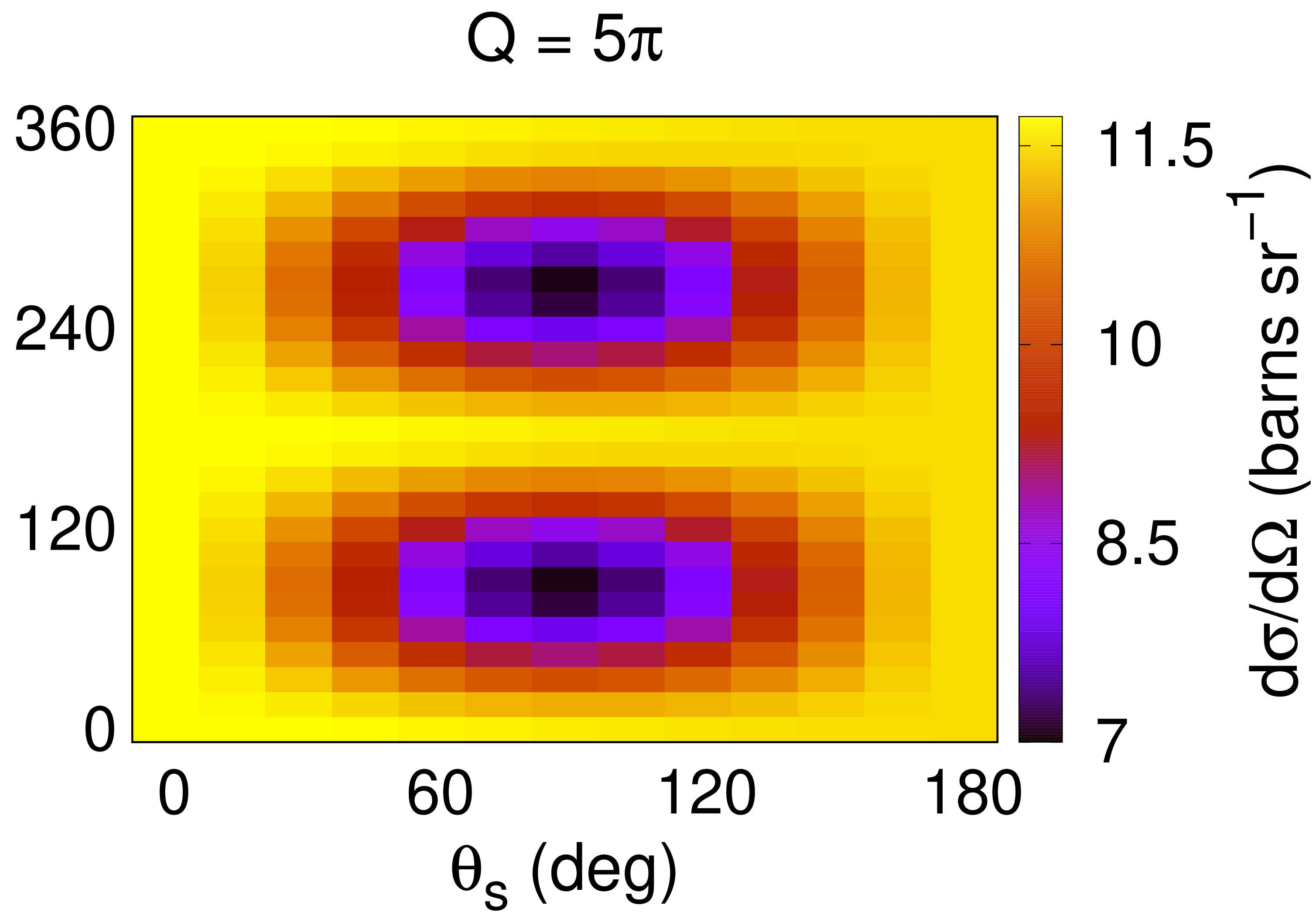}
  \caption{}
  \label{Fig_dcs_5pi}
\end{subfigure}\hfill 
\end{minipage}
\caption{Angular dependence of the DCS calculated for $\pi$-type pulses of different pulse area Q from (first column) elastic scattering, (second column) resonant fluorescence, (third column) incoherent sum and (fourth column) coherent sum. The incoherent and coherent sums are obtained using the sum of the elastic scattering and resonant fluorescence probabilities and amplitudes, respectively. In all the above plots, $\omega_{in} = 849.8$ eV, $t_{wid} = 0.25$~fs. For each pulse area, the range of the color bars is chosen based on the minimum and maximum of the coherent sum, except for the elastic channel, in which a constant range has been chosen. Note that the black color (Res. Fluor) corresponds to the minimum of that color map.
}
\label{Fig_dcs_channels_Pi}
\end{figure*}






\begin{figure*} 
\begin{minipage}[t][][t]{\textwidth}
       \hspace{0.2cm}
       \textbf{Elastic}
       \hspace{2.7cm}
       \textbf{Res. Fluor.}
       \hspace{1.8cm}
       \textbf{Incoherent sum}
       \hspace{1.6cm}
       \textbf{Coherent sum}
\end{minipage}
\begin{minipage}[b][][b]{\textwidth}
\vspace{0.3cm}
\begin{subfigure}{.25\linewidth}
  \includegraphics[width=\linewidth]{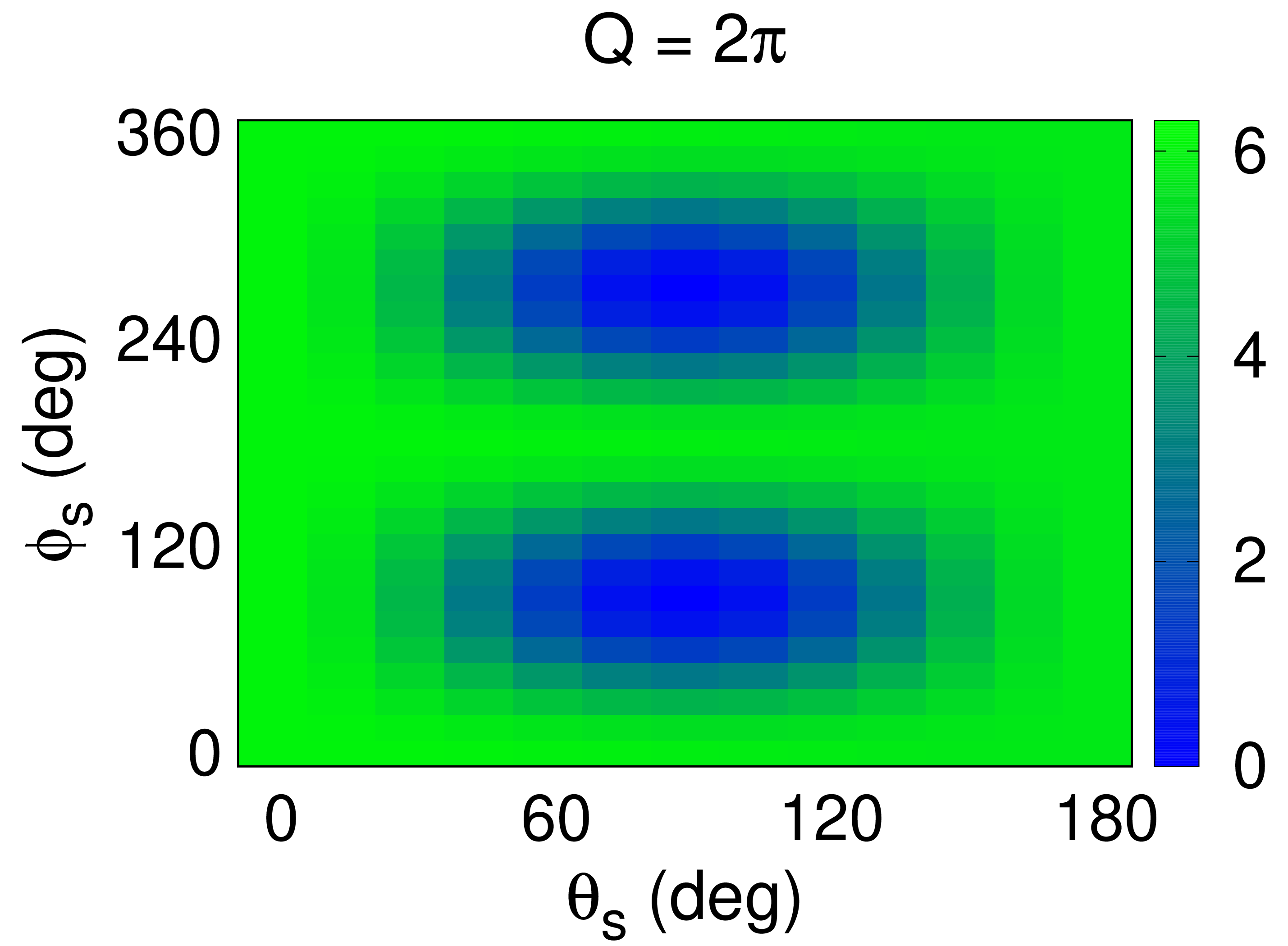}
  \label{Fig_DCS_Aconly_2pi}
\end{subfigure}
\begin{subfigure}{.235\linewidth}
  \includegraphics[width=\linewidth]{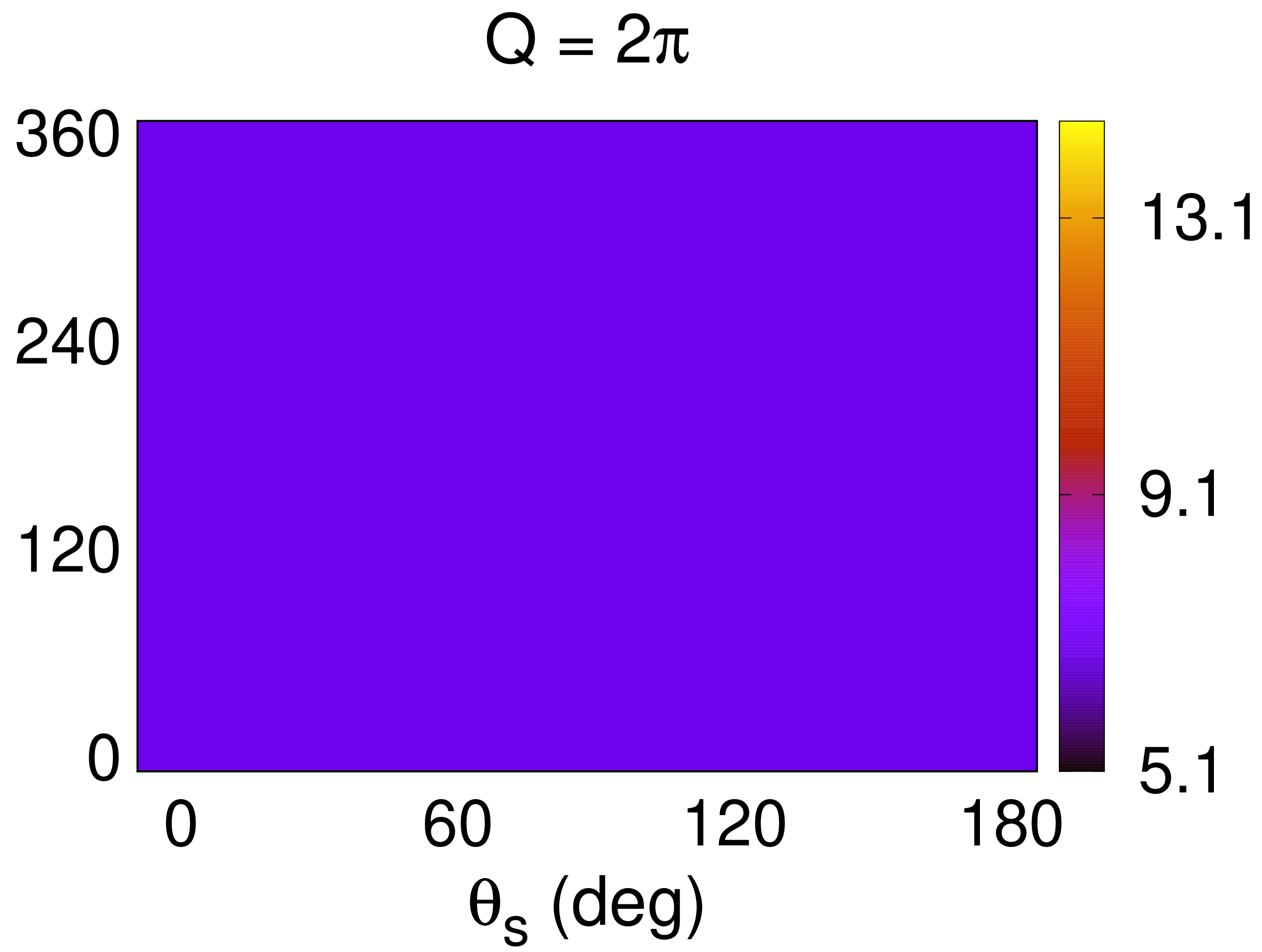}
  \label{Fig_DCS_AcOFF_2pi}
\end{subfigure} 
\begin{subfigure}{.235\linewidth}
  \includegraphics[width=\linewidth]{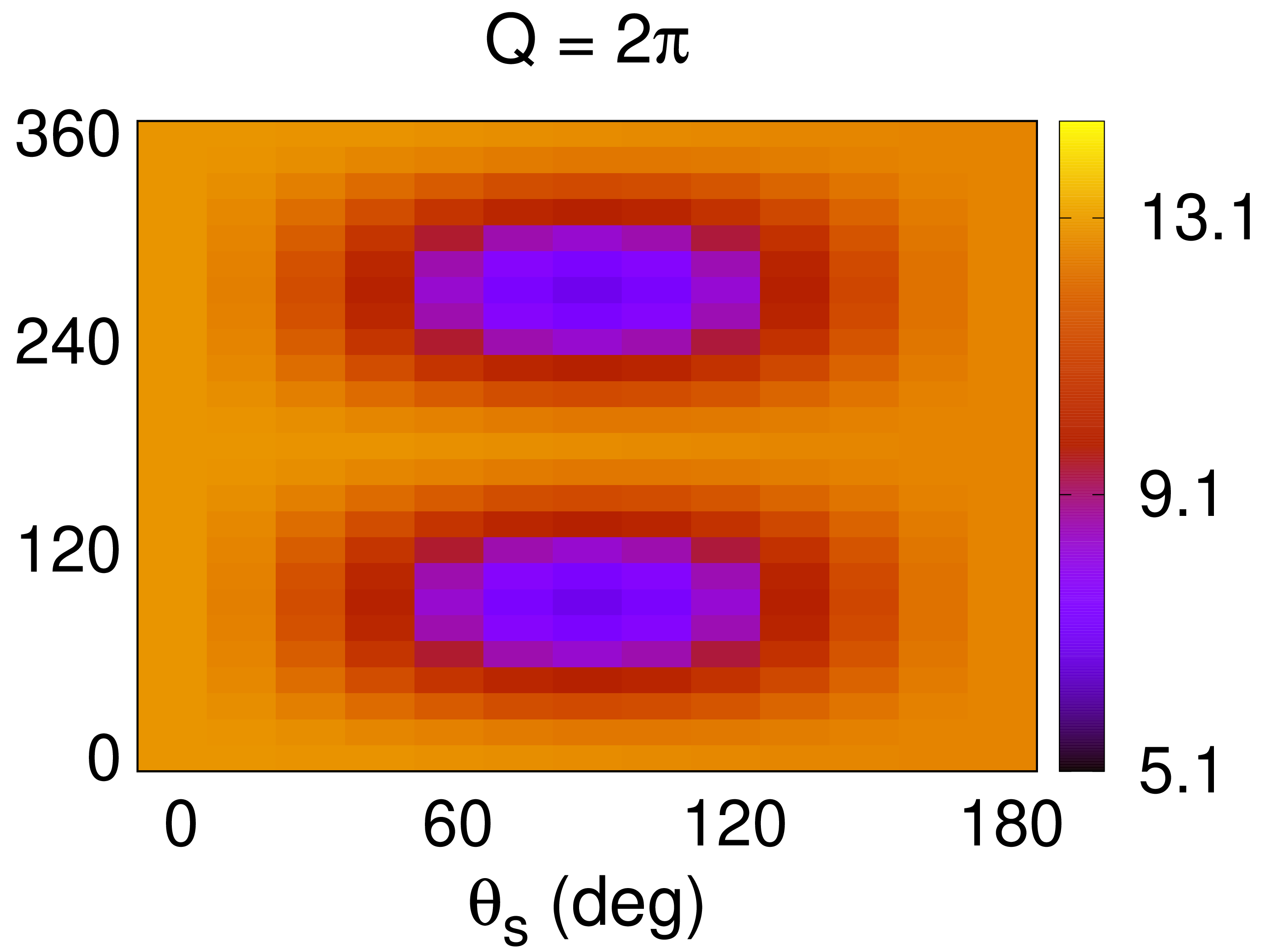}
  \label{Fig_DCS_incoherentsum_2pi}
\end{subfigure}
\begin{subfigure}{.25\linewidth}
  \includegraphics[width=\linewidth]{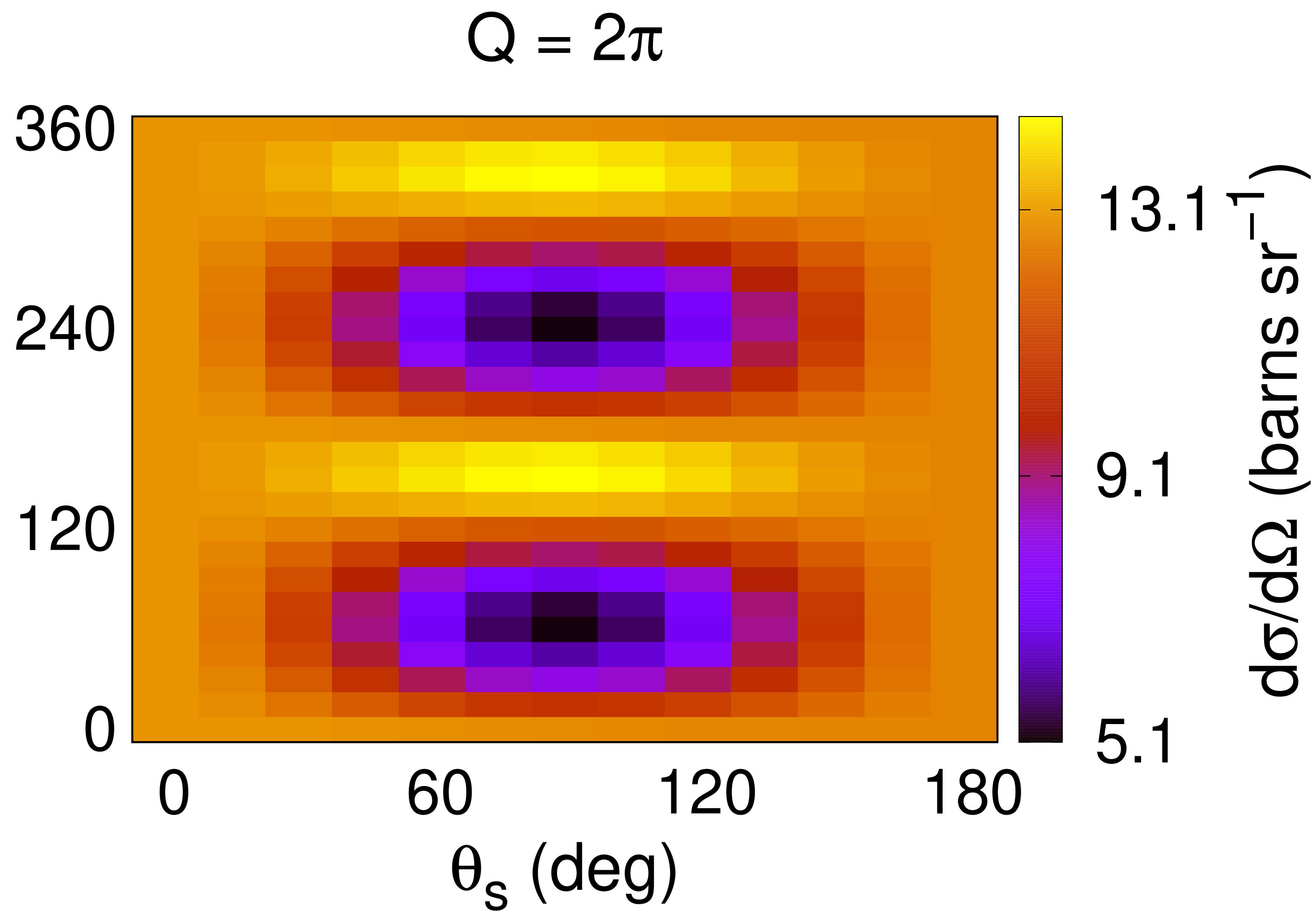}
  \label{Fig_dcs_2pi}
\end{subfigure}

\medskip 
\begin{subfigure}{.25\linewidth}
  \includegraphics[width=\linewidth]{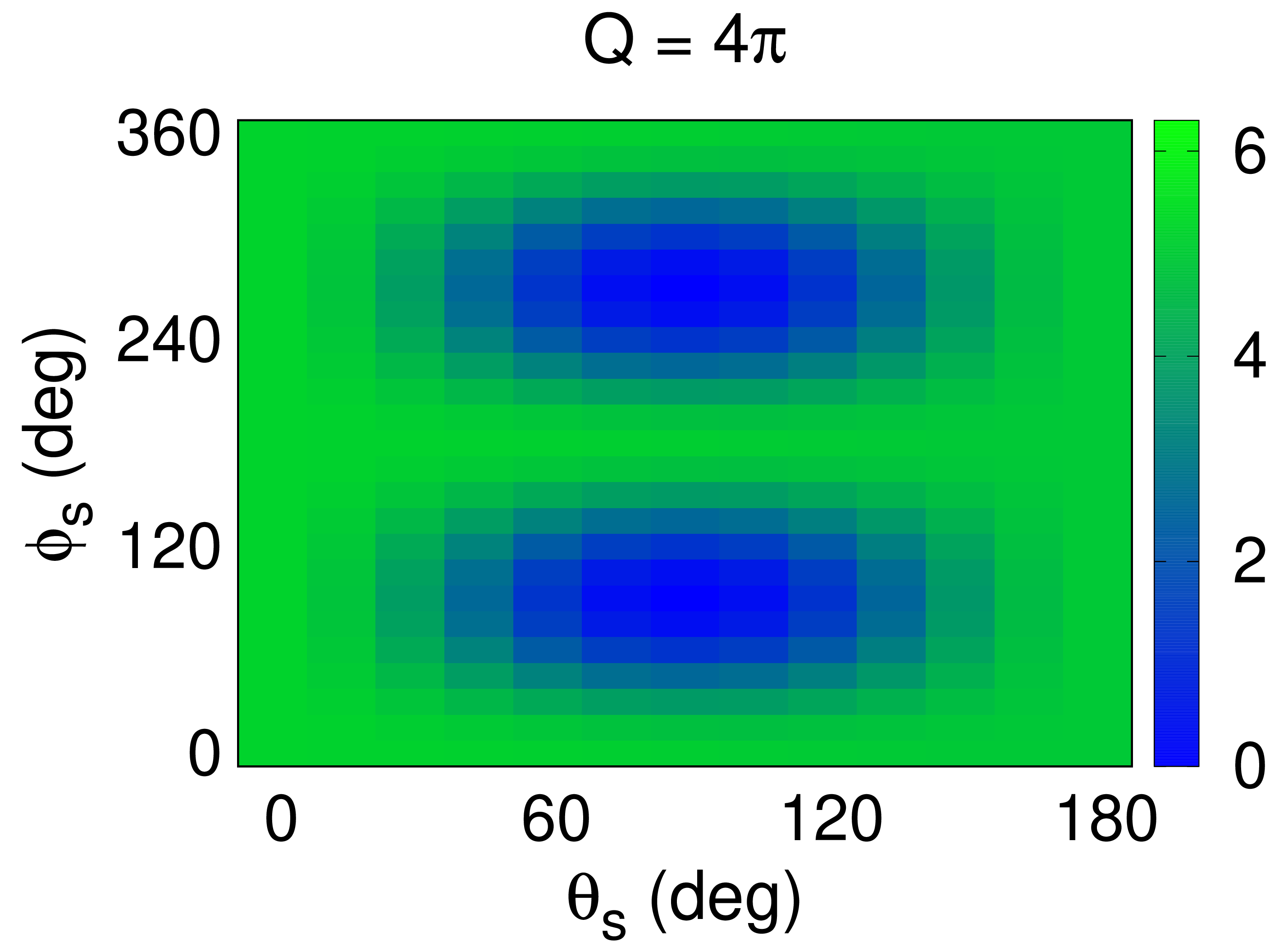}
  \label{Fig_DCS_Aconly_4pi}
\end{subfigure}
\begin{subfigure}{.235\linewidth}
  \includegraphics[width=\linewidth]{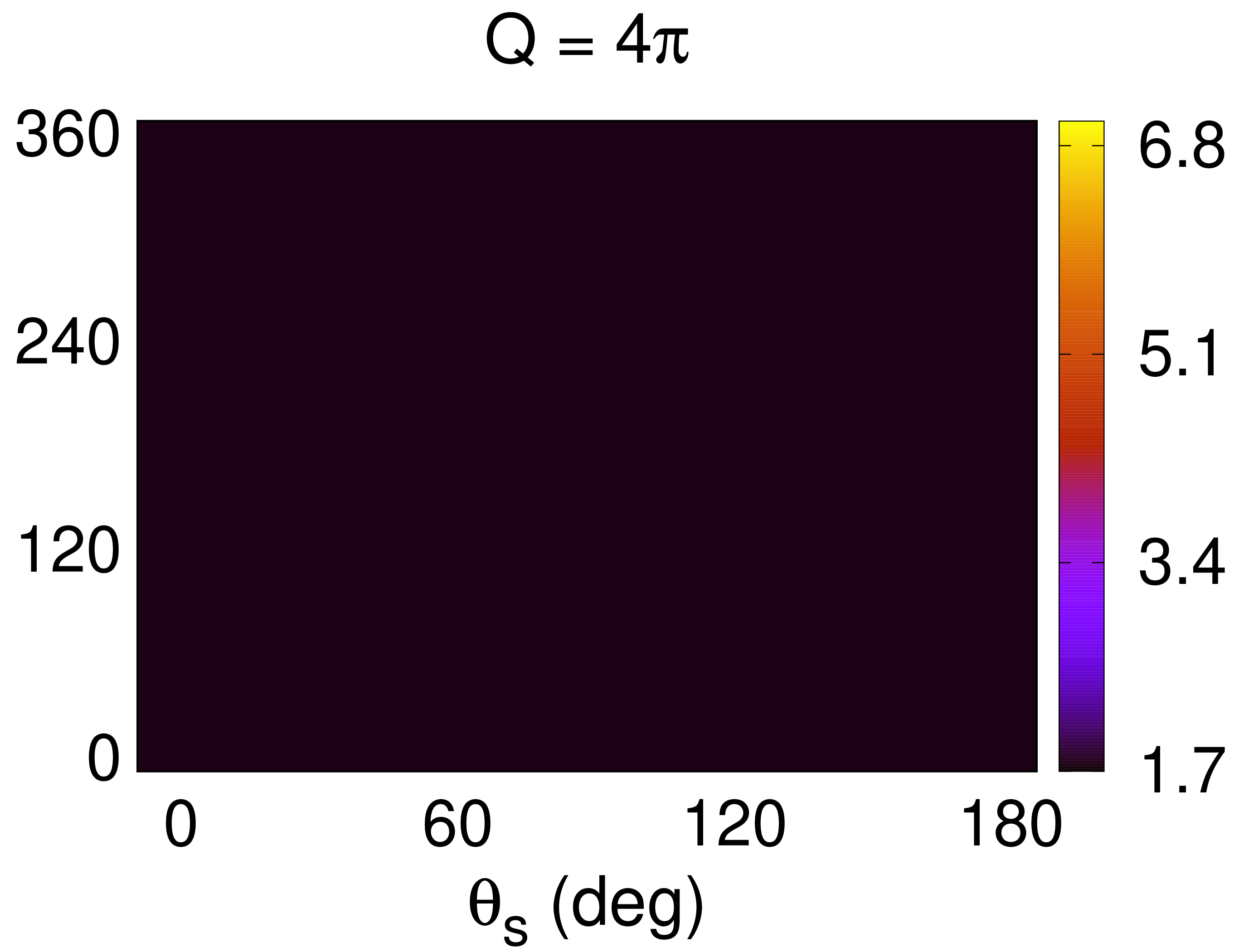}
  \label{Fig_DCS_AcOFF_4pi}
\end{subfigure} 
\begin{subfigure}{.23\linewidth}
  \includegraphics[width=\linewidth]{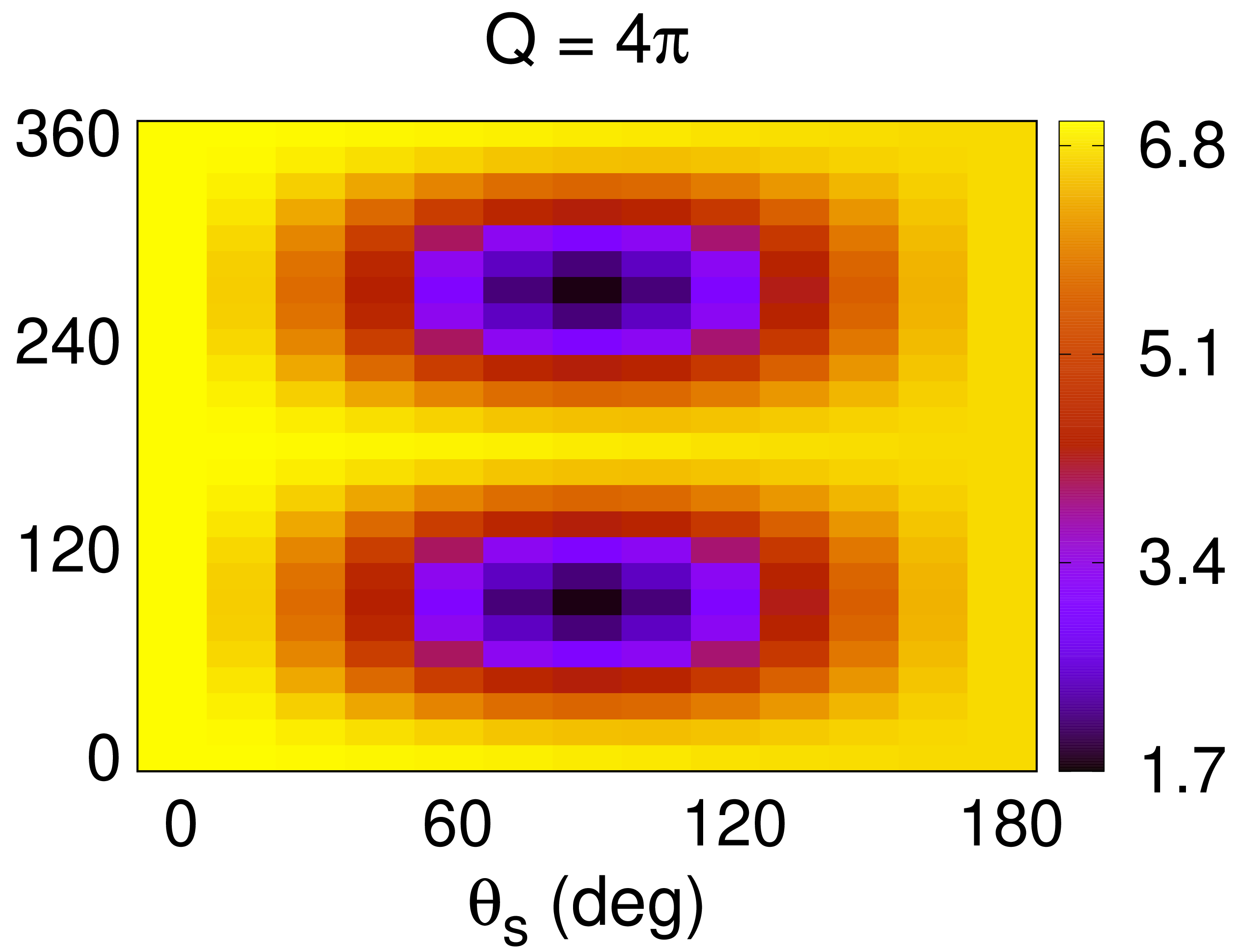}
  \label{Fig_DCS_incoherentsum_4pi}
\end{subfigure}
\begin{subfigure}{.255\linewidth}
  \includegraphics[width=\linewidth]{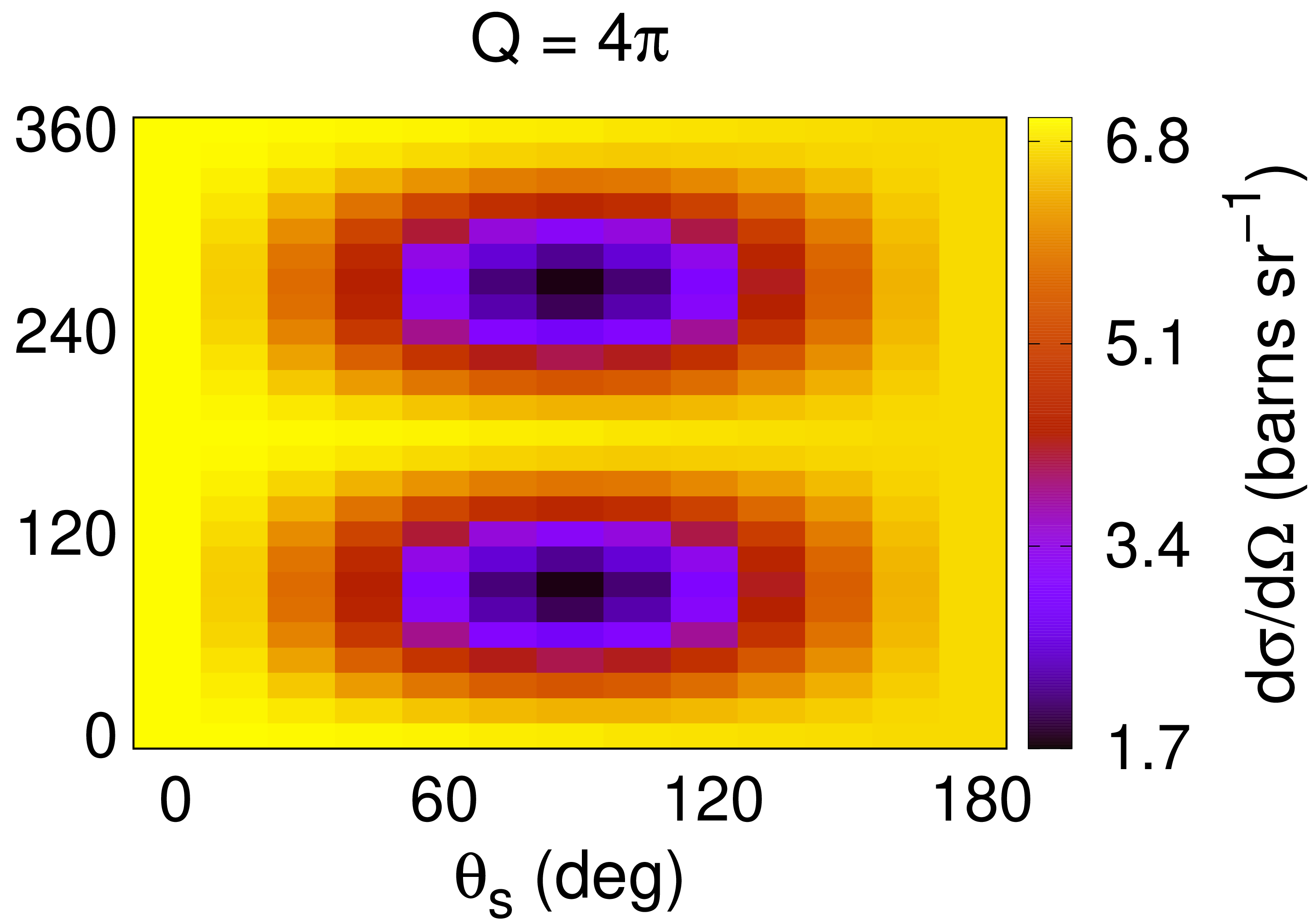}
  \label{Fig_dcs_4pi}
\end{subfigure}

\medskip 
\begin{subfigure}{.25\linewidth}
  \includegraphics[width=\linewidth]{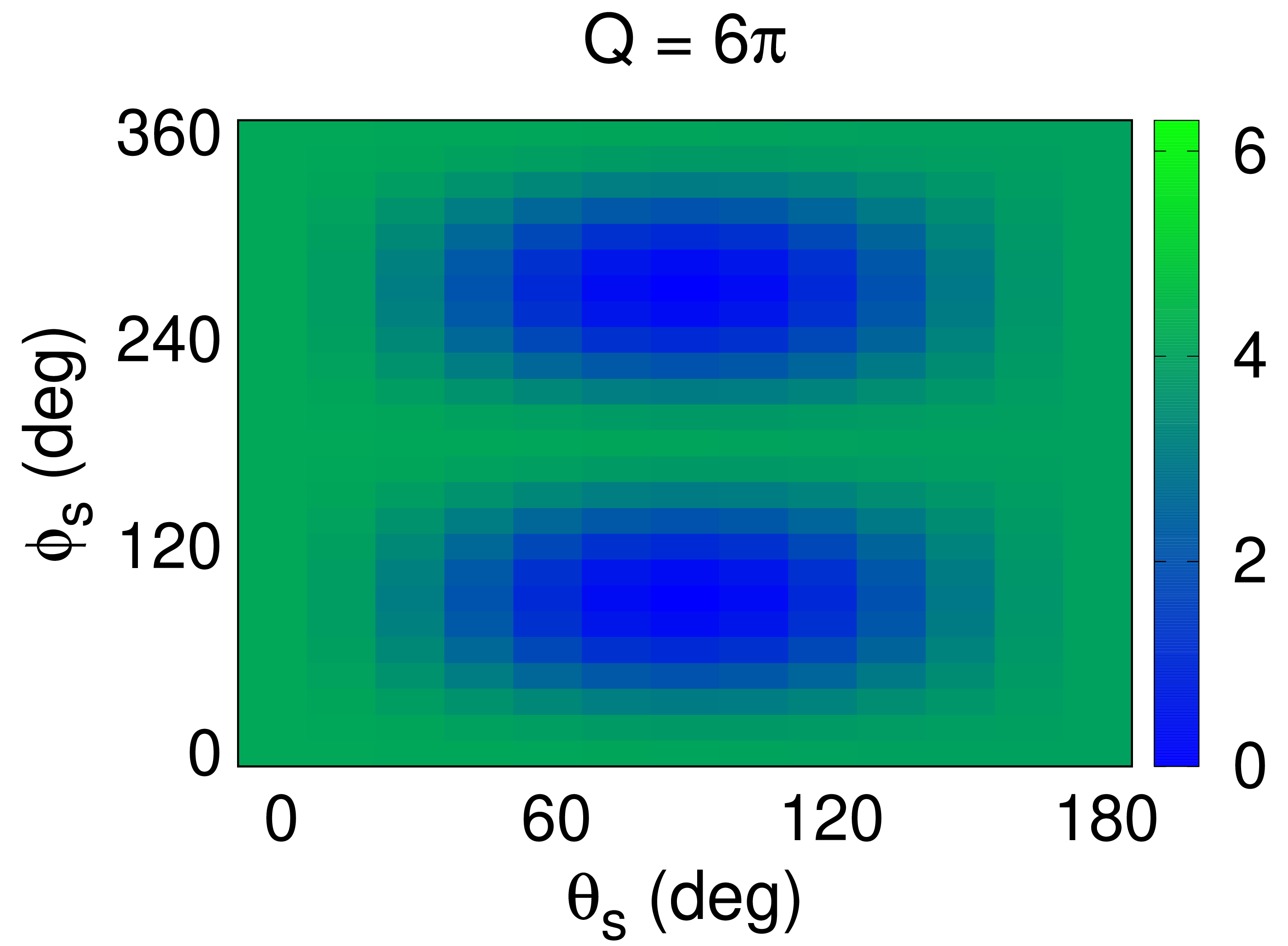}
  \caption{}
  \label{Fig_DCS_Aconly_6pi}
\end{subfigure}
\begin{subfigure}{.24\linewidth}
    \includegraphics[width=\linewidth]{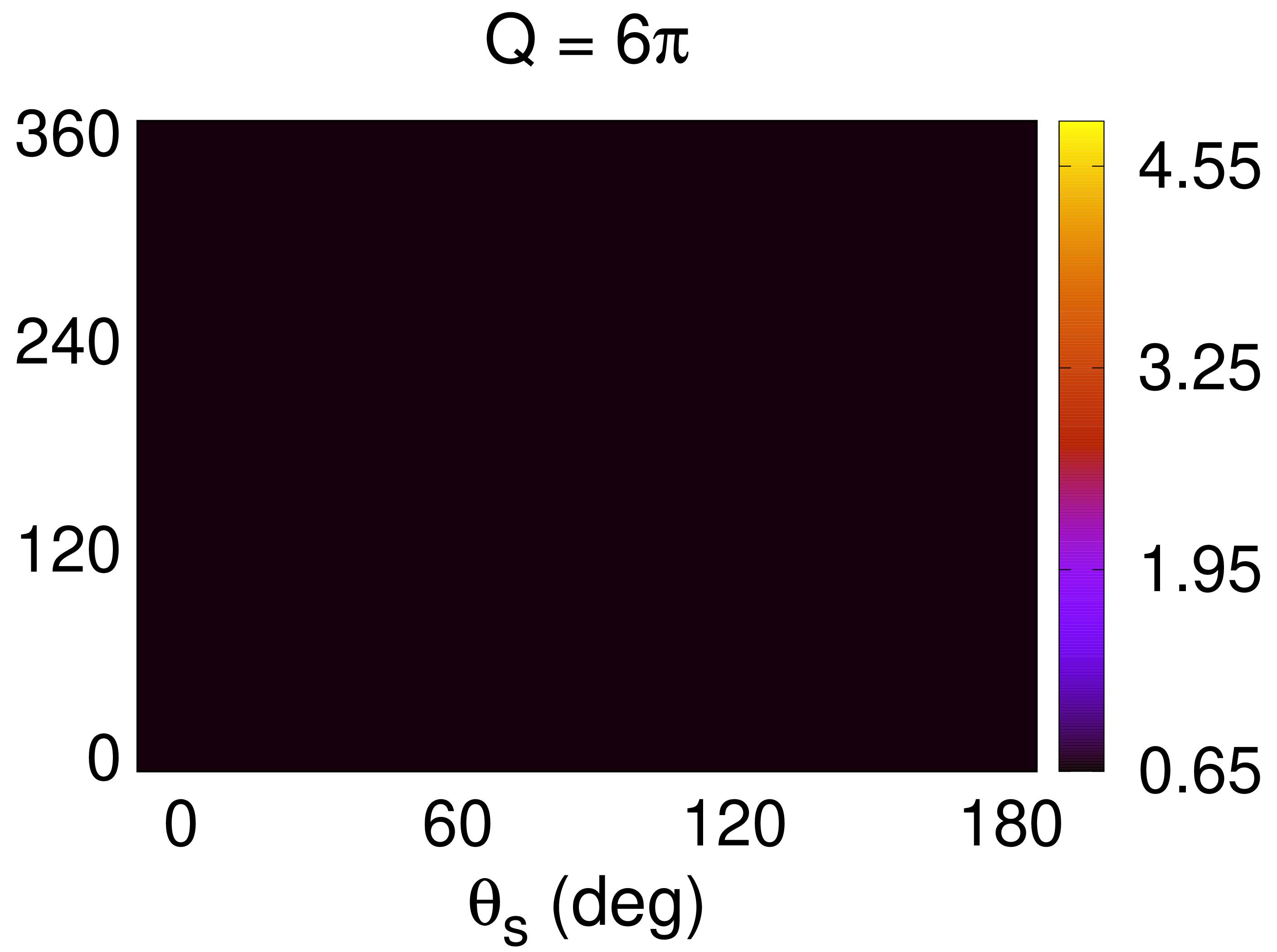}
  \caption{}
  \label{Fig_DCS_AcOFF_6pi}
\end{subfigure} 
\begin{subfigure}{.23\linewidth}
  \includegraphics[width=\linewidth]{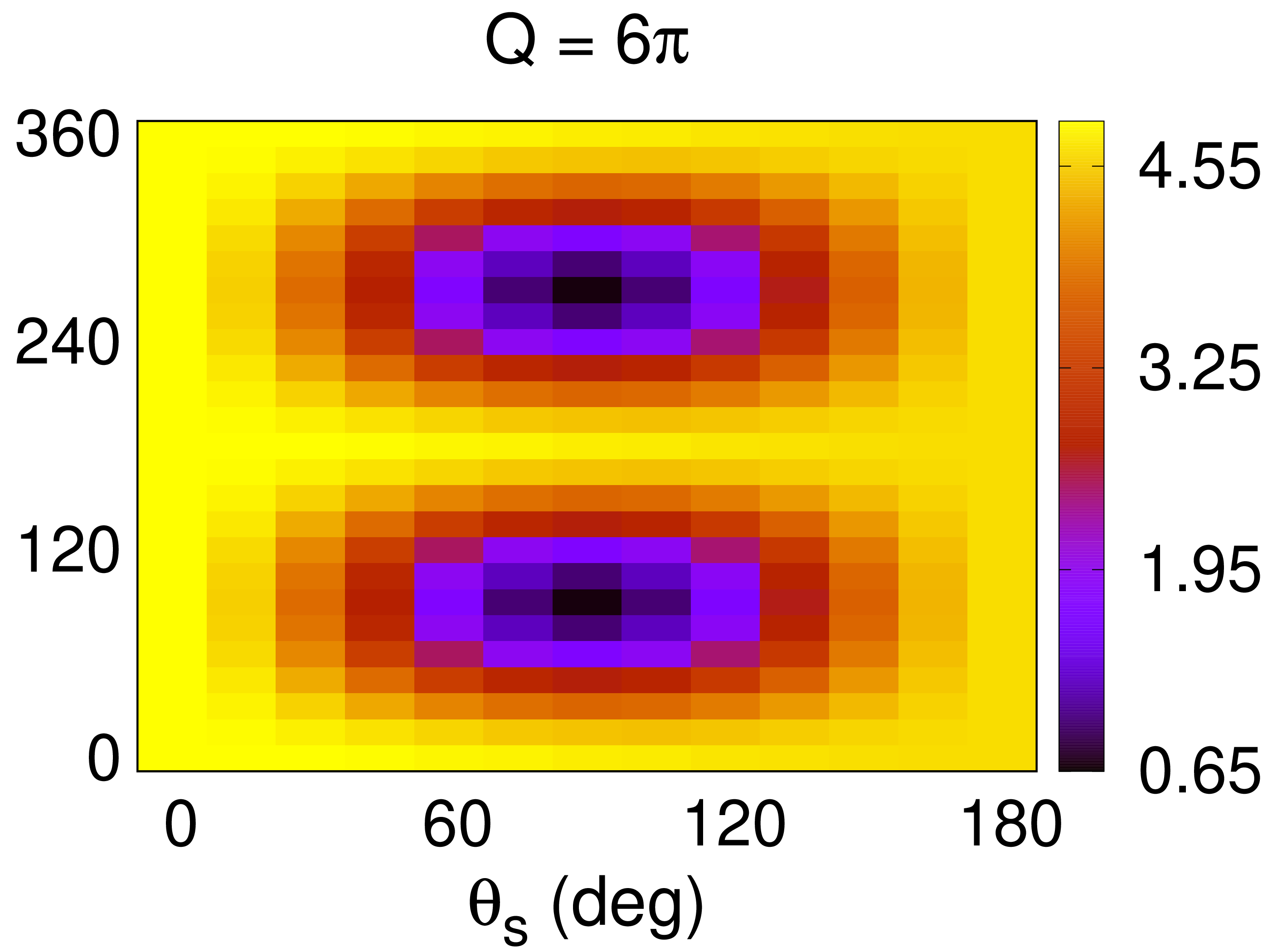}
  \caption{}
  \label{Fig_DCS_incoherentsum_6pi}
\end{subfigure}
\begin{subfigure}{.255\linewidth}
  \includegraphics[width=\linewidth]{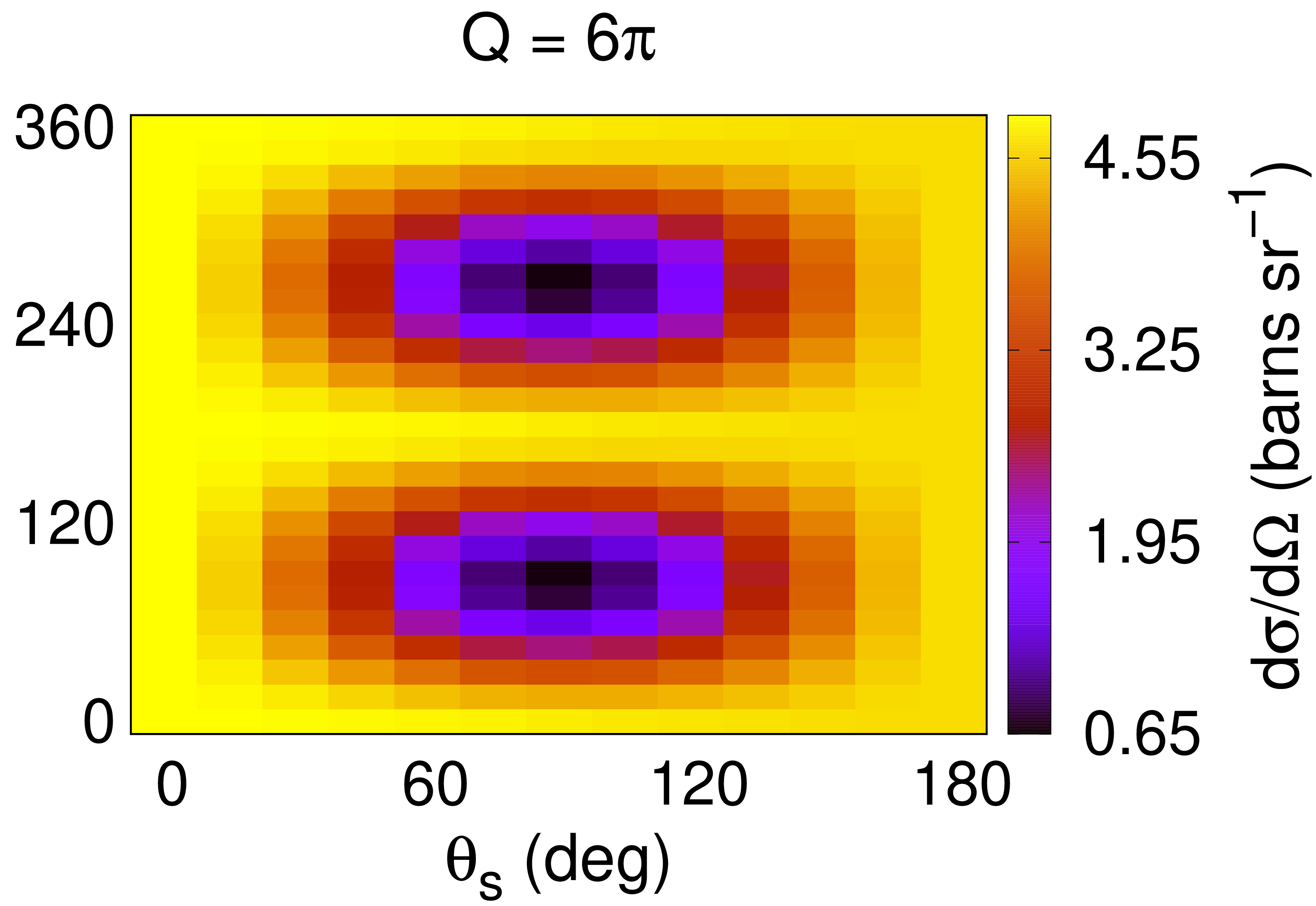}
  \caption{}
  \label{Fig_dcs_6pi}
\end{subfigure}
\end{minipage}
\caption{
Angular dependence of the DCS calculated for $2\pi$-type pulses of different pulse area Q from (first column) elastic scattering, (second column) resonant fluorescence, (third column) incoherent sum and (fourth column) coherent sum. The other parameters are the same as in Fig.~\ref{Fig_dcs_channels_Pi}.
}
\label{Fig_dcs_channels_2Pi}
\end{figure*}


\begin{figure}
\resizebox{90mm}{!}{\includegraphics{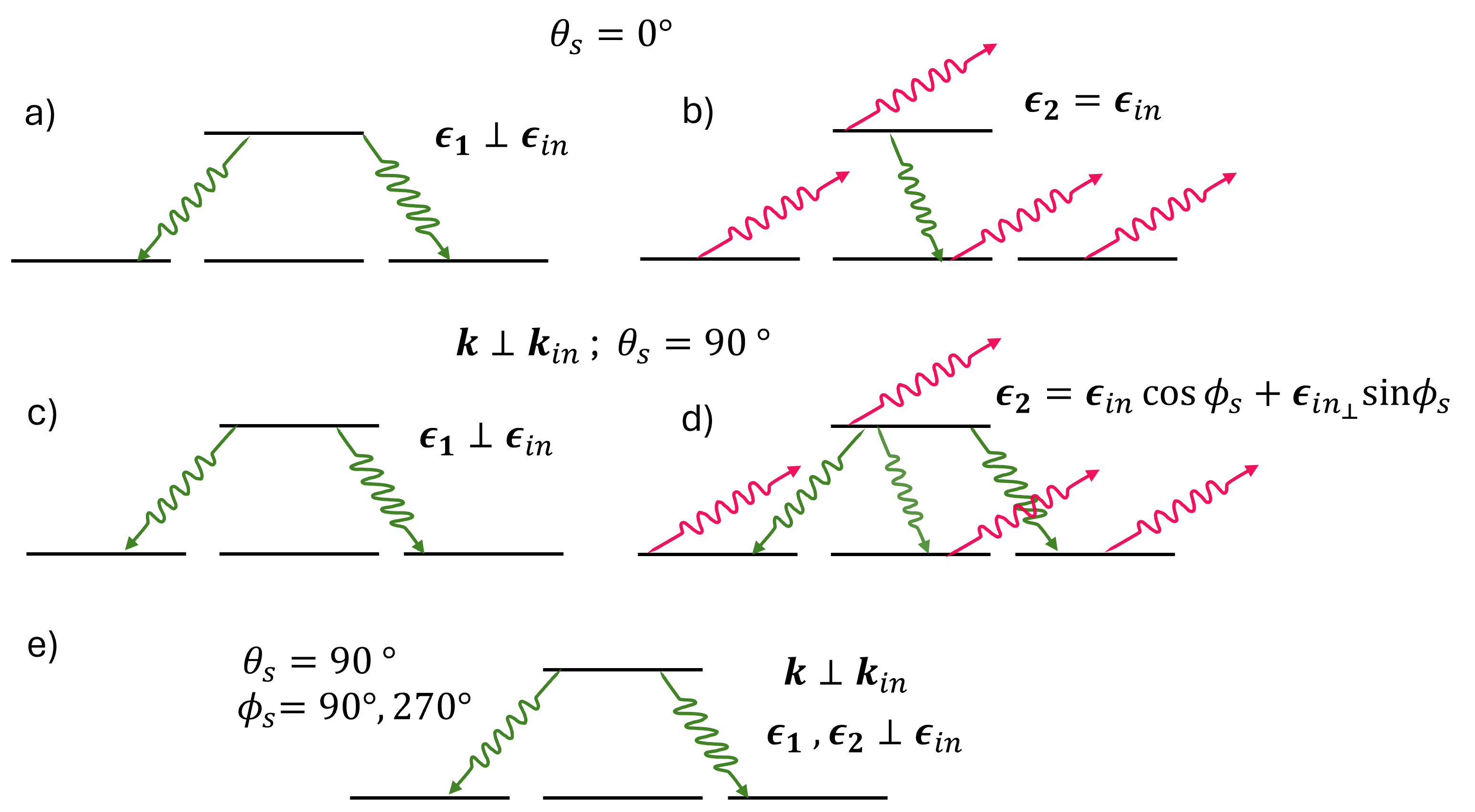}}
\caption{\label{Channels_Interference_scheme}
Schematic diagram showing the pathways for interference between elastic scattering and resonant fluorescence for different scattering angles and scattered photon polarizations. Each row of diagrams is for a particular choice of scattering angle. Each scattering angle allows for two possible choices of outgoing photon polarization ($\boldsymbol{\epsilon_1}, \boldsymbol{\epsilon_2}$) for which the allowed channels are shown in the sketch. The green squiggly lines represent resonant fluorescence from transition to that particular final state (same as Fig.~\ref{Schematic_diagram}). The red squiggly lines represent elastic Thomson scattering from the bound electrons in that state.
}
\end{figure}


\begin{figure*} [hbt!] 
\begin{minipage}[b][][b]{\textwidth}
\begin{subfigure}{.24\linewidth}
  \includegraphics[width=\linewidth]{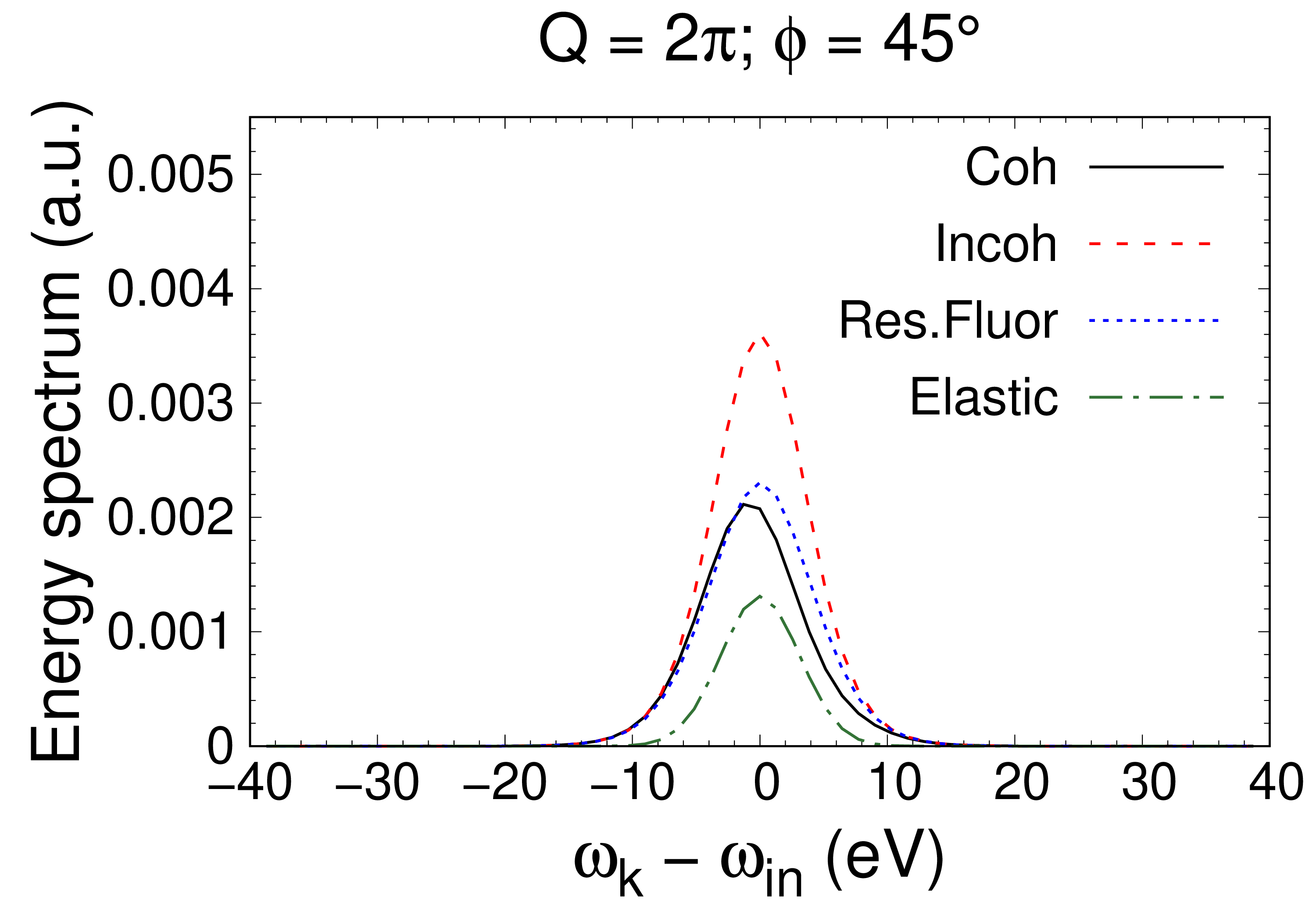}
  \label{Fig_DDSP_p25fs_2pi_90_45}
\end{subfigure}
\begin{subfigure}{.24\linewidth}
  \includegraphics[width=\linewidth]{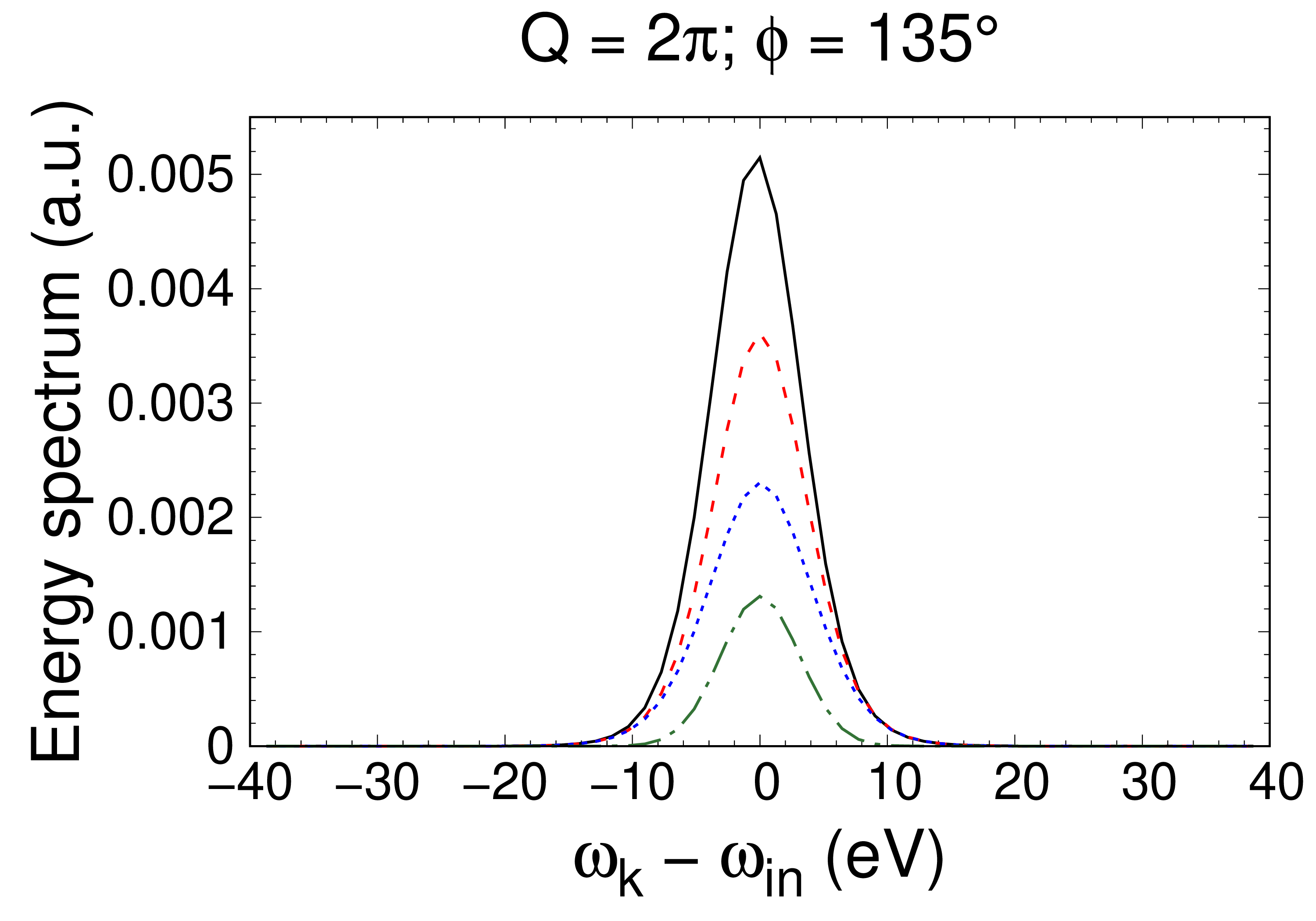}
  \label{Fig_DDSP_p25fs_2pi_90_135}
\end{subfigure} 
\begin{subfigure}{.24\linewidth}
  \includegraphics[width=\linewidth]{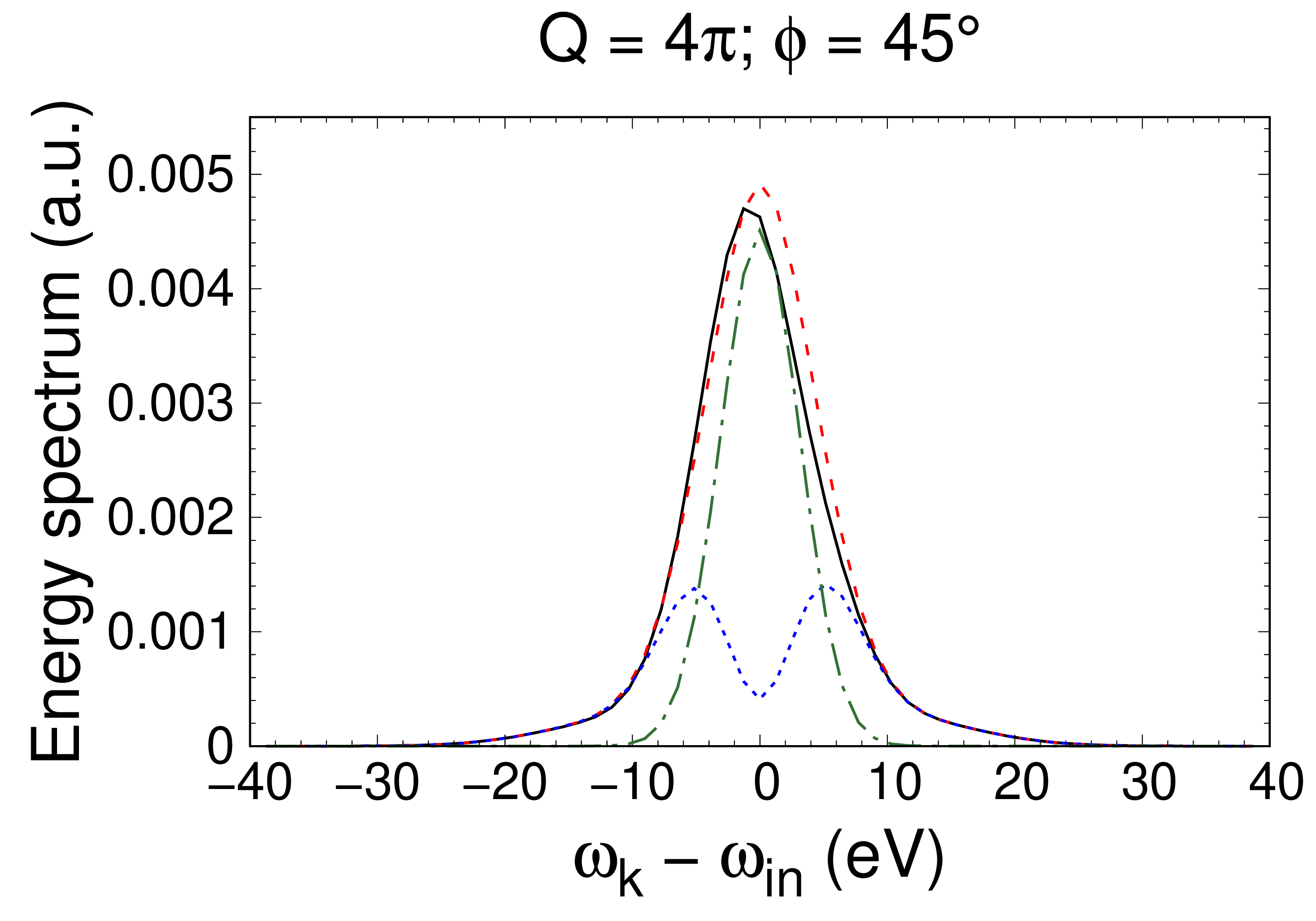}
  \label{Fig_DDSP_p25fs_4pi_90_45}
\end{subfigure}
\begin{subfigure}{.24\linewidth}
    \includegraphics[width=\linewidth]{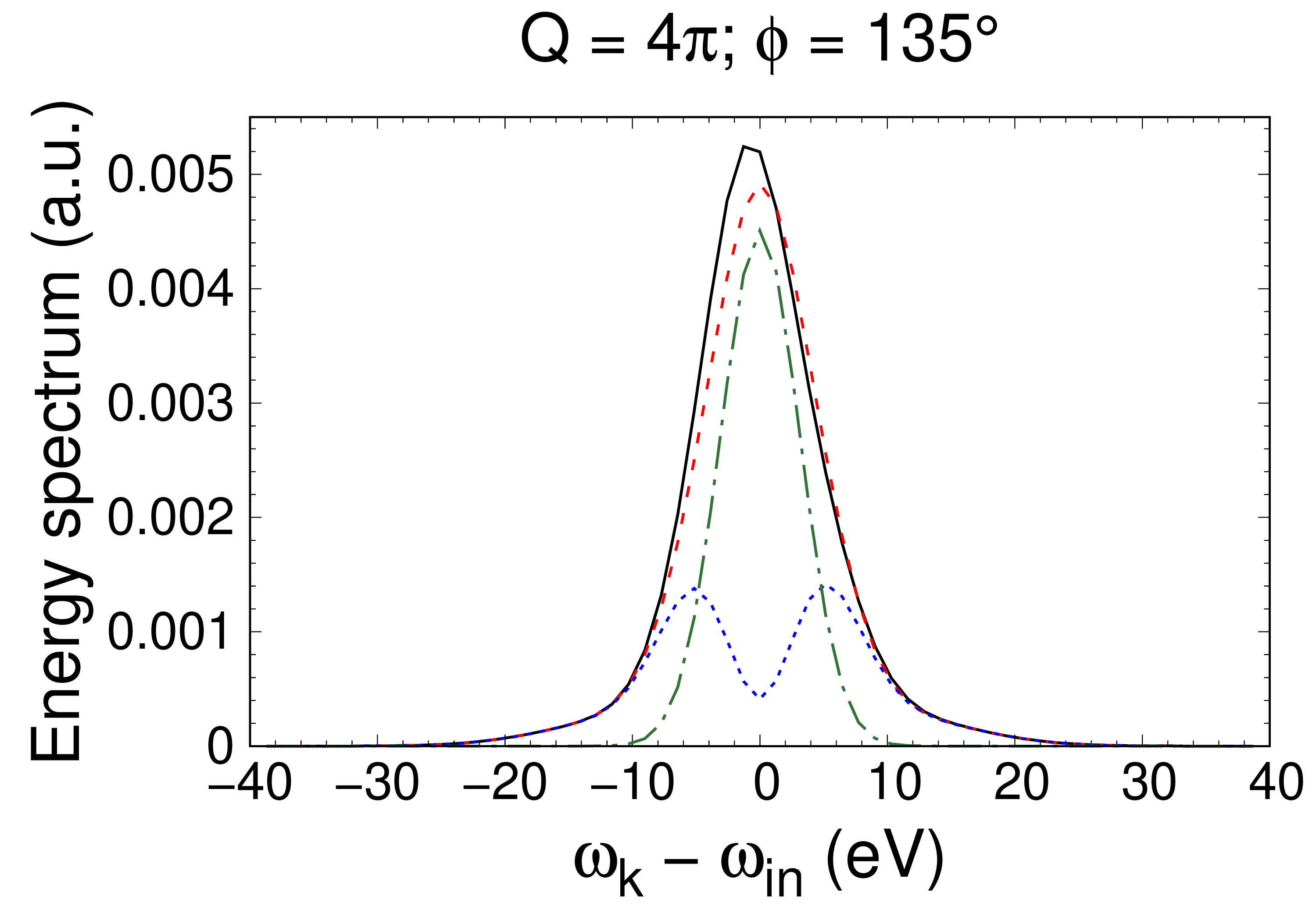}
  \label{Fig_DDSP_p25fs_4pi_90_135}
\end{subfigure} 


\begin{subfigure}{.245\linewidth}
  \includegraphics[width=\linewidth]{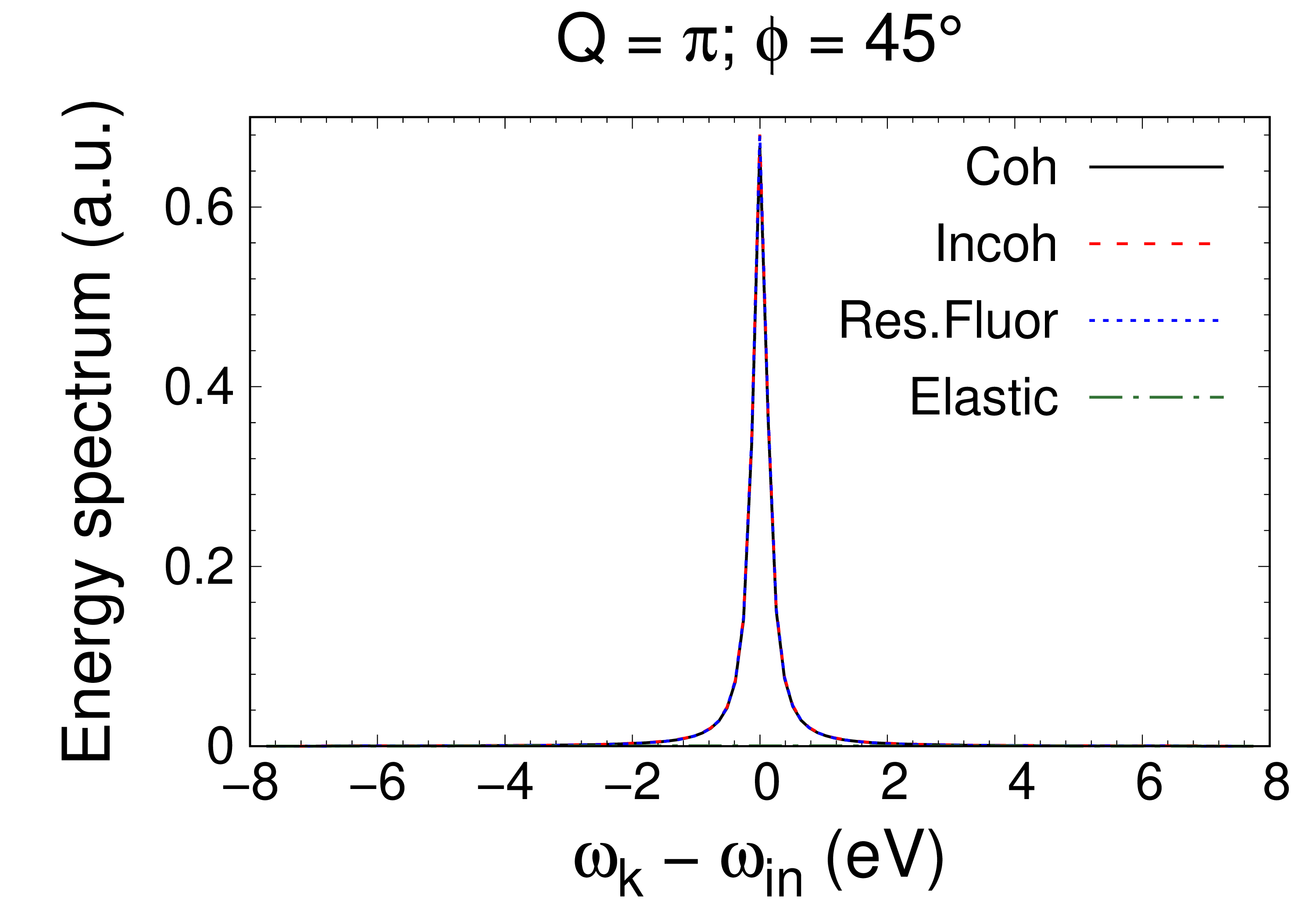}
  \label{Fig_DDSP_p25fs_Pi_90_45}
\end{subfigure}
\begin{subfigure}{.245\linewidth}
  \includegraphics[width=\linewidth]{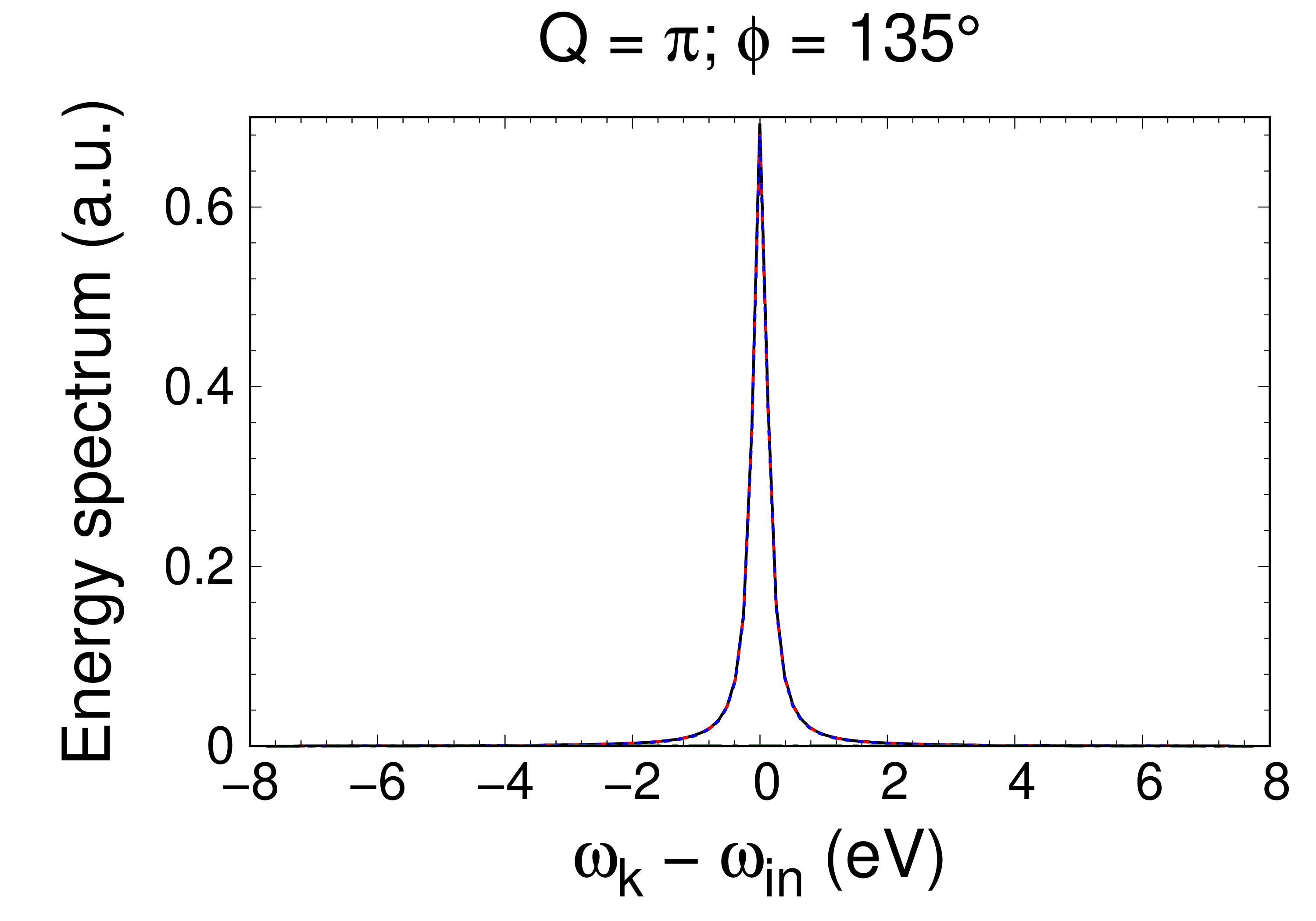}
  \label{Fig_DDSP_p25fs_Pi_90_135}
\end{subfigure} 
\begin{subfigure}{.245\linewidth}
  \includegraphics[width=\linewidth]{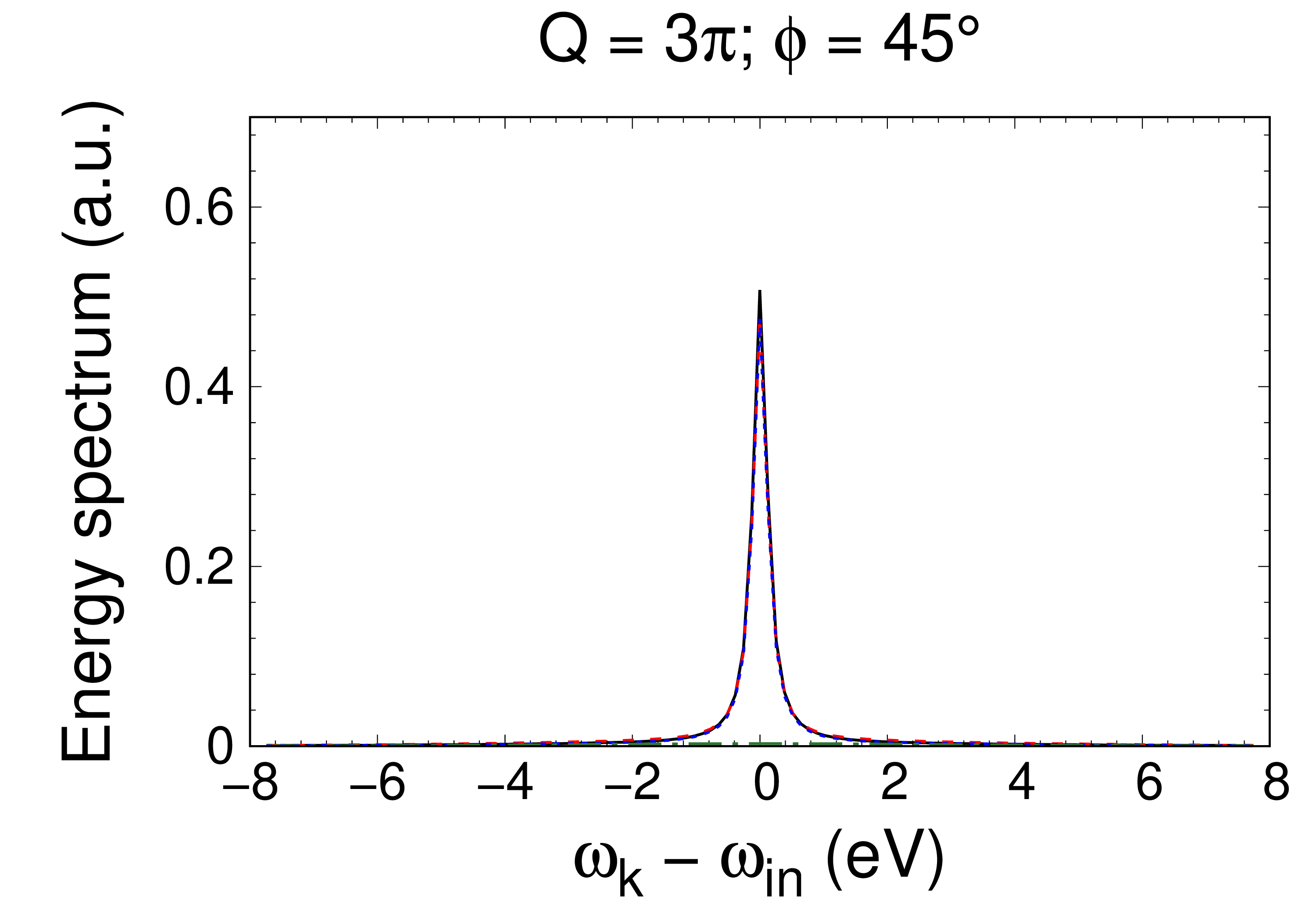}
  \label{Fig_DDSP_p25fs_3Pi_90_45}
\end{subfigure}
\begin{subfigure}{.245\linewidth}
    \includegraphics[width=\linewidth]{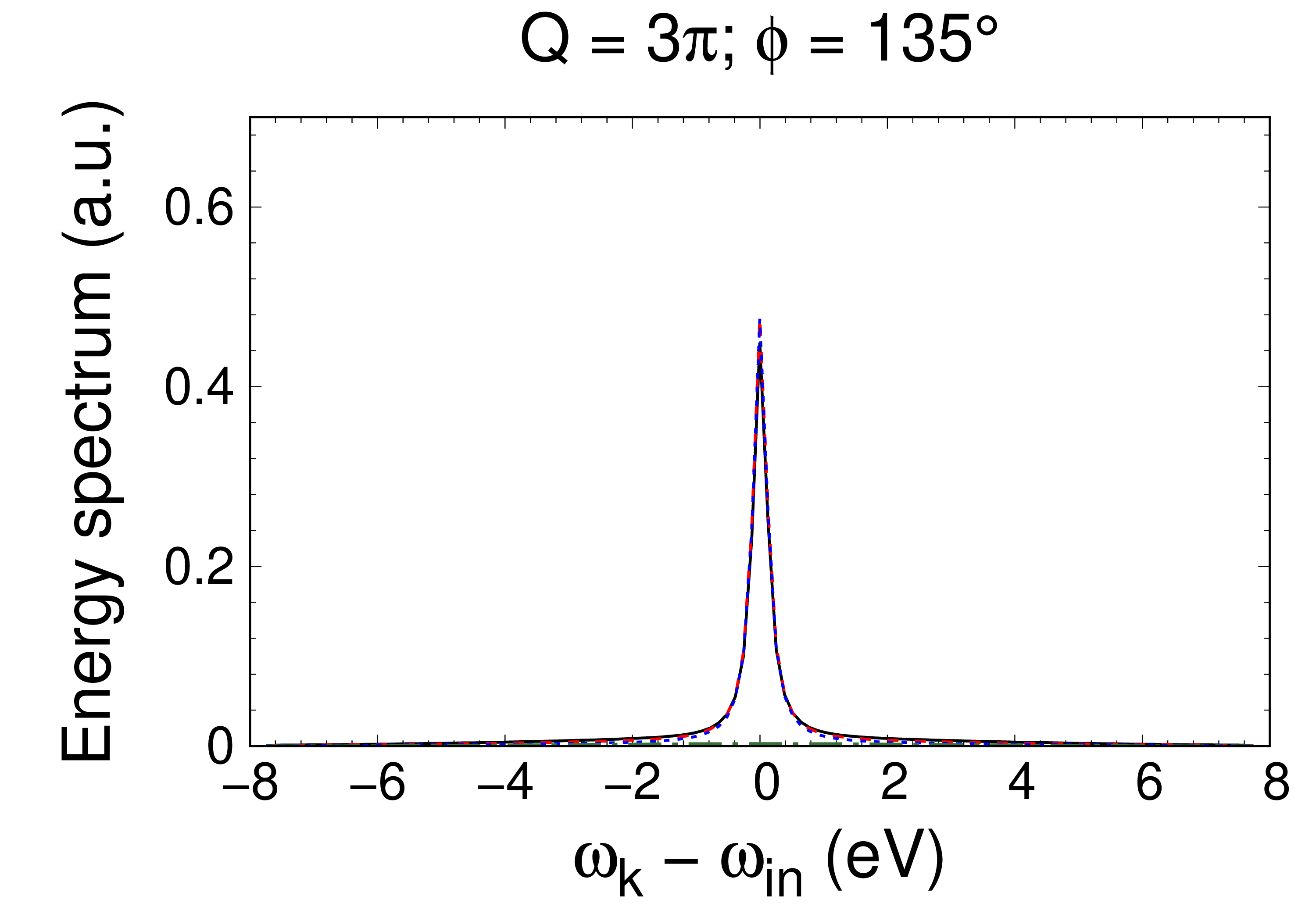}
  \label{Fig_DDSP_p25fs_3Pi_90_135}
\end{subfigure} 
\end{minipage}
\caption{ Energy spectrum of the two channels, their coherent and incoherent sum are all shown for $2\pi$-type pulses (top row) and $\pi$-type pulses (bottom row) in the region of minimum ($\phi_s$ = 45$\degree$) and maximum overall interference ($\phi_s$ = 135$\degree$) between the elastic scattering and resonant fluorescence channel. Here $t_{wid}$ = 0.25 fs and $\theta_s$ = 90$\degree$.
}
\label{Fig_ES_p25fs}
\end{figure*}




\begin{figure*}
\begin{minipage}[t][][t]{\textwidth}
       \hspace{0.2cm}
       \textbf{Elastic}
       \hspace{2.7cm}
       \textbf{Res. Fluor.}
       \hspace{1.8cm}
       \textbf{Incoherent sum}
       \hspace{1.6cm}
       \textbf{Coherent sum}
\end{minipage}
\hspace{-0.3cm}
\begin{minipage}[b][][b]{\textwidth}
\centering
\vspace{0.3cm}
\begin{subfigure}{0.24\linewidth}
  \includegraphics[width=\linewidth]{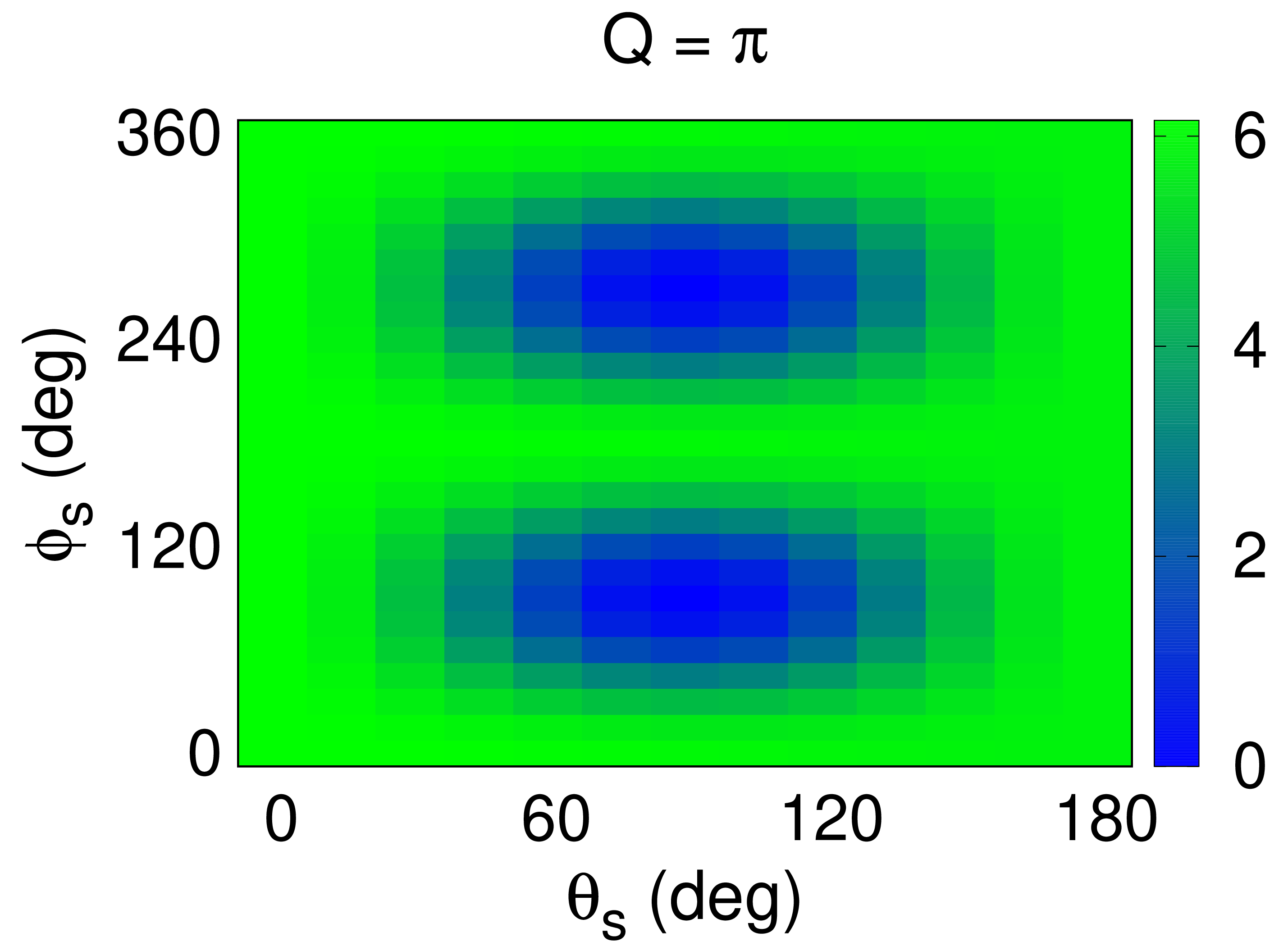}
  \label{Fig_2fs_DCS_Aconly_pi}
\end{subfigure}
\begin{subfigure}{0.24\linewidth}
  \includegraphics[width=\linewidth]{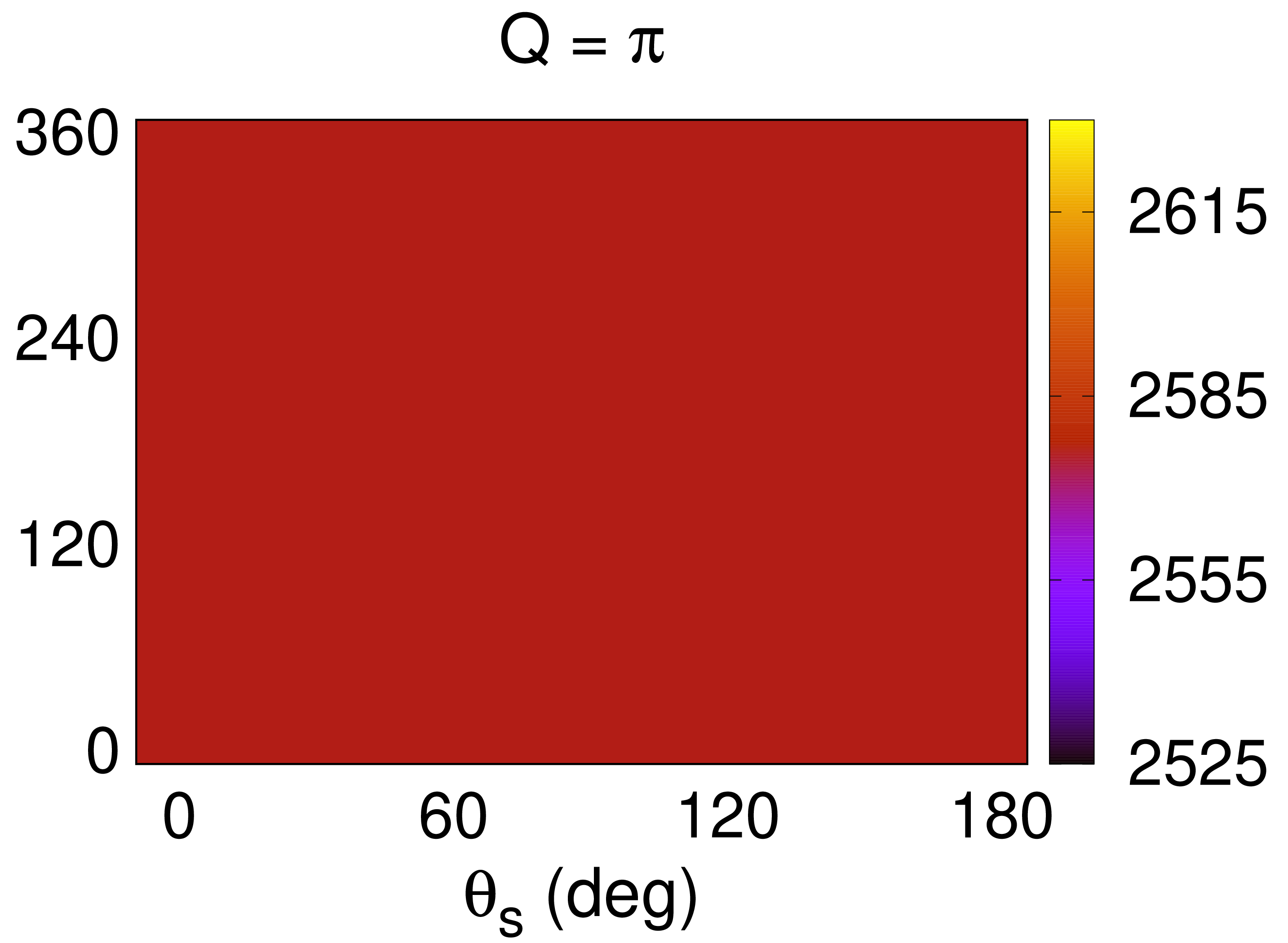}
  \label{Fig_2fs_DCS_AcOFF_pi}
\end{subfigure} 
\begin{subfigure}{0.24\linewidth}
  \includegraphics[width=\linewidth]{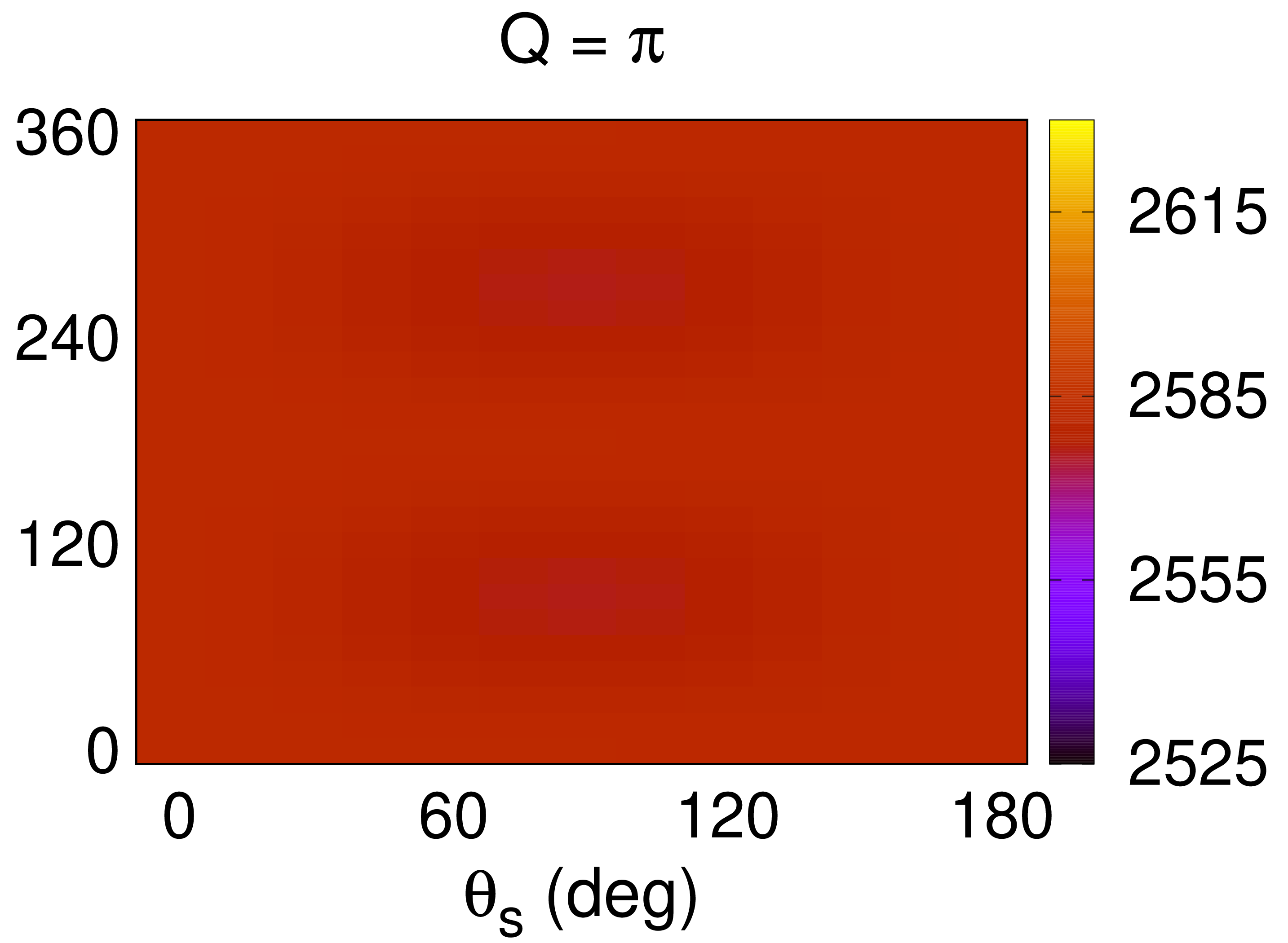}
  \label{Fig_2fs_DCS_incoherentsum_pi}
\end{subfigure} 
\begin{subfigure}{0.25\linewidth}
  \includegraphics[width=\linewidth]{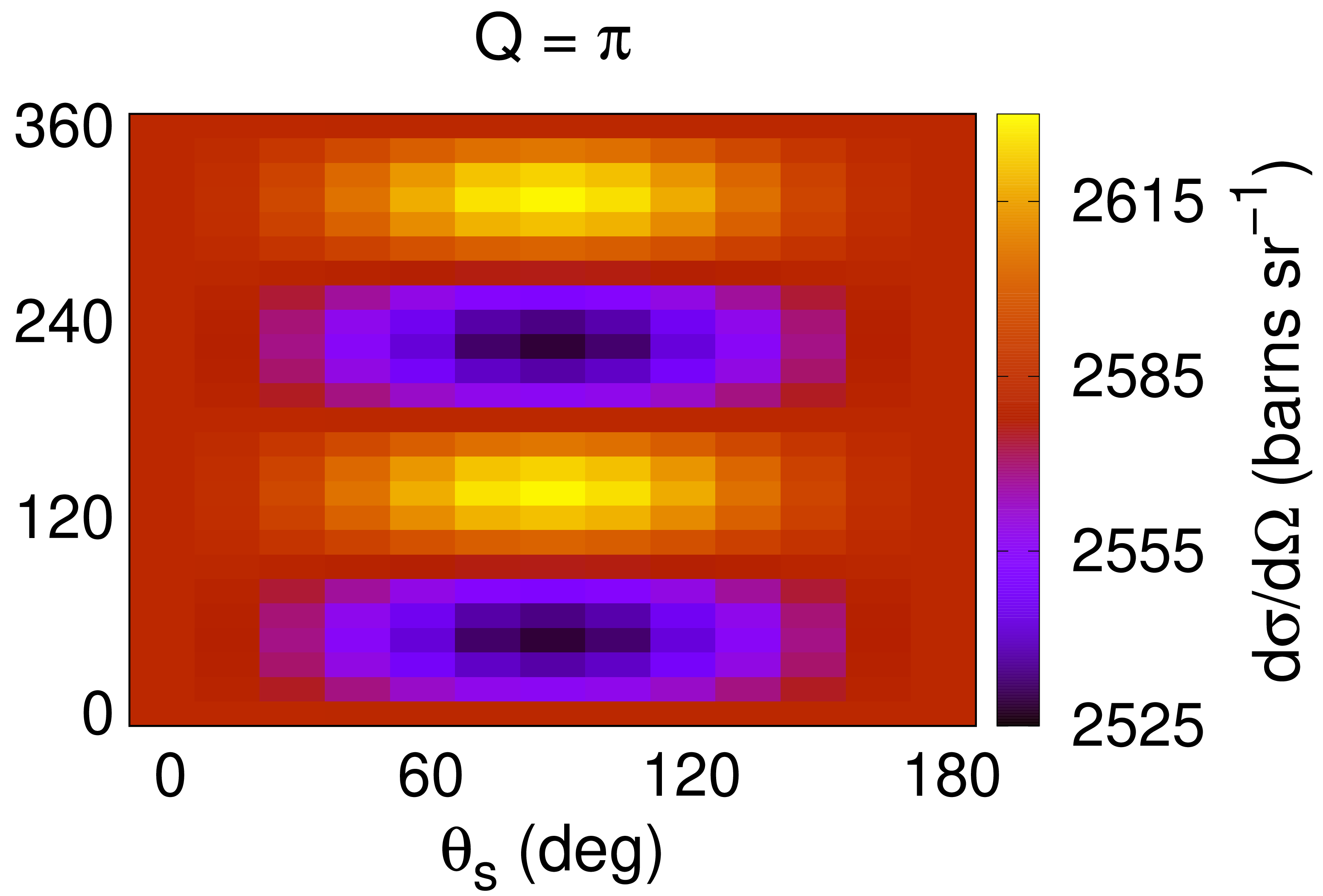}
  \label{Fig_2fs_dcs_pi}
\end{subfigure} 

\medskip 
\begin{subfigure}{.24\linewidth}
  \includegraphics[width=\linewidth]{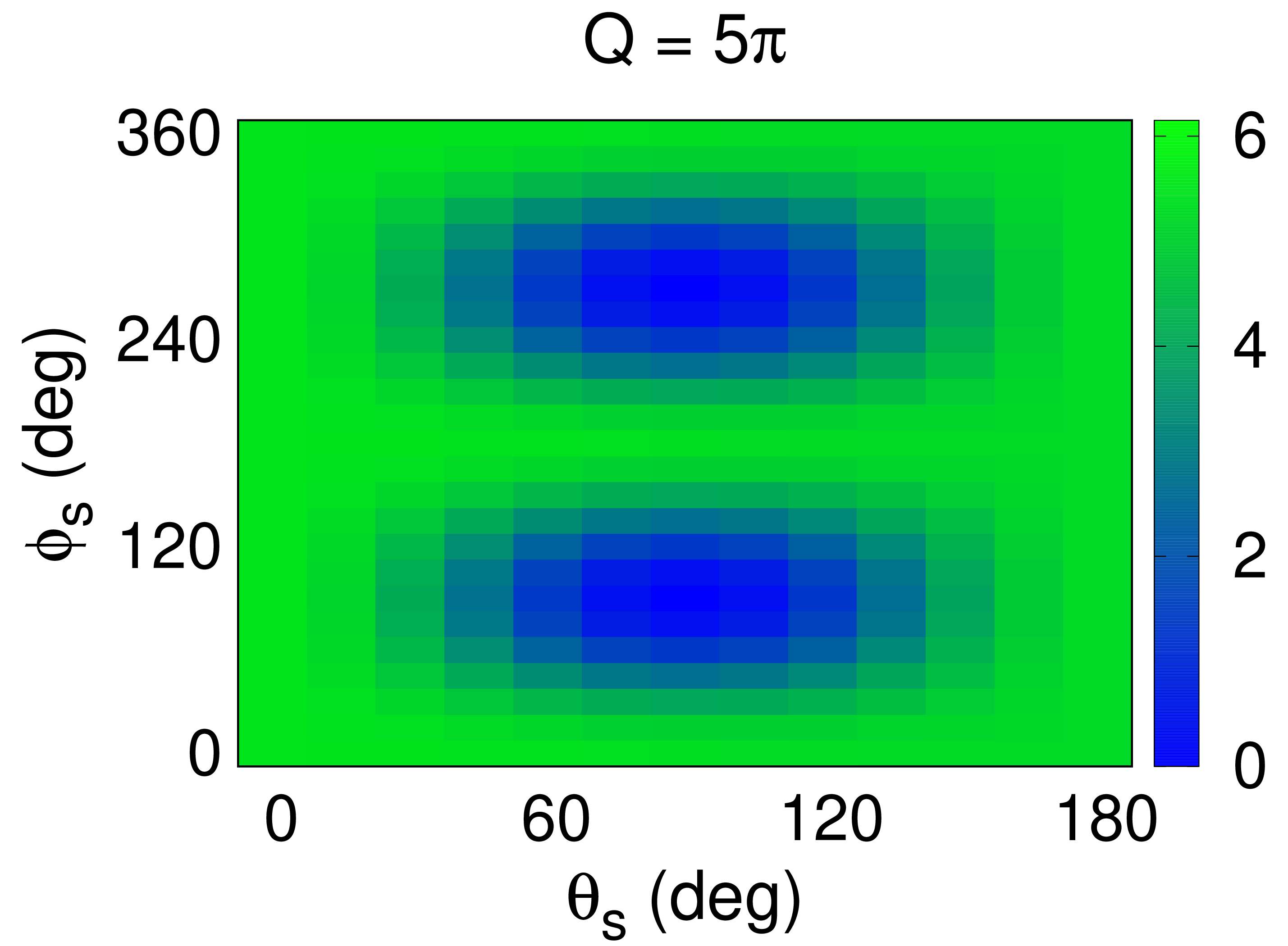}
  \label{Fig_2fs_DCS_Aconly_5pi}
\end{subfigure}
\hspace{0.05cm}
\begin{subfigure}{.235\linewidth}
    \includegraphics[width=\linewidth]{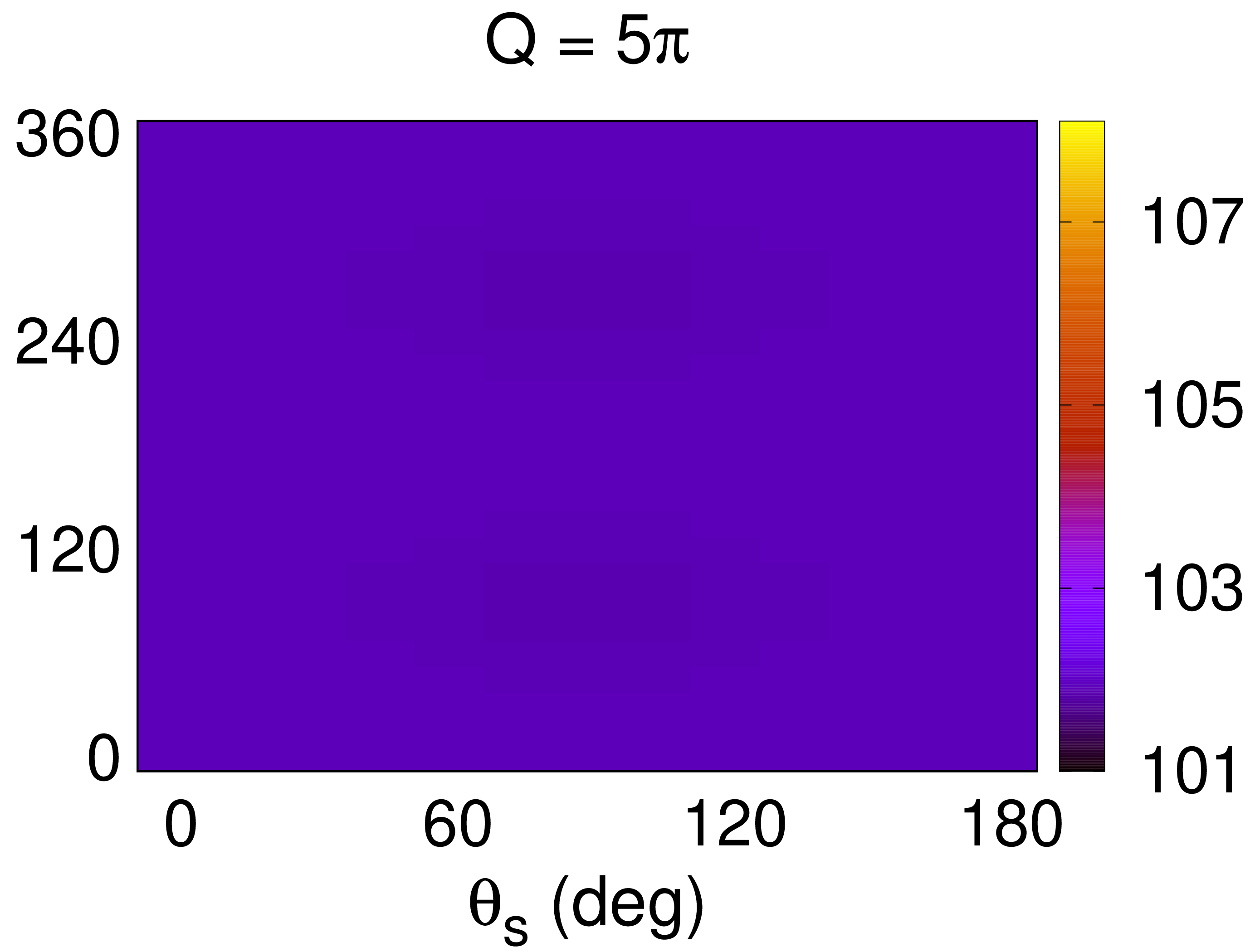}
  \label{Fig_2fs_DCS_AcOFF_5pi}
\end{subfigure}\hfill 
\begin{subfigure}{.23\linewidth}
    \includegraphics[width=\linewidth]{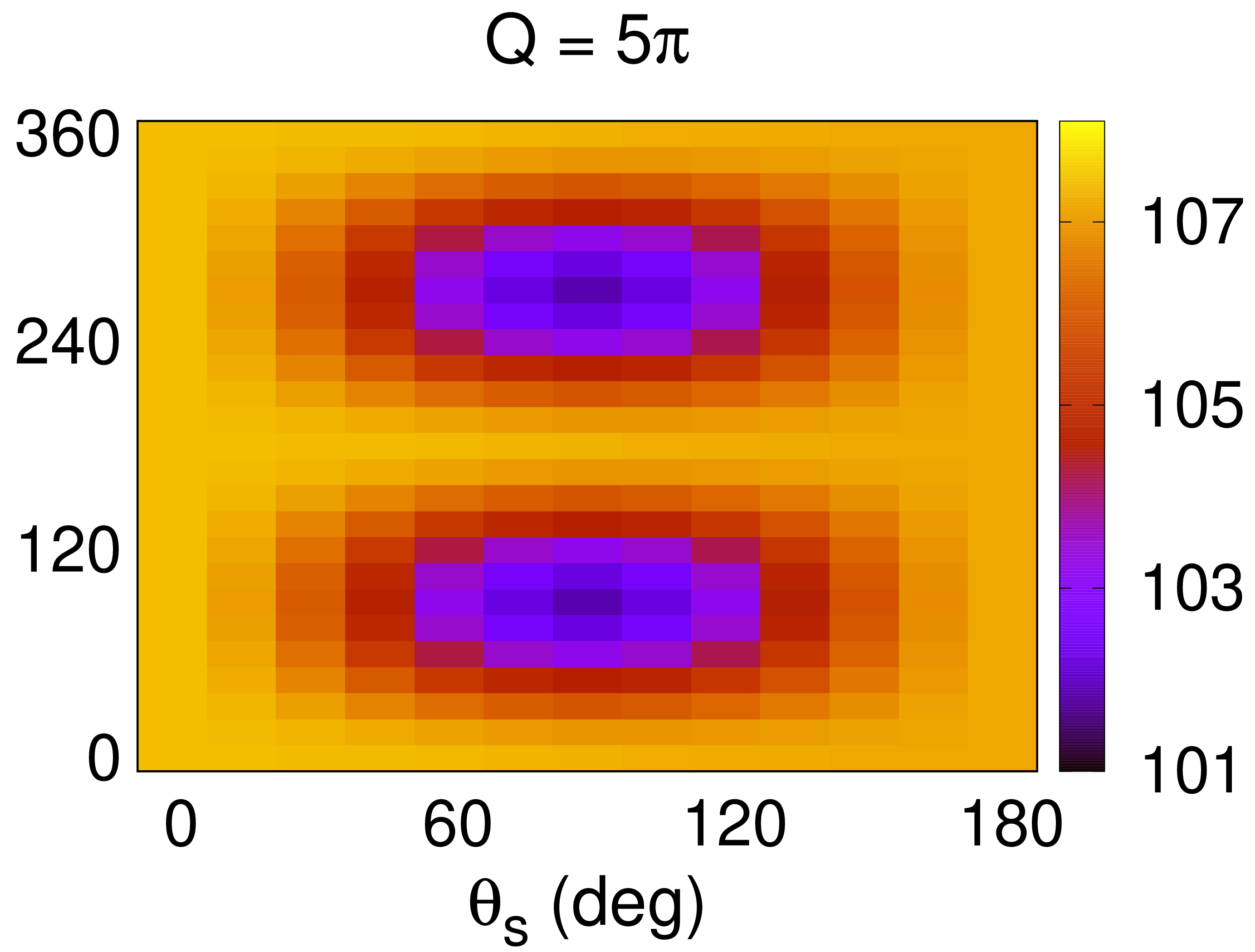}
  \label{Fig_2fs_DCS_incoherentsum_5pi}
\end{subfigure} 
\hspace{0.1cm}
\begin{subfigure}{.255\linewidth}
    \includegraphics[width=\linewidth]{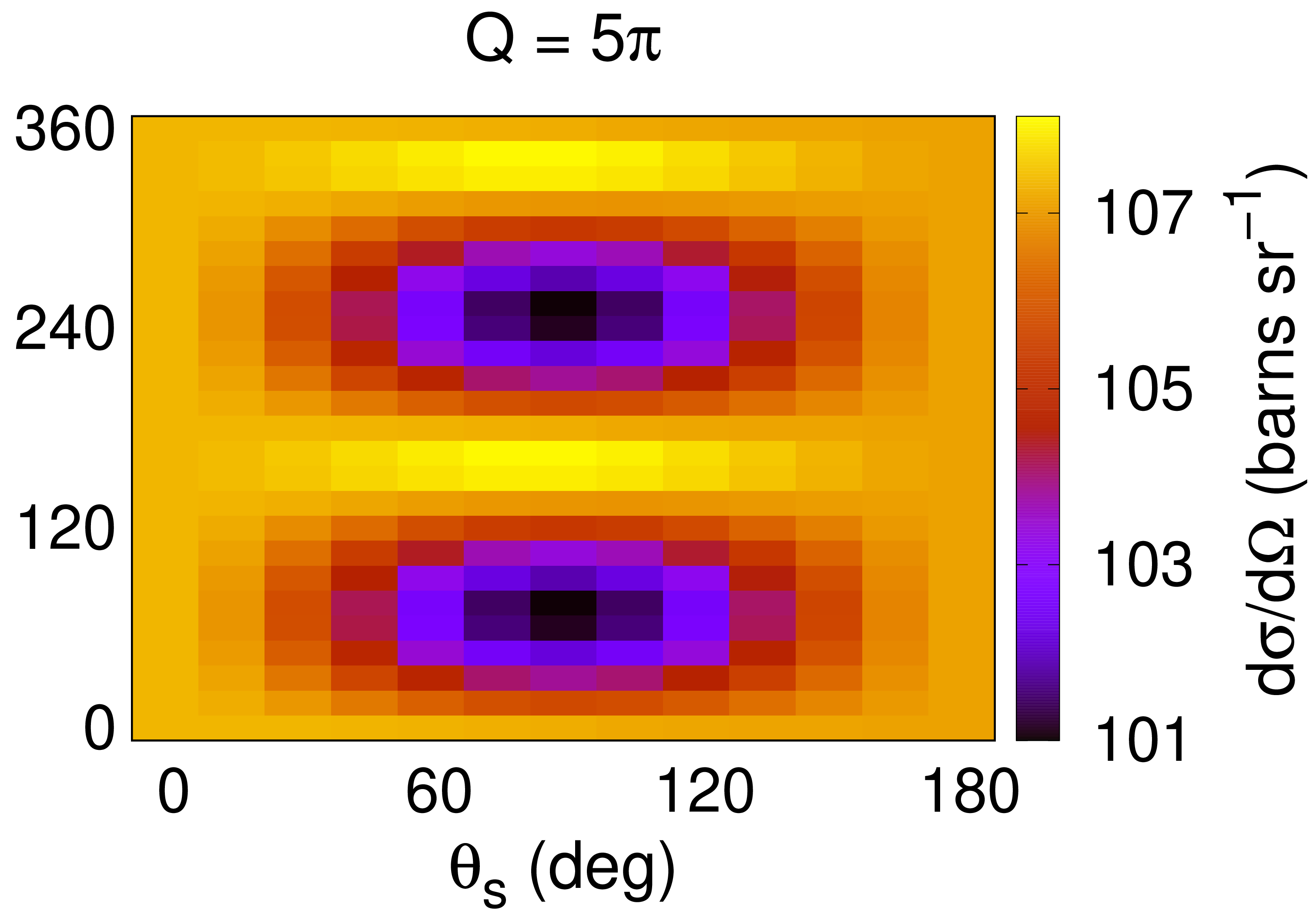}
  \label{Fig_2fs_dcs_5pi}
\end{subfigure}\hfill 

\medskip 
\begin{subfigure}{.24\linewidth}
  \includegraphics[width=\linewidth]{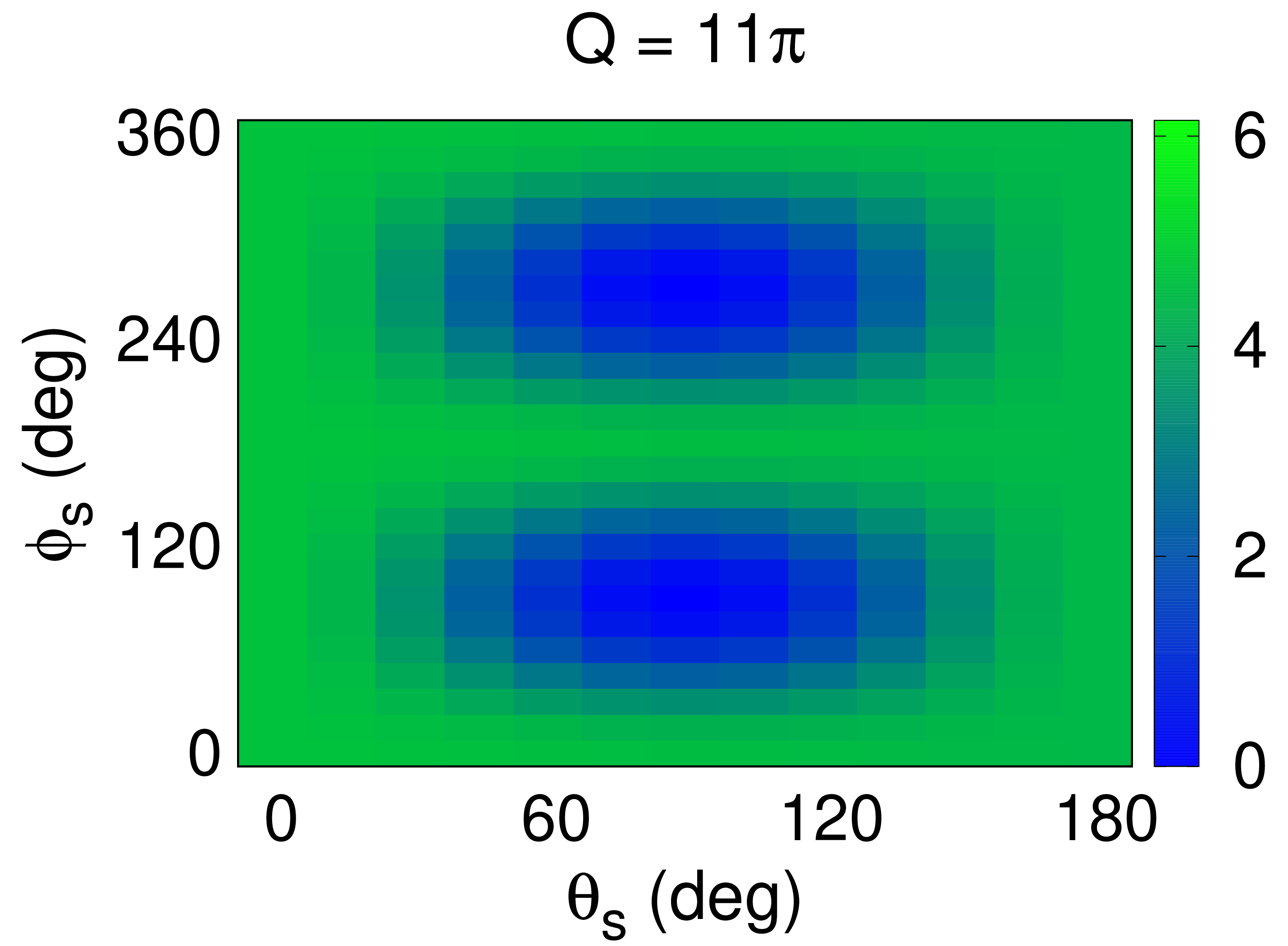}
  \caption{}
  \label{Fig_2fs_DCS_Aconly_11pi}
\end{subfigure}
\hspace{0.05cm}
\begin{subfigure}{.235\linewidth}
  \includegraphics[width=\linewidth]{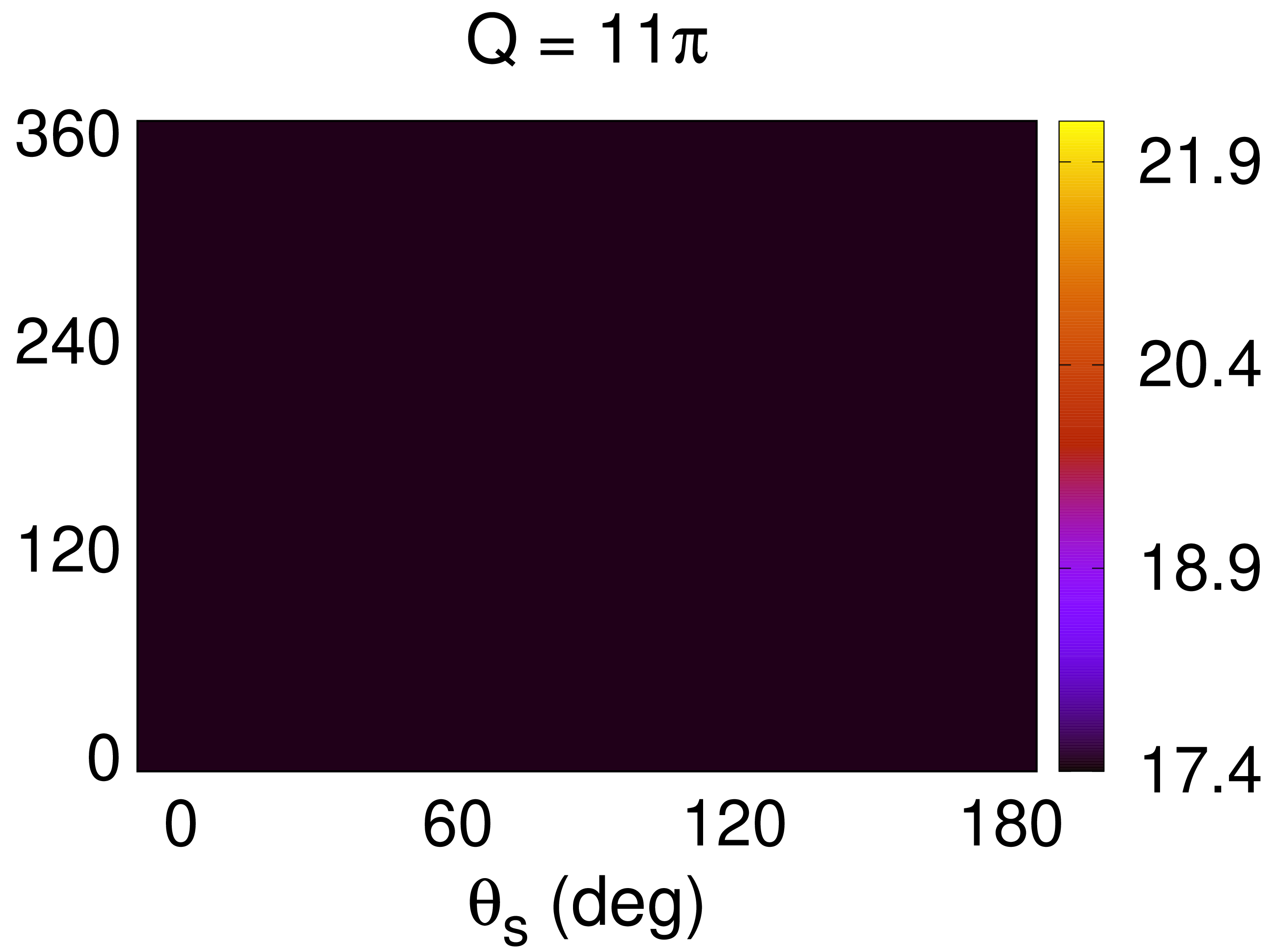}
  \caption{}
  \label{Fig_2fs_DCS_AcOFF_11pi}
\end{subfigure} 
\begin{subfigure}{.235\linewidth}
  \includegraphics[width=\linewidth]{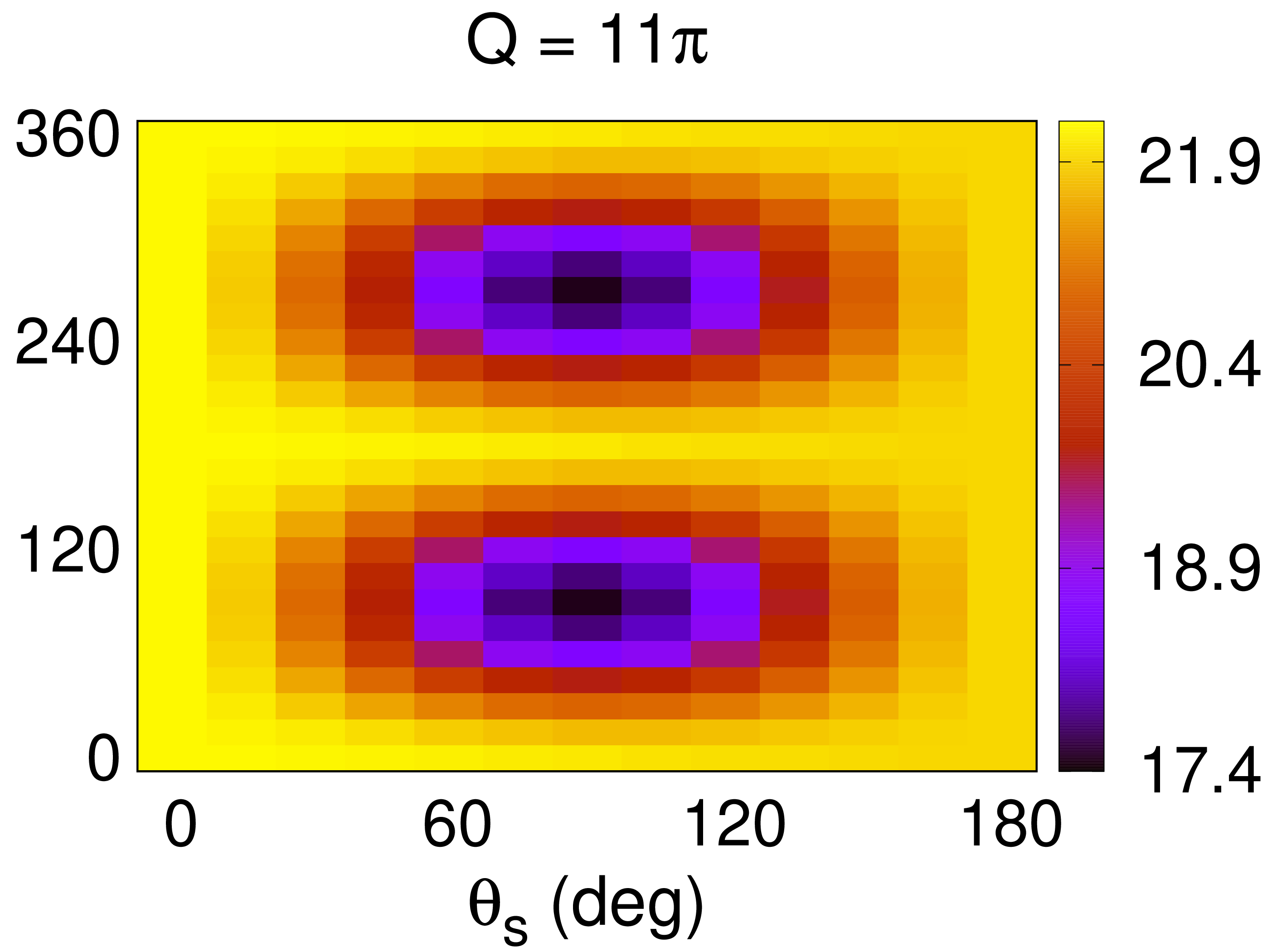}
  \caption{}
  \label{Fig_2fs_DCS_incoherentsum_11pi}
\end{subfigure}\hfill 
\begin{subfigure}{.265\linewidth}
  \includegraphics[width=\linewidth]{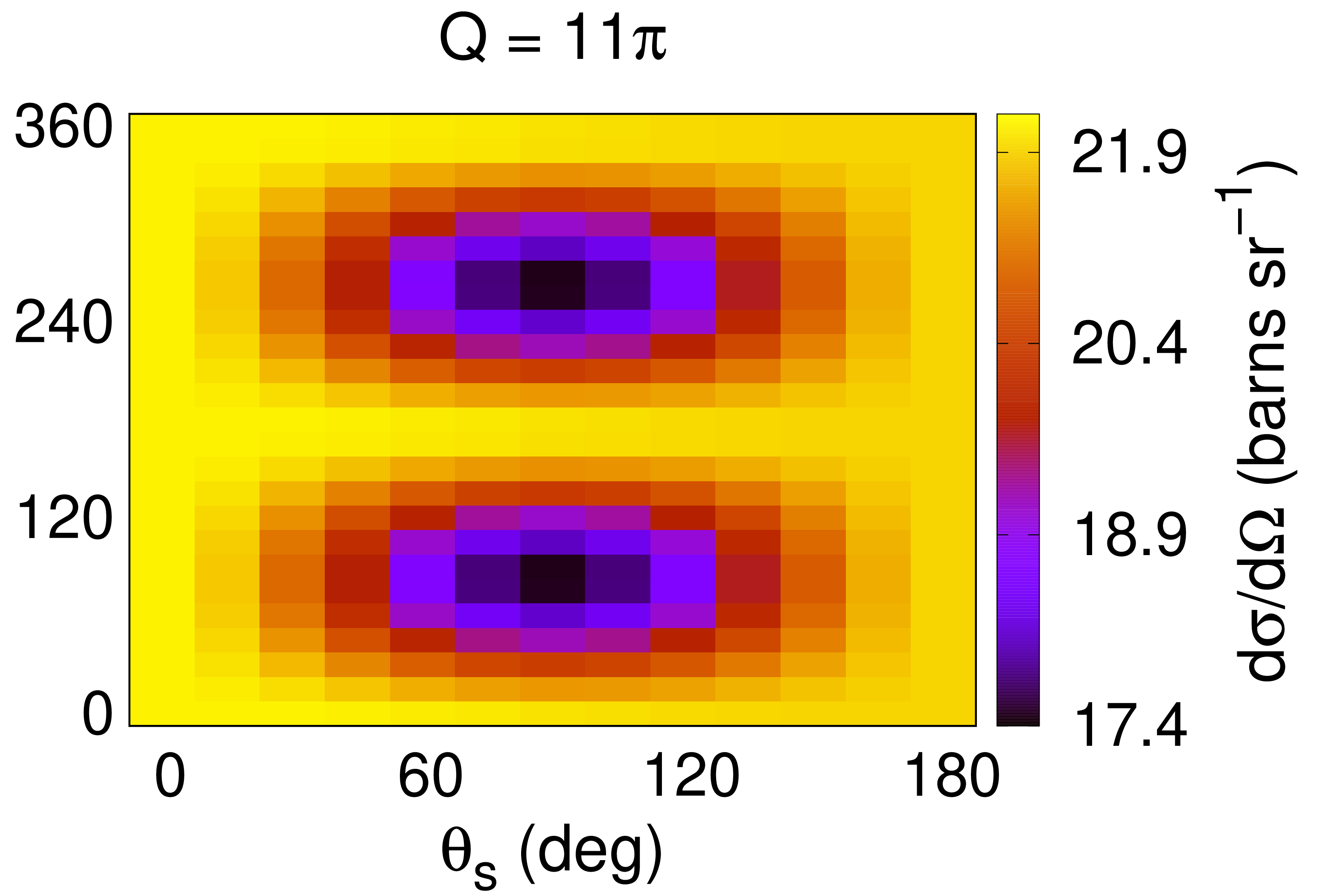}
  \caption{}
  \label{Fig_2fs_dcs_11pi}
\end{subfigure} 

\end{minipage}
\caption{Angular distribution of the DCS calculated for a pulse duration of $t_{wid} = 2$~fs for $\pi$-type pulses from only the elastic scattering channel (Fig.~\ref{Fig_2fs_DCS_Aconly_11pi}), from only the resonant fluorescence channel (Fig.~\ref{Fig_2fs_DCS_AcOFF_11pi}), an incoherent sum of the two channels (Fig.~\ref{Fig_2fs_DCS_incoherentsum_11pi}), and a coherent sum of the two channels (Fig.~\ref{Fig_2fs_dcs_11pi}).
}
\label{Fig_2fs_dcs_channels_Pi}
\end{figure*}




\begin{figure*} [hbt!] 
\begin{minipage}[t][][t]{\textwidth}
       \hspace{0.2cm}
       \textbf{Elastic}
       \hspace{2.7cm}
       \textbf{Res. Fluor.}
       \hspace{1.8cm}
       \textbf{Incoherent sum}
       \hspace{1.4cm}
       \textbf{Coherent sum}
\end{minipage}
\begin{minipage}[b][][b]{\textwidth}
\vspace{0.3cm}
\begin{subfigure}{.24\linewidth}
  \includegraphics[width=\linewidth]{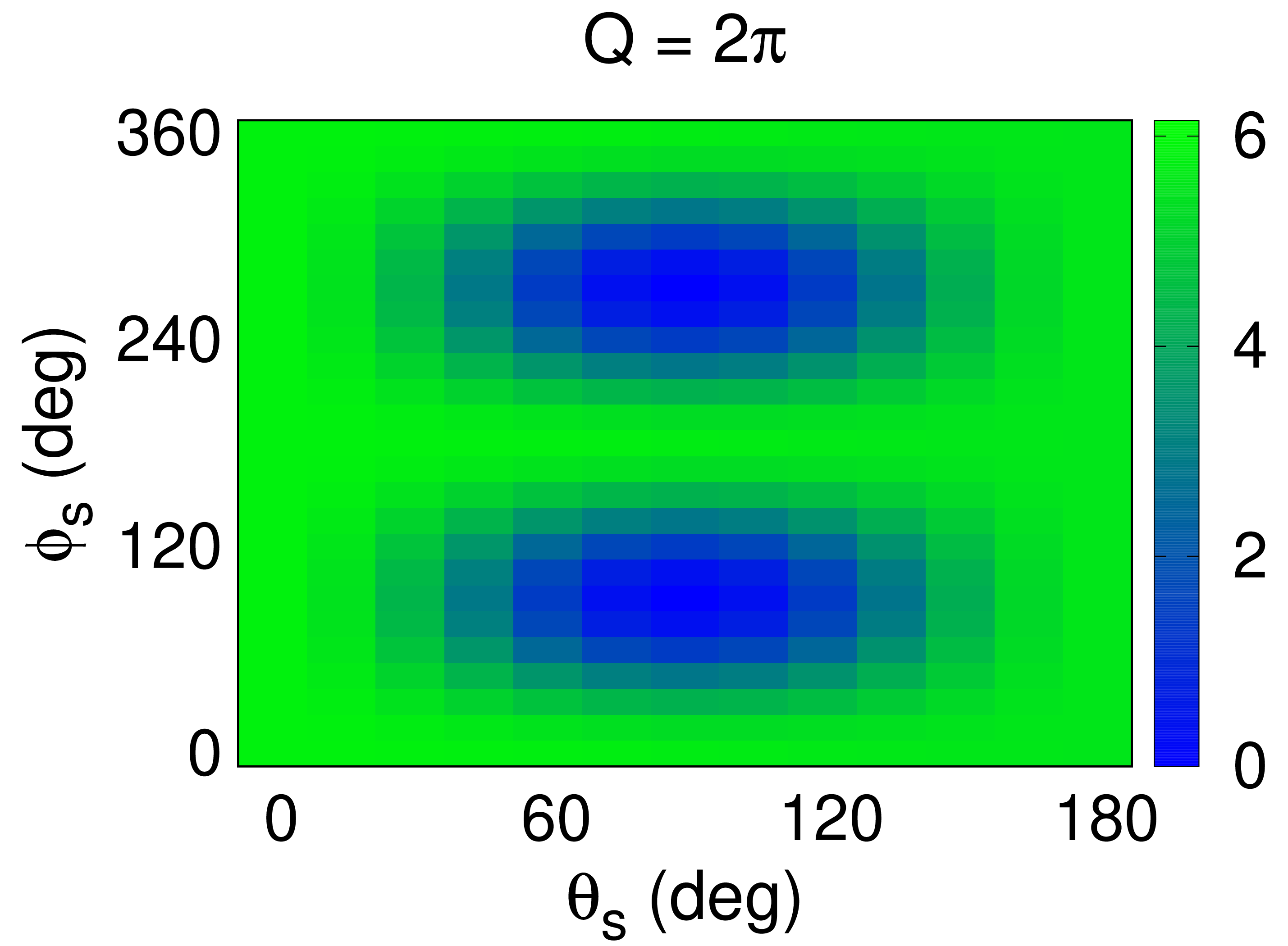}
  \label{Fig_2fs_DCS_Aconly_2pi}
\end{subfigure}
\begin{subfigure}{.235\linewidth}
  \includegraphics[width=\linewidth]{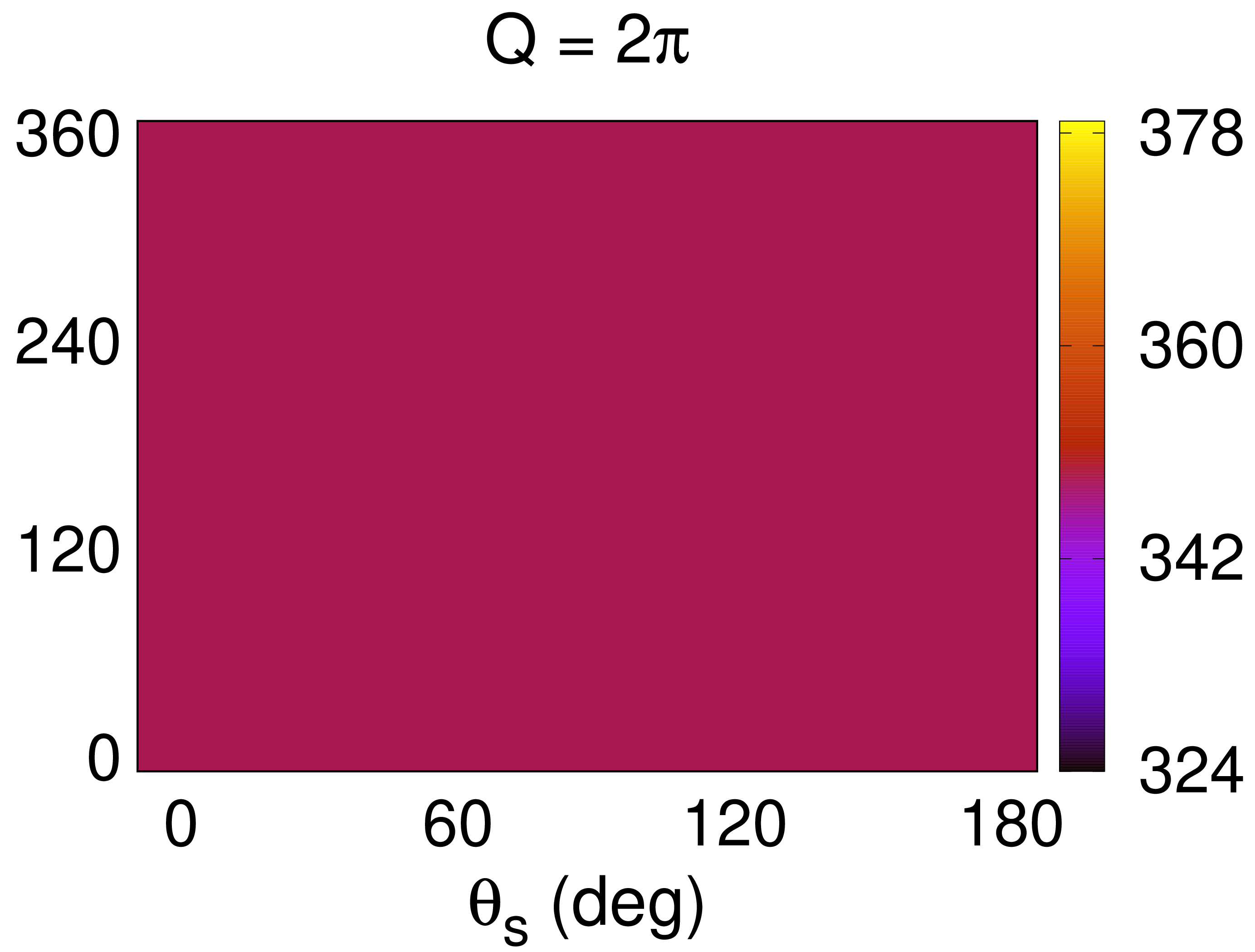}
  \label{Fig_2fs_DCS_AcOFF_2pi}
\end{subfigure} 
\begin{subfigure}{.23\linewidth}
  \includegraphics[width=\linewidth]{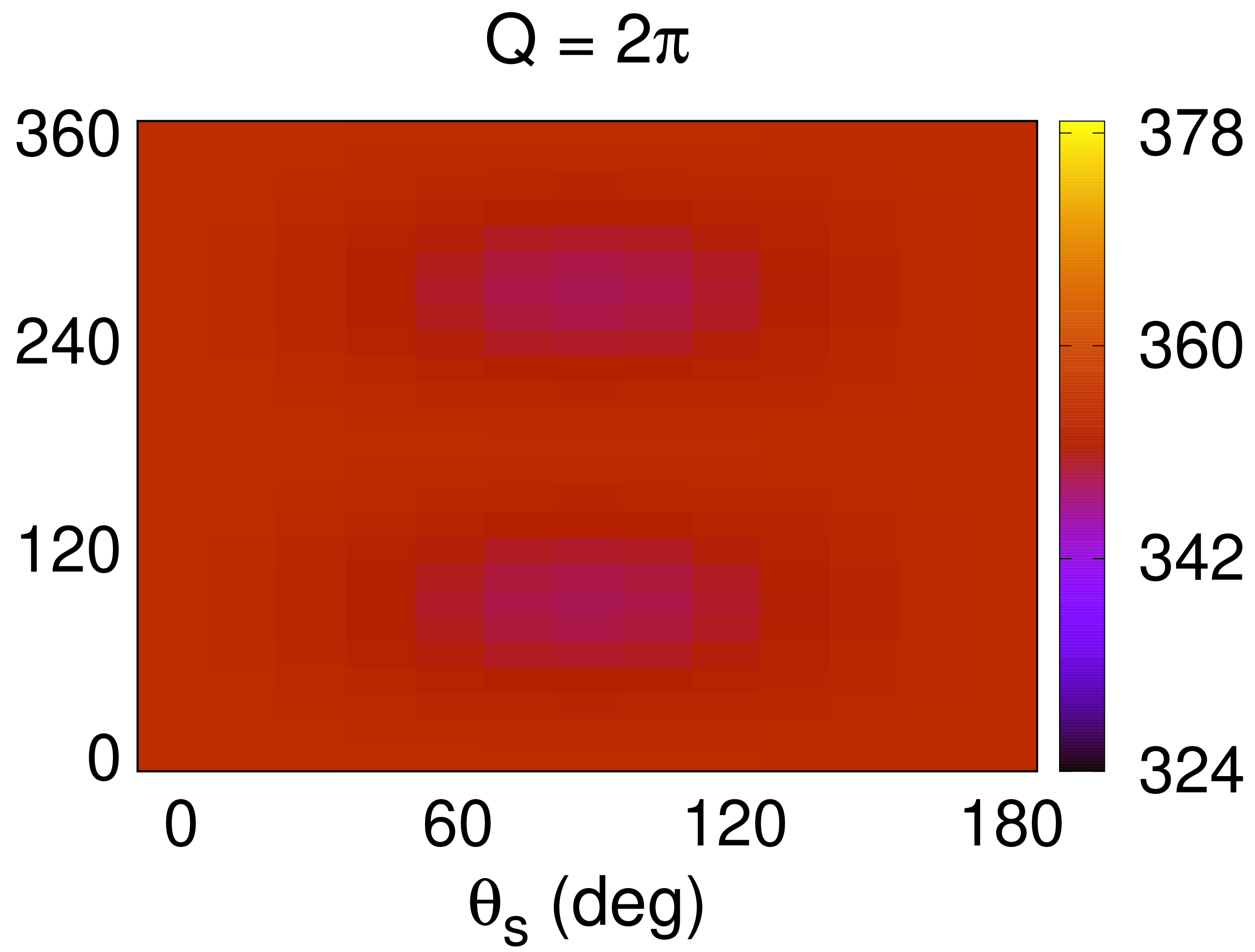}
  \label{Fig_2fs_DCS_incoherentsum_2pi}
\end{subfigure}
\begin{subfigure}{.25\linewidth}
  \includegraphics[width=\linewidth]{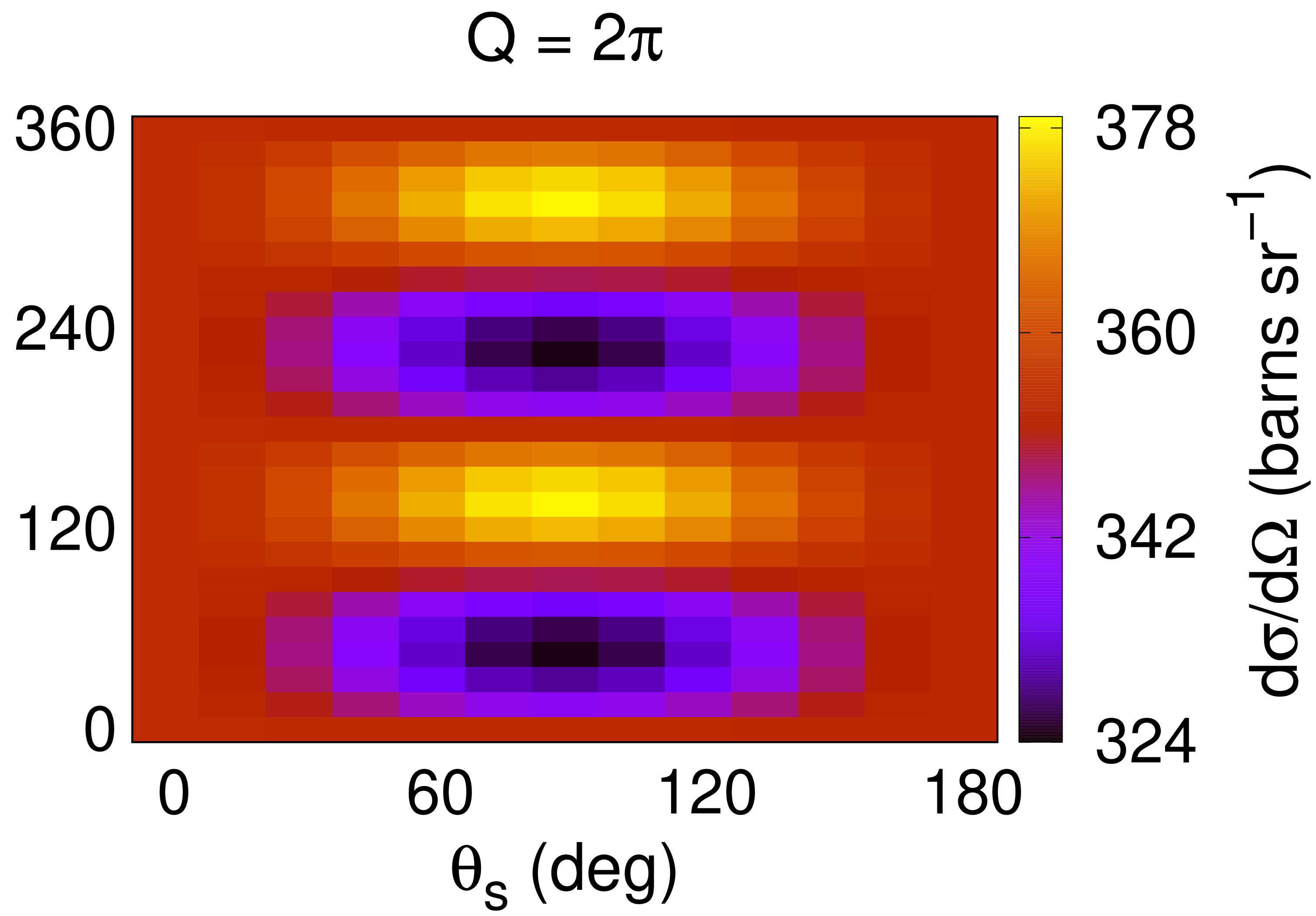}
  \label{Fig_2fs_dcs_2pi}
\end{subfigure}

\medskip 
\begin{subfigure}{.24\linewidth}
  \includegraphics[width=\linewidth]{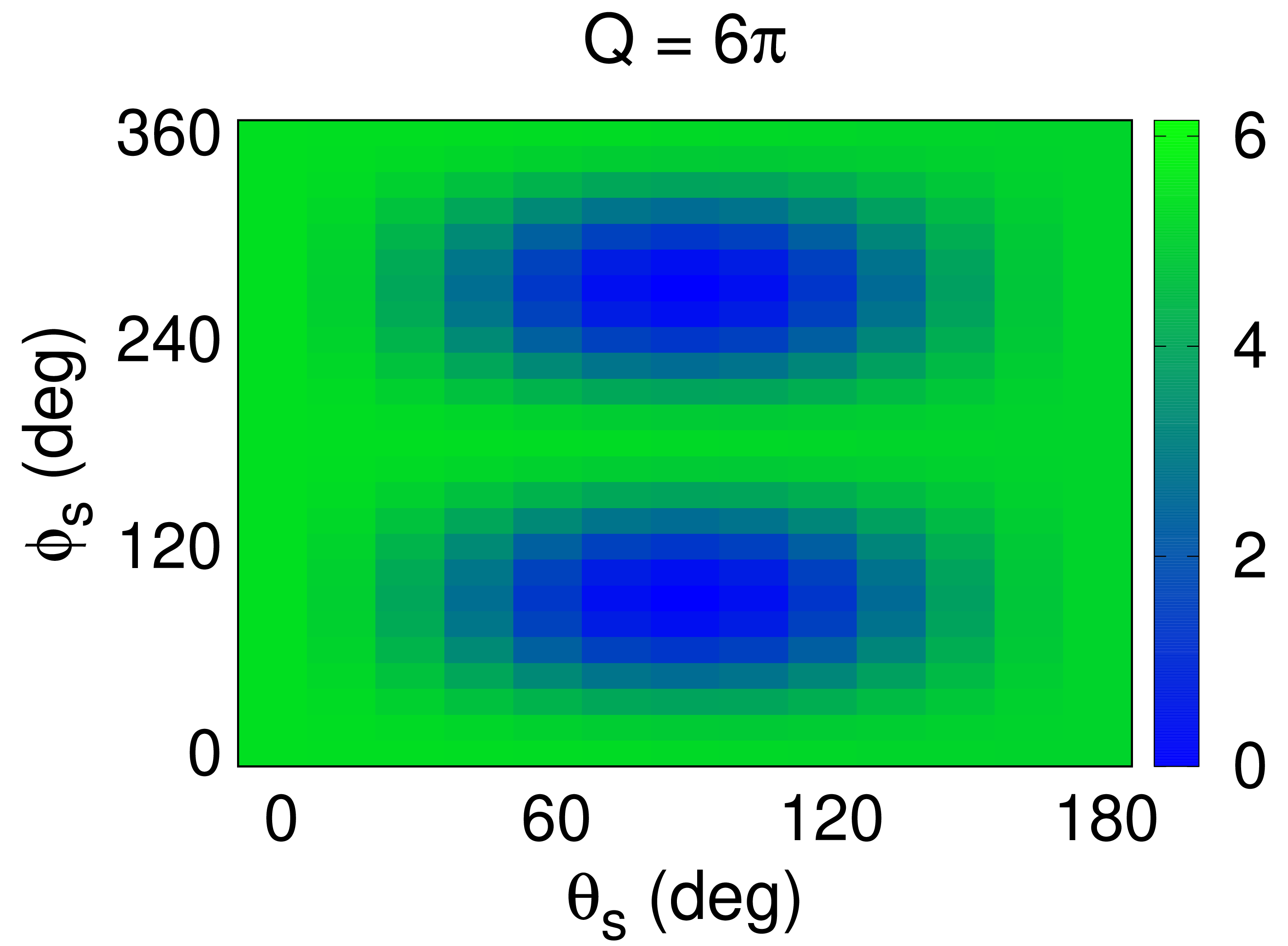}
  \label{Fig_2fs_DCS_Aconly_6pi}
\end{subfigure}
\begin{subfigure}{.23\linewidth}
  \includegraphics[width=\linewidth]{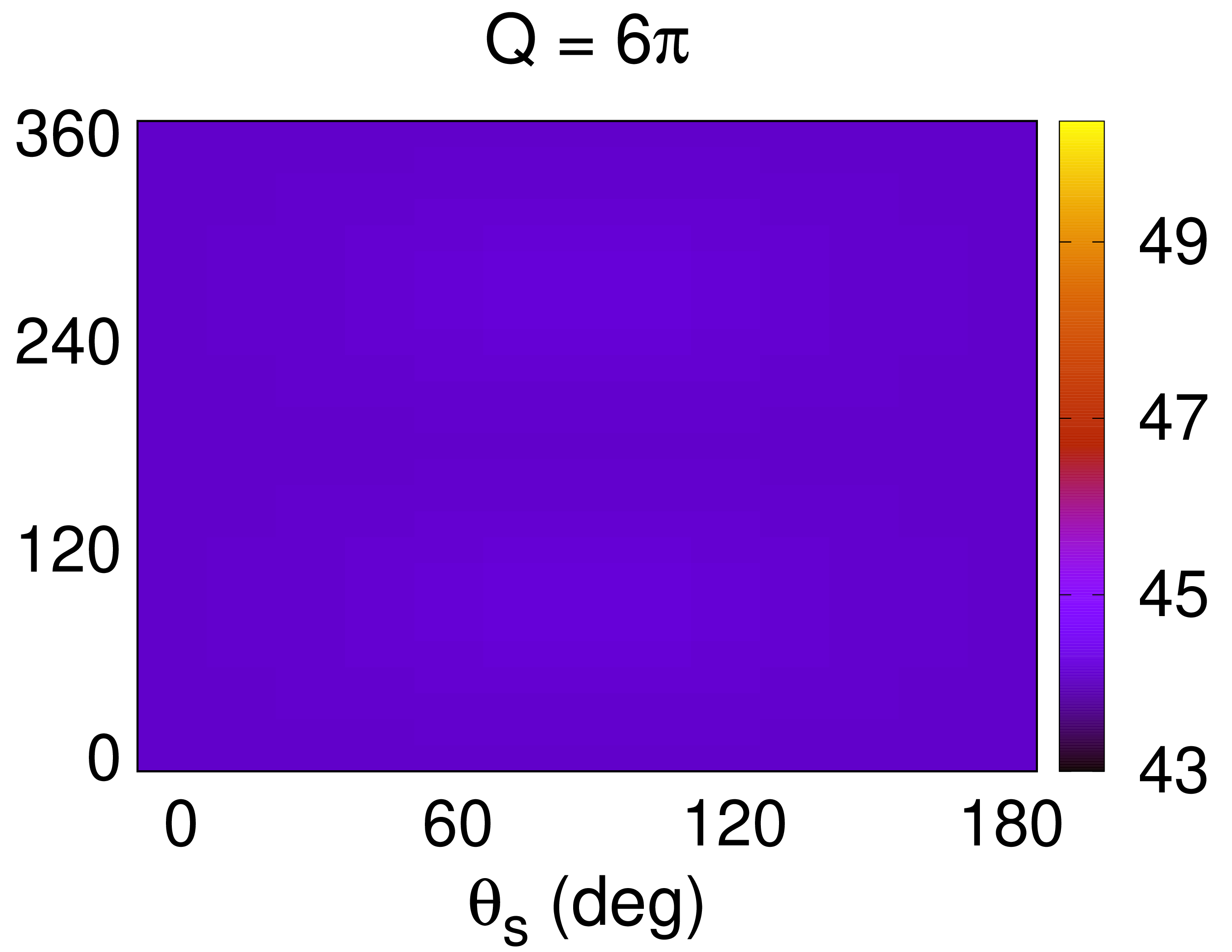}
  \label{Fig_2fs_DCS_AcOFF_6pi}
\end{subfigure} 
\begin{subfigure}{.23\linewidth}
  \includegraphics[width=\linewidth]{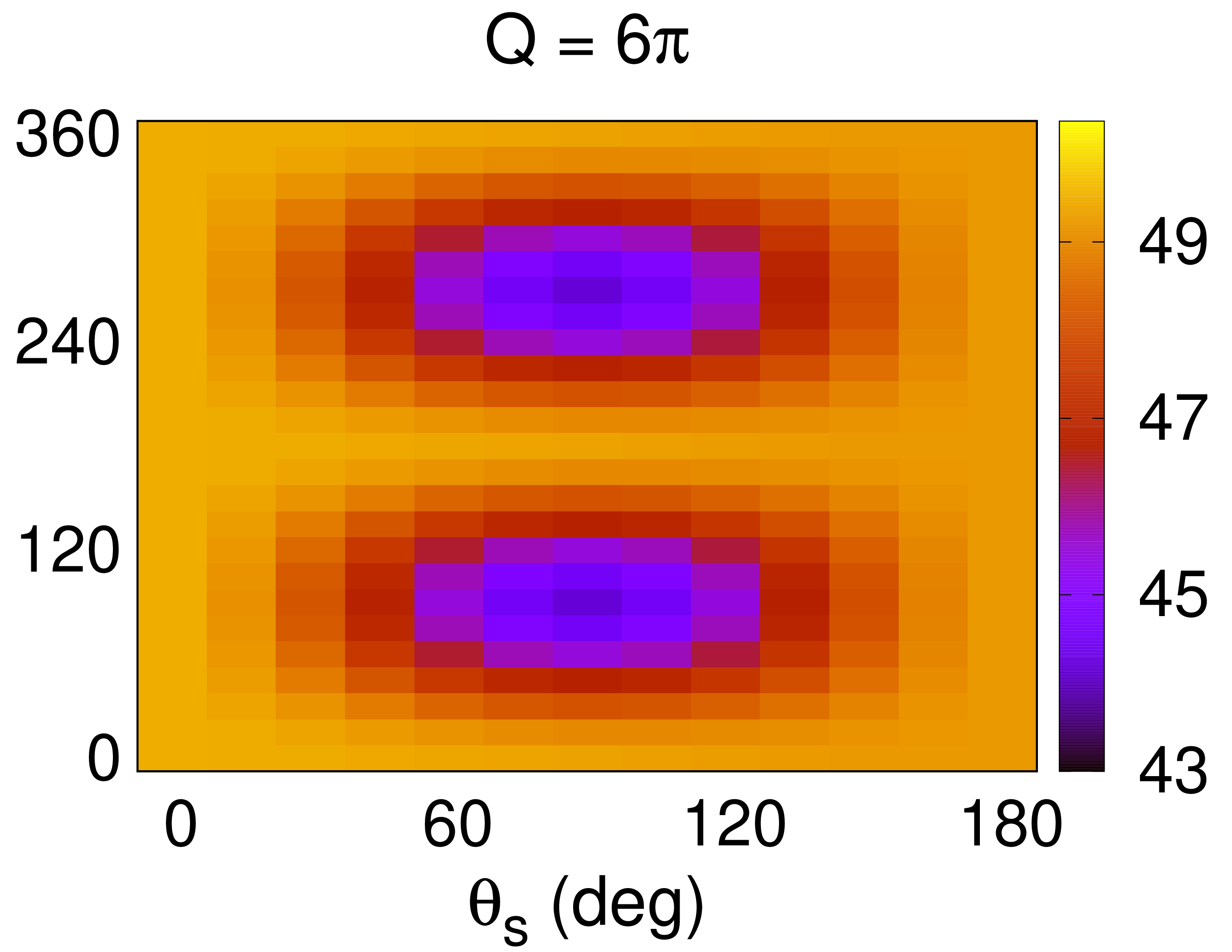}
  \label{Fig_2fs_DCS_incoherentsum_6pi}
\end{subfigure}
\begin{subfigure}{.25\linewidth}
  \includegraphics[width=\linewidth]{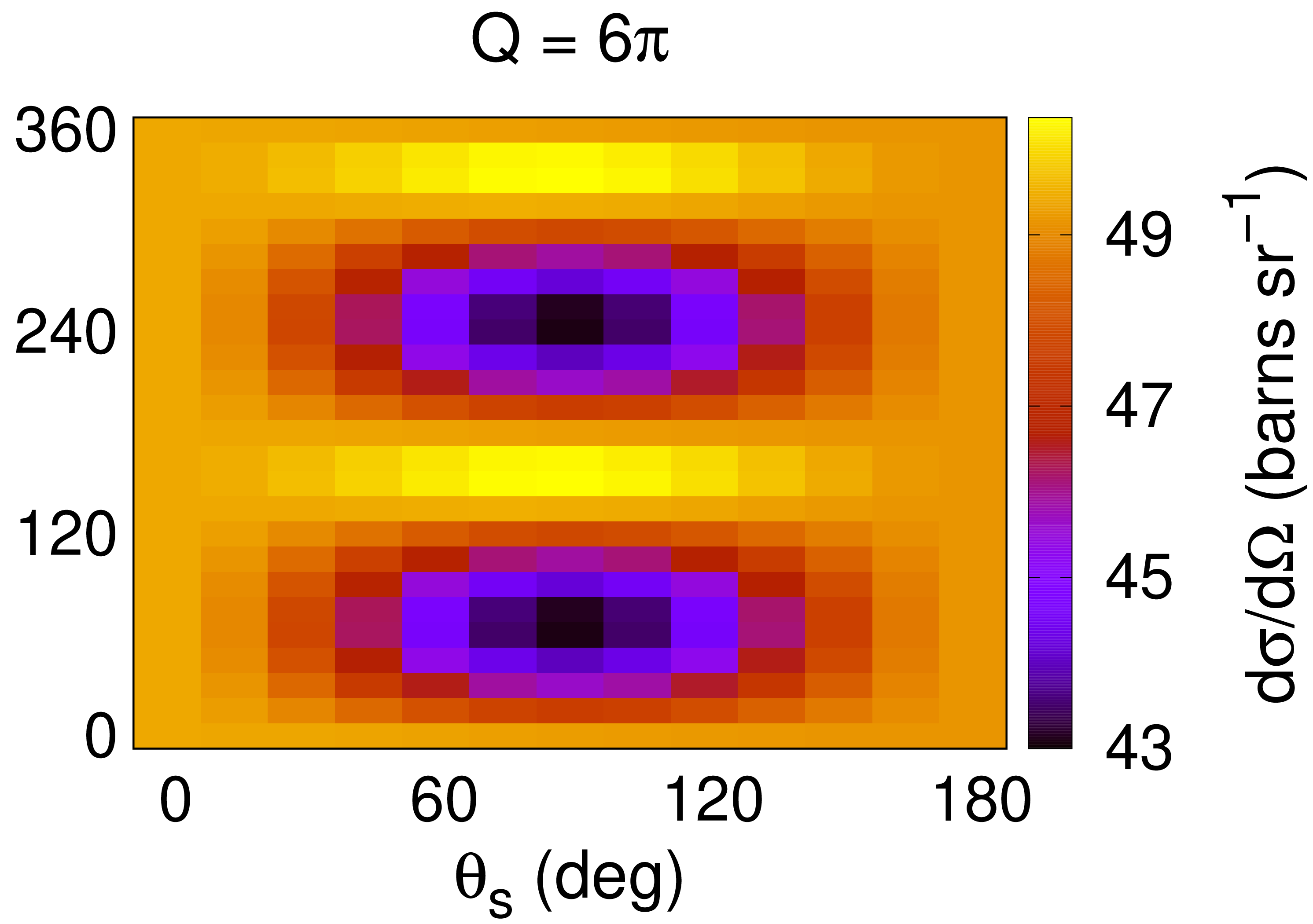}
  \label{Fig_2fs_dcs_6pi}
\end{subfigure}

\medskip 
\begin{subfigure}{.24\linewidth}
  \includegraphics[width=\linewidth]{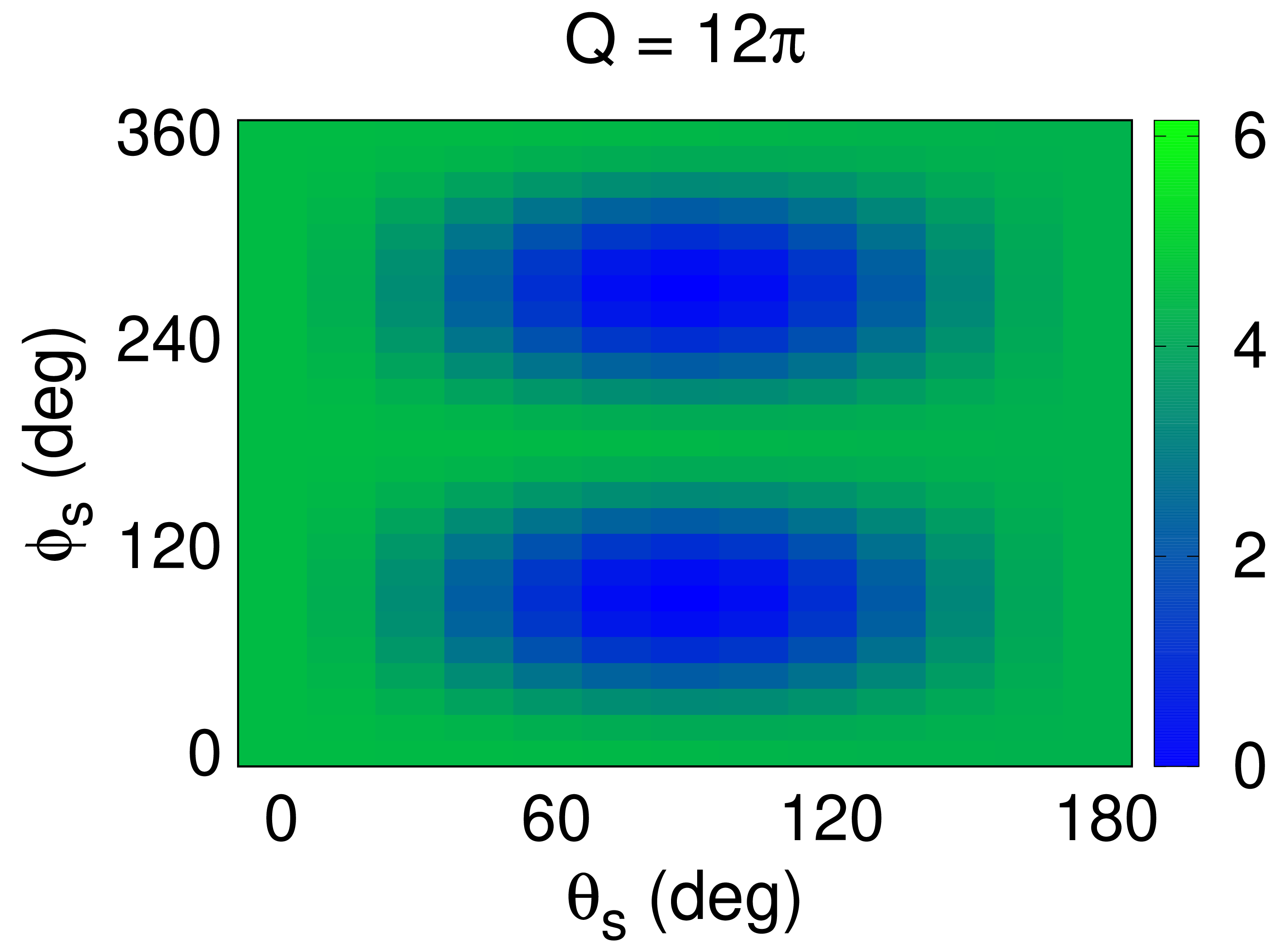}
  \caption{}
  \label{Fig_2fs_DCS_Aconly_12pi}
\end{subfigure}
\begin{subfigure}{.23\linewidth}
    \includegraphics[width=\linewidth]{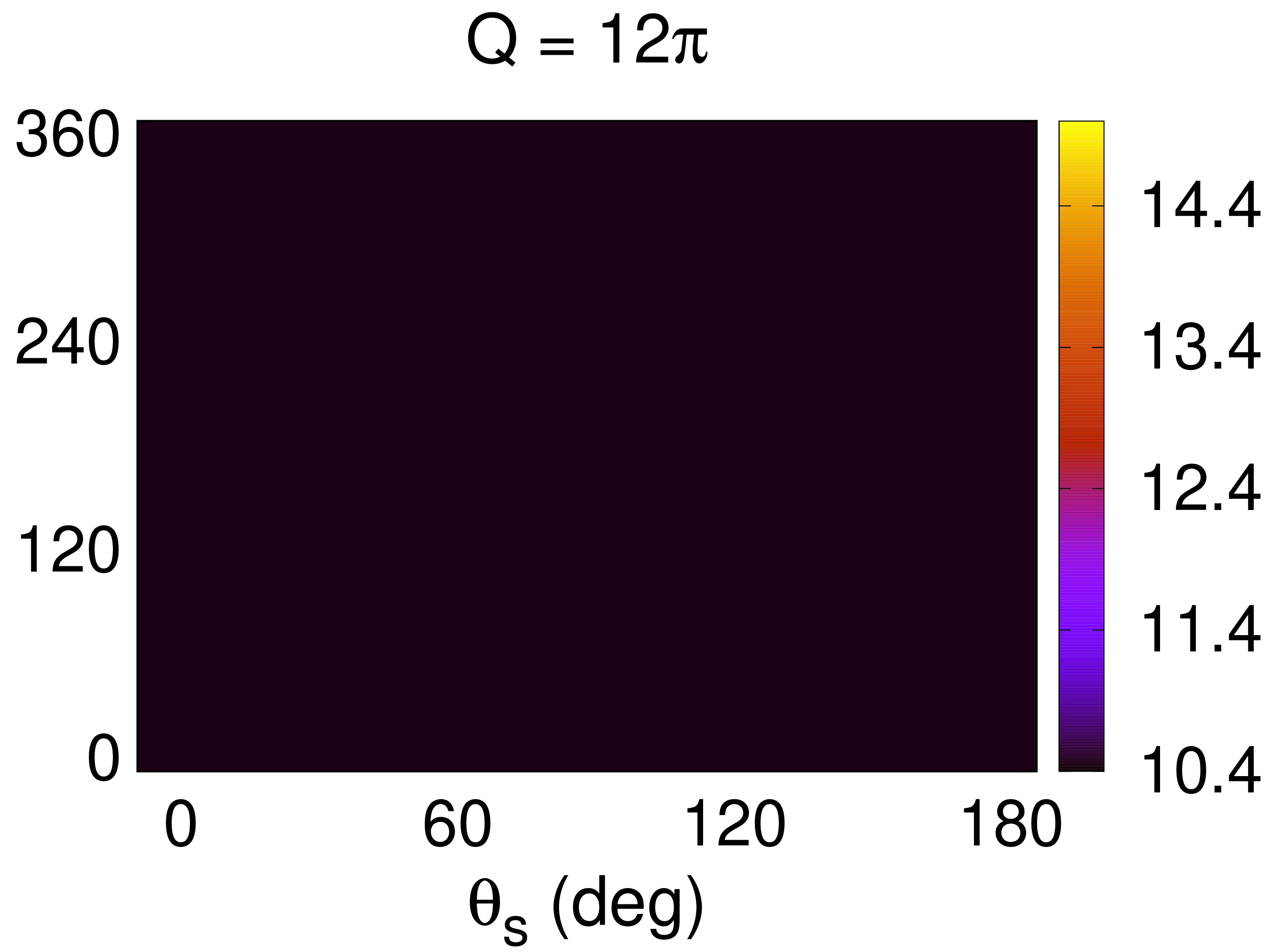}
  \caption{}
  \label{Fig_2fs_DCS_AcOFF_12pi}
\end{subfigure} 
\begin{subfigure}{.23\linewidth}
  \includegraphics[width=\linewidth]{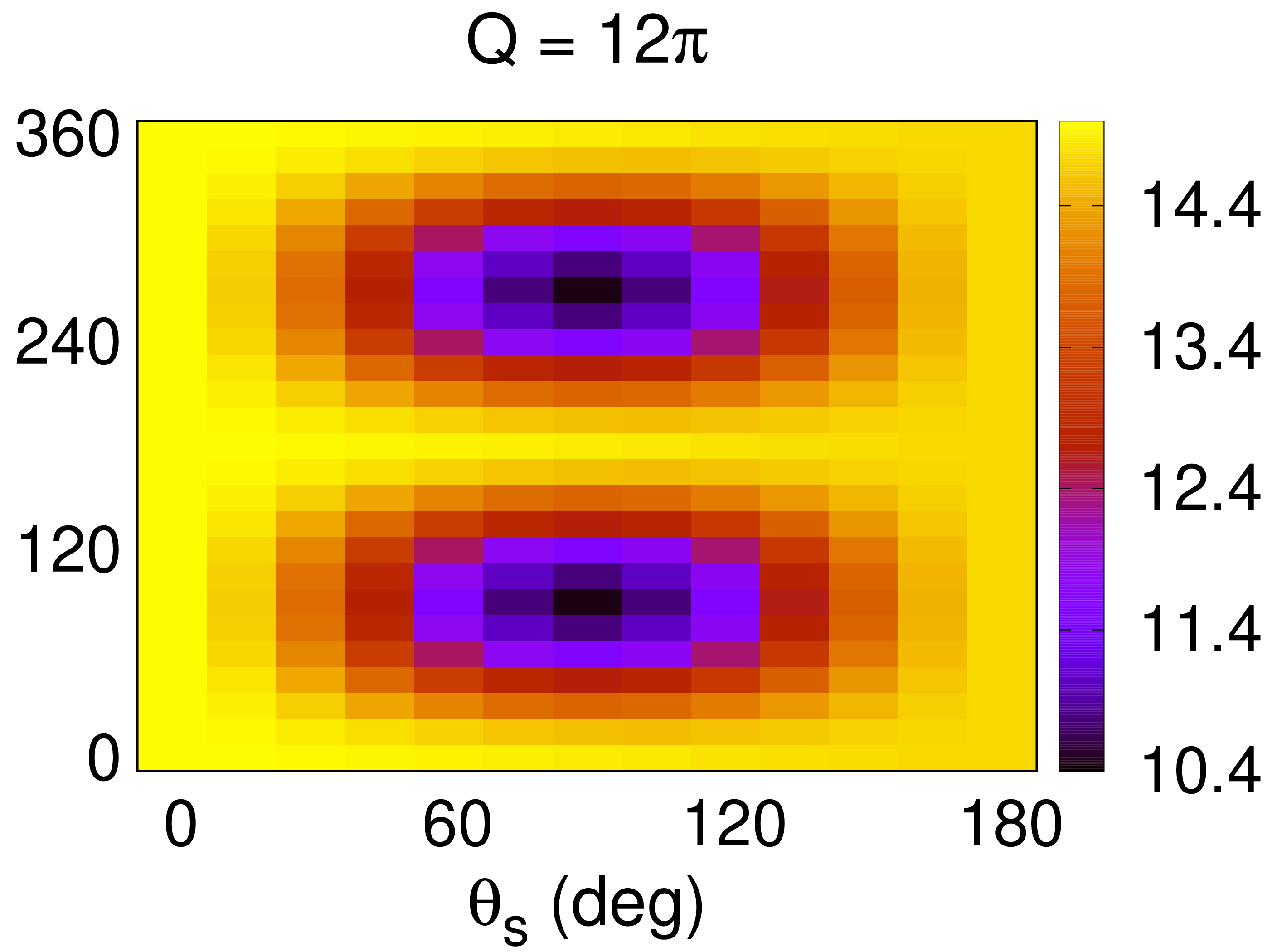}
  \caption{}
  \label{Fig_2fs_DCS_incoherentsum_12pi}
\end{subfigure}
\begin{subfigure}{.25\linewidth}
  \includegraphics[width=\linewidth]{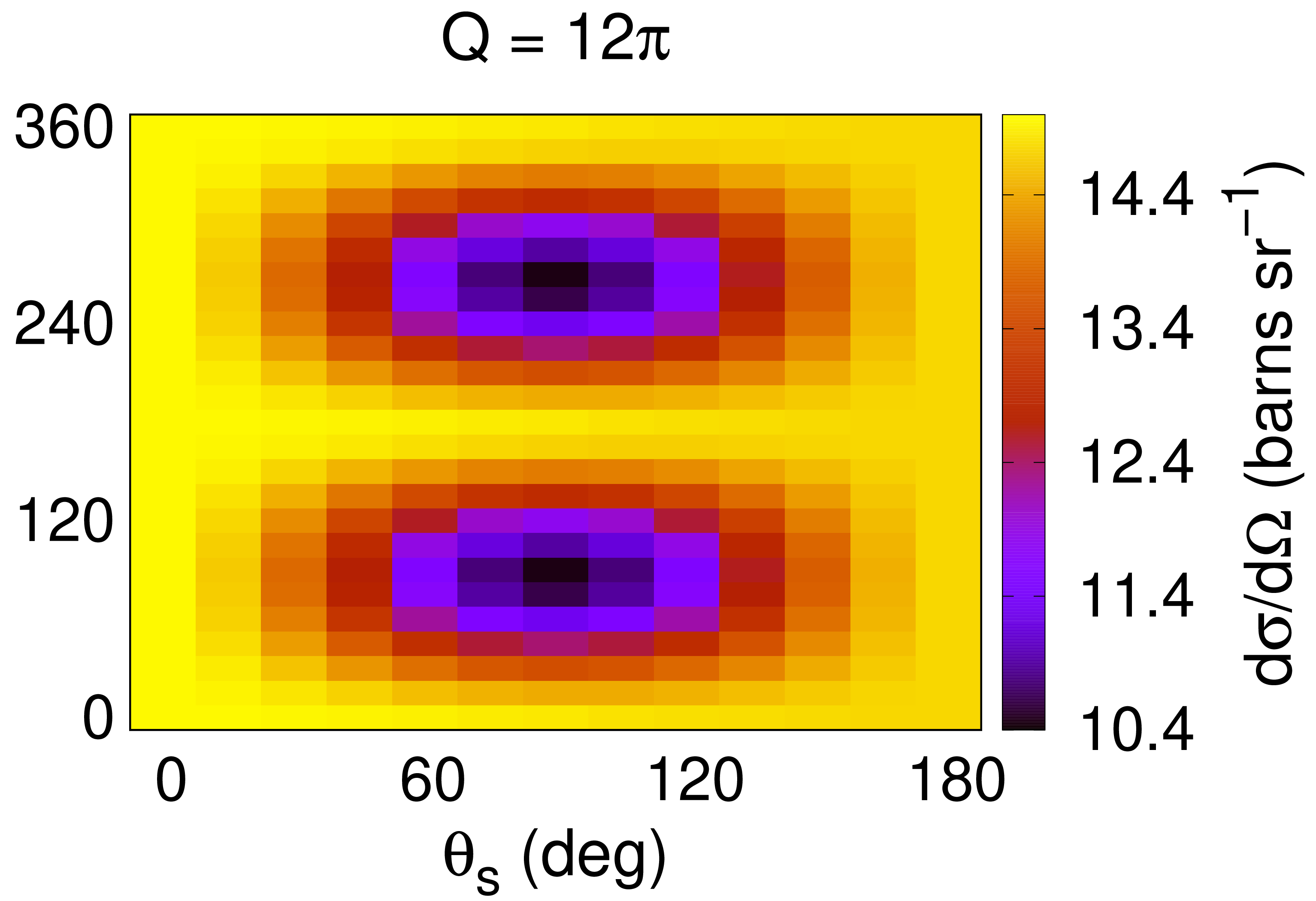}
  \caption{}
  \label{Fig_2fs_dcs_12pi}
\end{subfigure}
\end{minipage}
\caption{Angular distribution of the DCS calculated for a pulse duration of $t_{wid} = 2$~fs for $2\pi$-type pulses from only the elastic scattering channel (Fig.~\ref{Fig_2fs_DCS_Aconly_12pi}), from only the resonant fluorescence channel (Fig.~\ref{Fig_2fs_DCS_AcOFF_12pi}), an incoherent sum of the two channels (Fig.~\ref{Fig_2fs_DCS_incoherentsum_12pi}), and a coherent sum of the two channels (Fig.~\ref{Fig_2fs_dcs_12pi}).
}
\label{Fig_2fs_dcs_channels_2Pi}
\end{figure*}



Next, we examine the effects of Rabi oscillations on the scattering response for different pulse areas and two pulse durations of 0.25 and 2 fs.  To determine the contributions from elastic scattering and resonant fluorescence channels, we compute the DCSs [Eq.~\ref{dcs}] of these channels individually and compare them with the DCSs from the complete scattering response that includes both channels coherently.

\subsubsection{Elastic scattering channel}

We first examine the contribution from the elastic scattering channel for both the $\pi$- and $2\pi$-type pulses.  Figs.~\ref{Fig_DCS_Aconly_5pi} and \ref{Fig_DCS_Aconly_6pi} show that the DCS from the elastic scattering has a strong angular dependence, which is independent of the pulse parameter. The elastic scattering response can be understood from a simple description,
\begin{equation} \label{eqn:elastic_dcs}
\left( \frac{d \sigma}{d \Omega} \right)_{\boldsymbol{A}^2} = \frac{1}{\omega_{in}}\frac{d \sigma_{th}}{d \Omega} \int_{-\infty}^{+ \infty}\!\! dt ~ I(t) \sum_{m} |f_{m} (\boldsymbol{q}) |^2 N_{m}(t) 
\end{equation}
where 
\begin{equation}
\frac{d \sigma_{th}}{d \Omega} = r^2_0 \left( \cos^2\theta_s \sin^2\phi_s + \cos^2\phi_s \right)
\end{equation}
is the Thomson scattering cross section summed over the two outgoing field polarizations with $\theta_s$ and $\phi_s$ defined in Fig. \ref{Schematic_diagram}~\cite{Companion_letter}.  $N_{m}(t)$ and  $f_m(\boldsymbol{q})$ are the population and form factor of the state $m$ respectively, and $\boldsymbol{q} = \boldsymbol{k}_{in} - \boldsymbol{k} $. In the forward direction, the form factor is equal to the number of electrons (9) in Ne\textsuperscript{+}.  From Eq.~(\ref{eqn:elastic_dcs}), it is clear that the angular distribution comes from the Thomson scattering DCS and the form factor, and is independent of the pulse intensity and duration.

Figs.~\ref{Fig_DCS_Aconly_5pi} and \ref{Fig_DCS_Aconly_6pi} also reveal that the differential cross-section decreases slightly as the pulse area (or pulse intensity) increases due to increasing photoionization. However, this effect is relatively small for the range of intensities (see Fig.~\ref{Fig_Rabi_dynamics_p25fs}d) explored in the Figs.~\ref{Fig_dcs_5pi} and \ref{Fig_dcs_6pi}.
We note that our calculated scattering response excludes the scattering from free electrons and ionized species beyond 1+ . Interestingly, for very large pulse areas ($\sim10^{20}$~W/cm\textsuperscript{2}), we find that the scattering probability nearly saturates rather than completely vanishes. This occurs because elastic scattering from the bound electrons happens during the rising part of the pulse envelope, before the higher intensity from the later part of the pulse causes substantial photoionization \cite{Barty-NatPho-2012, Ho-2016-PRA}.  
For such intensities, the scattering from the ionized species and the free electrons will contribute significantly to the background \cite{Ho-2016-PRA}.

\subsubsection{Resonant Fluorescence channel}

Next, we examine the angular distribution of the resonant fluorescence channel, also referred to as resonant scattering~\cite{Sakurai_adv}.  A resonant fluorescence photon can originate from the de-excitation of the core-hole state to any one of the degenerate states $1s(2p_{-1})^{-1}$, $1s(2p_{0})^{-1}$, $1s(2p_{1})^{-1}$ (see Fig.~\ref{Schematic_diagram}).  However, the fluorescence matrix elements constrain which final electronic states are allowed for a given scattering angle and outgoing photon polarization. For instance, in the case that $\theta_s = 90\degree , \phi_s = 0 \degree$ and $\epsilon = \hat{z}$ (the example used for benchmarking in Sec.~\ref{sec_benchmarking}),  $1s(2p_{0})^{-1}$ is the only final state from fluorescence. This justifies a 2-state description for resonant fluorescence for those angles and choice of polarization. For other sets of ($\theta_s$ , $\phi_s$, $\epsilon$), a 4-state description is needed. In addition, the scattered electrons in one of these final states can subsequently absorb another photon to reach the core-hole state during the pulse.
 
It is evident from Fig.~\ref{Fig_DCS_AcOFF_5pi} that the angular distribution of the resonant fluorescence channel is effectively isotropic to about 1\% which is the convergence limit of our calculated results. This behaviour is a consequence of the summing over all the final electronic states, outgoing photon polarizations and integrating over the outgoing photon energies. We note that, unlike the elastic scattering channel, the resonant fluorescence yield does not scale near-linearly with intensity in this parameter regime. Therefore, its DCS is not independent of the pulse intensity. 
For a given Ne\textsuperscript{+} atom, the amount of fluorescence is determined by the population of the core-hole excited state, which scales linearly with incident intensity only for weak-fields \cite{Ho-2021-SD}. As the field intensity increases, the resonant fluorescent yield exhibits sublinear increase until maximum occupancy of the core-excited state, which occurs for a $Q=\pi$ pulse.   At higher $Q$, the resonant fluorescence yield decreases from this maxima due to Rabi oscillations, photoionization and Auger channels. This is supported by the results for the DCS of the resonant fluorescence channel, which monotonically decreases with increase in pulse area.

The calculations show that the resonant fluorescent yields in $2\pi$-type pulses are lower than those to $\pi$-type pulses. This is because resonant fluorescence can occur only during the pulse for $2\pi$-type pulses, whereas for $\pi$-type pulses the fluorescence can take place also after the pulse.  For 2$\pi$-type pulses (Fig.~\ref{Fig_DCS_AcOFF_6pi}), the resonant fluorescence can occur only for the duration of the incident pulses because there exists no population in the core-excited state at the end of the pulse.  This is revealed by a comparison of the differential cross section of $Q = \pi$ and $Q = 2\pi$ (Figs.~\ref{Fig_DCS_AcOFF_5pi} \& \ref{Fig_DCS_AcOFF_6pi} and Figs.~\ref{Fig_2fs_DCS_AcOFF_11pi} \& \ref{Fig_2fs_DCS_AcOFF_12pi}). The $Q = 2\pi$ case has a lower DCS despite the increased incident intensity, showing substantially reduced resonant fluorescent yield. Interestingly, increasing the number of Rabi oscillations (increasing $Q$) for $2\pi$-type pulses increases the resonant fluorescence yield. This is because the time-averaged probability of the core-excited state increases with increase in pulse area. Since for a given pulse duration, the intensity increases quadratically with the pulse area, the DCS (Fig.~\ref{Fig_DCS_AcOFF_6pi}) nevertheless decreases as the pulse area increases from $Q = 2\pi$ to $Q = 6\pi$.


\subsubsection{Total coherent response}

To describe the complete single-atom response, we include contributions coherently from both resonant fluorescence and elastic scattering channels. During the pulse, an outgoing photon ($\boldsymbol{k}, \boldsymbol{\epsilon}$) can come from either elastic scattering from an occupied state or from resonant fluorescence. If the final electronic state of both pathways are the same, then these pathways are indistinguishable and can interfere. In the weak-field regime, the elastic scattering amplitude is typically negligible~\cite{Sakurai_adv} compared to the resonant fluorescence amplitude, and their ratio is intensity-independent [Eq.~(\ref{eqn_KH_dcs})]. Still, some interference between the elastic and resonant fluorescence channels appears to have been observed experimentally in spectroscopic measurements~\cite{Thomson_resonant_interference2012}. However, in the strong-field regime, this ratio of elastic to resonant fluorescence yield depends on intensity and the yields from both the channels can become comparable.  As a result, the interference between them can become substantial.

To illustrate the interference effect, Figs.~\ref{Fig_dcs_5pi} \& \ref{Fig_dcs_6pi} compares the results from the coherent sum (sum of amplitudes) and incoherent sum (sum of probabilities) of the two channels as a function of pulse area.  A comparison between the results of incoherent and coherent sums shows that they have different angular distributions and ranges for the DCSs.  For some values of $\theta_s$ and $\phi_s$, the total scattering response can be smaller or larger than pure resonant fluorescence.  Also, the angular distribution of the DCS in the coherent sum is different from that of elastic scattering channel alone. These differences result from the interference between the two channels, which depend on the scattering angles.  In the weak-field monochromatic regime, the interference can be understood from the Kramers-Heisenberg expression [Eq.~(\ref{eqn_KH_dcs})]. The phase difference between the elastic scattering amplitude and resonant fluorescence amplitude depends on the sign of the matrix elements and the direction of the incident and outgoing photon polarizations. Beyond the weak-field regime, when Rabi oscillations occur, both the phase and the amplitude of the two channels becomes important. This is evident from the source terms in Eq.~(\ref{eqn_psi1_nstate}) contributing to the total scattering probability amplitude. Since the total interference term is obtained by summing over all possible final electronic  states, $\boldsymbol{k}$, and $\boldsymbol{\epsilon}$, the interference also depends on the number of interfering pathways that exist (see Fig.~\ref{Channels_Interference_scheme}).

To illustrate the different interfering pathways, consider the case of forward scattering ($\theta_s$ = 0), as shown in Figs.~\ref{Channels_Interference_scheme}a and ~\ref{Channels_Interference_scheme}b.  For the two possible choices of outgoing photon polarizations ($\boldsymbol{\epsilon_1}, \boldsymbol{\epsilon_2}$ ), non-zero elastic scattering occurs only when $\boldsymbol{\epsilon} = \boldsymbol{\epsilon_{in}}$. By examining the different interfering pathways, it is clear that for this polarization configuration there can be interference for only two of the four possible final electronic states ($1s2p_{0}^{-1}$ and $1s^{-1}2p$) as the resonant fluorescence contribution to the other final states $1s2p_{+1}^{-1}$ and $1s2p_{-1}^{-1}$ are zero (Fig.~\ref{Channels_Interference_scheme}b). Similarly for orthogonal scattering $\theta_s$ = $90\degree$ (Figs.~\ref{Channels_Interference_scheme}c and \ref{Channels_Interference_scheme}d), it can be seen that interference for all the final states is possible for the choice of polarization, shown in Fig.~\ref{Channels_Interference_scheme}d. For the special cases of orthogonal scattering $\theta_s$ = $90\degree$ and $\phi_s = 90\degree, 270\degree$ (Fig.~\ref{Channels_Interference_scheme}e), there occurs no elastic scattering for any choice of outgoing photon polarization and hence no interference, making it the ideal angle for measuring pure resonant fluorescence. The phase difference between the two channels for a given final state depends on the probability amplitude of the states participating in the Rabi dynamics. Further details on the phase difference between the two channels is discussed in the Appendix~\ref{App_phasediff}.

From the figures showing the total coherent DCS, the effect of the interference term is clearly evident especially for $\theta_s$ = $90 \degree$ along the azimuthal direction. However the peaks in the total coherent DCS (Figs.~\ref{Fig_dcs_5pi}, \ref{Fig_dcs_6pi}, \ref{Fig_2fs_dcs_11pi}, and ~\ref{Fig_2fs_dcs_12pi}) is slightly shifted from the maximum and the minimum of the interference term (Appendix~\ref{App_phasediff}). This can be attributed to the fact that the coherent sum comprises both the incoherent sum and the interference term, and the incoherent sum also has an angular dependence due to the elastic scattering contribution. Another observed trend is that for large pulse areas, the size of the interference terms decreases. This is because during the pulse, the fraction of the resonant fluorescence yield outside the pulse bandwidth increases, reducing the amount of resonant fluorescence yield that is indistinguishable from the elastic scattering channel, whose bandwidth is similar to that of the incident x-ray pulse.  Our results suggest that for large pulse areas, the computation of the total coherent DCS can be approximated by calculating the DCS of elastic scattering and resonant fluorescence channels individually and summing them incoherently.

The effect of the interference is also evident in the energy spectrum at a given scattering angle. The results of the energy spectrum for a pulse duration of 0.25 fs for various pulse area are shown in Fig.~\ref{Fig_ES_p25fs}. The interference between the elastic scattering and resonant fluorescence channels is substantial, particularly for $2\pi$-type pulses. The calculations indicate that the maximum destructive interference and maximum constructive interference manifests for azimuthal angles of  $\phi_s = 45\degree$ and $\phi_s = 135 \degree$, respectively (see Appendix~\ref{App_phasediff} for explanation). The interference is more difficult to observe in the energy spectrum for $\pi$-type pulses, as the resonant fluorescence is 2 orders of magnitude larger than that elastic scattering yield. Conversely, we find a notable asymmetry in the total coherent response for $2\pi$-type pulses. This asymmetry is a direct consequence of the interference between the elastic scattering and resonant fluorescence channels. Interestingly, an asymmetry arising from interference of photoionization pathways during Rabi dynamics in the ultrafast regime has been reported recently in the XUV domain~\cite{nandi2022_observationRabidynamics}.

\subsubsection{Pulse Duration Dependence}

In addition to the calculations with 0.25-fs pulses, we also examine the scattering processes with a 2-fs pulse as a function of pulse areas (see Figs. \ref{Fig_2fs_dcs_11pi} and \ref{Fig_2fs_dcs_12pi}).  We find that the DCSs for the elastic scattering are largely independent of the pulse duration, whereas the resonant fluorescence channel shows a strong dependence on pulse duration. 

The role of pulse duration on the resonant fluorescence channel can be understood in the following manner: the resonant fluorescence yield depends on the time-averaged excited state population, which effectively depends on the pulse area. For a given pulse area of $\pi$-type pulse, the resonant fluorescence yield is nearly independent of the pulse duration, provided the pulse duration is smaller than the lifetime of the excited state and photoionization is minimal. However, a shorter pulse requires a larger incident intensity to achieve the same pulse area, resulting in a lower DCS for the resonant fluorescence channel. In the frequency domain, this is equivalent to a shorter pulse having a larger bandwidth, leading to a lower fraction of the pulse envelope being resonant and thus requiring higher intensity to achieve the same excited state population. 

For $2\pi$-type pulses, the resonant fluorescence occurs only during the pulse. Therefore, for a given pulse area of $2\pi$-type pulse, the resonant fluorescence yield tends to increase with the incident pulse duration. The results for the total DCS (coherent sum) are qualitatively similar to the results from 0.25 fs, although with a lower fraction of elastic scattering to resonant fluorescence. One notable difference is that the interference still manifests for relatively larger pulse areas, due to the smaller bandwidth of the pulse.

\subsubsection{Experimental outlook}

To observe the interference effects in energy spectrum, pulses that can produce elastic and resonant fluorescence with comparable amplitudes are ideal.  Fig.~\ref{Fig_ES_p25fs} suggests that, in comparison to $\pi$-type pulses, 2$\pi$-type pulses are better as they produce smaller fluorescence yield with their temporal emission restricted to the duration of the pulse, thus having higher elastic scattering to resonant fluorescence signal ratios. Additionally, for a given pulse area, a shorter pulse is preferred as it has a higher elastic scattering to resonant fluorescence signal ratio and reduces electronic and structural damage in molecules and extended samples~\cite{imagingsucrose_phay, largescale_rateequation_montecarlo_phay}.

To experimentally study the interference effects, temporally coherent x-ray pulses with some control over the pulse area (intensity) are required. For example, current XFEL facilities~\cite{attoXLEAP2020,Malyzhenkov-2020-PRR,Trebushinin-Photonics-2023}, which can produce near transform-limited coherent x-ray pulses, have the potential to enable such studies. Pulse area control can be achieved in seeded FELs. In unseeded FELs, the challenge of pulse area control can be mitigated in principle by using millions of shots with the approximate intended intensity. These shots can then be sorted post-measurement based on their pulse area, which can be estimated from the measured energy spectrum of the scattered photons. For a given pulse duration, the single-atom response strongly depends on the pulse area. 
 
For forward and small-angle scattering from a gas jet or clusters, multi-atom interference effects plays a role. However, for large angle scattering, as shown in the energy spectrum plots (Fig.~\ref{Fig_ES_p25fs}), a single-atom response can be sufficient. For example, if the total DCS can be written as, 
\begin{equation}
\begin{split}
\frac{d \sigma}{d \Omega} &= |\sum_{j} \mathcal{F}_{j} (q) e^{i \boldsymbol{q} \cdot \boldsymbol{R}_j}|^2 \\
&= \sum_{j} |\mathcal{F}_{j}|^2 + \sum_{j \neq m} \mathcal{F}^*_{m} \mathcal{F}_{j} e^{i \boldsymbol{q} \cdot (\boldsymbol{R}_j - \boldsymbol{R}_m) }
\end{split}
\end{equation}
where $\mathcal{F}_{j}$ and $\boldsymbol{R}_j$ are the effective form factor and the position of the $j$-th atom in the sample, respectively. When the separation between atoms is greater than the incident field wavelength, the second term vanishes for large scattering angles and a large number of atoms, resulting in an incoherent sum of contributions from each atom.  This occurs because for every constructive interference, there is a corresponding destructive interference. Thus, for large-angle scattering, a single-atom response may suffice.

\section{Conclusion and summary} \label{conclusion_summary}

In this work, we present an approach to model the effects of coherent electron dynamics, specifically Rabi oscillations, on resonant x-ray scattering from intense x-ray pulses. Our description goes beyond the weak-field Kramers-Heisenberg model for monochromatic fields and the widely adopted rate equation methods for nonlinear x-ray interactions. This approach for describing the coherent light-matter interaction calculates both the resonant fluorescence and elastic scattering probability amplitudes while accounting for competing processes such as Auger decay and photoionization.

We benchmark our results against the previous work on resonant fluorescence and rate equation approaches in the intense x-ray field regime, as well as the Kramers-Heisenberg DCS in the weak-field monochromatic regime. As an example, we apply the method to calculate the complete single-atom response for near-elastic scattering from Ne\textsuperscript{+}. This approach allows us to investigate the dependence of the total scattering response on pulse area (intensity), pulse duration, initial electronic state, final state of the scattered electron, and scattered photon momentum and polarization.

Using this approach, we examine the contributions from both the elastic scattering and resonant fluorescence channels and the role of interference between these channels in the strong-field limit. The interference exhibits a non-trivial dependence on intensity, leading to an intensity-dependent differential scattering cross section. We observe a notable difference in the DCS between coherent and incoherent sums at small pulse areas (Q), which diminishes at larger pulse areas. This observation suggests a simplified approach for the computation of the total scattering response at large pulse areas.

The interference is also found to cause an asymmetry in the energy spectrum. This interference can potentially be exploited to enhance the elastic scattering signal for x-ray imaging applications.

\section{Acknowledgements}
We are grateful to S. Cavaletto for valuable discussions on their work. A.V. thanks E. Pelimanni for helpful discussions on the experimental challenges involved. This work was supported by the U.S. Department of Energy, Office of Basic Energy Sciences, Division of Chemical Sciences, Geosciences, and Biosciences through Argonne National Laboratory. Argonne is a U.S. Department of Energy laboratory managed by UChicago Argonne, LLC, under Contract No. DE-AC02-06CH11357. We gratefully acknowledge the computing resources provided on Improv, a high-performance computing cluster operated by the Laboratory Computing Resource Center at Argonne National Laboratory.

\appendix
\section{Characterizing the phase difference between resonant fluorescence and elastic scattering amplitudes }\label{App_phasediff}

Here we discuss the phase difference between the elastic scattering and resonant fluorescence channels for different final states. Since Eq.~(\ref{eqn_psi1_nstate}) is a linear non-homogeneous differential equation, the solution when both source terms (non-homogeneous terms) are present is the sum of the solutions obtained with each source term. Therefore, the phase difference between the source terms provide insights into the phase difference between the amplitudes of the two channels. To simplify the notation, we denote the states in the following manner, $\ket{1} = \ket{ 1s2p_0^{-1} }$, $\ket{1_{\pm}} = \ket{ 1s2p_{\pm1}^{-1} }$, and $\ket{2} = \ket{ 1s^{-1}2p }$.

For the final scattered state $1s2p_0^{-1}$, Eq.~(\ref{eqn_psi1_nstate}) becomes,
\begin{widetext}
\begin{equation} \label{eqn_psi1_ex}
\begin{split}
i & \frac{\partial C^{(1)}_{1, \boldsymbol{k} \boldsymbol{\epsilon}} }{\partial t} - \bigg[ E_{1} C^{(1)}_{1, \boldsymbol{k} \boldsymbol{\epsilon}} + \boldsymbol{A}_C \cdot \sum\limits_{ j=1}^{n} C^{(1)}_{j, \boldsymbol{k} \boldsymbol{\epsilon}} \bra{ 1 }\sum\limits_{ b=1}^{N}  \boldsymbol{P}_b \ket{\psi_{j} } \bigg] \\
 & = ~\sqrt{\frac{2\pi}{ V\omega_{k}} } e^{i\omega_{k} t } \bigg[ C^{(0)}_{2}(t) \boldsymbol{\epsilon}^* \cdot \bra{ 1 }\sum\limits_{ b=1}^{N} \boldsymbol{P}_b \ket{ 2 } 
   - \frac{i}{2}  A_0(t) C^{(0)}_{1}(t) e^{-i\omega_{in} t } ~\boldsymbol{\epsilon}^* \cdot \boldsymbol{\epsilon}_{in} 
\bra{ 1 }\sum\limits_{ b=1}^{N} e^{i ( \boldsymbol{k}_{in}  - \boldsymbol{k})\cdot \boldsymbol{r}_b}  \ket{ 1 } \bigg] .
\end{split}
\end{equation}
\end{widetext}
In this case, the resonant fluorescence amplitude is non-zero only if the outgoing photon polarization has a non-zero component along $\epsilon_{in}$. Therefore, the source terms ($\mathit{s_0}$) can be simplified in the following manner,
\begin{equation} \label{eqn_sourceterm_step1}
\begin{split}
 \mathit{s_0} = & \sqrt{\frac{2\pi}{ V\omega_{k}} } \boldsymbol{\epsilon}^* \cdot \boldsymbol{\epsilon}_{in} e^{i\omega_{k} t } \bigg[ -i \mu ~\Delta E ~C^{(0)}_{2}(t)   \\
 &- \frac{i}{2}  A_0(t) C^{(0)}_{1}(t) e^{-i\omega_{in} t } f(\boldsymbol{q}) \bigg]. 
\end{split}
\end{equation}
Here $\mu$ is the transition dipole moment which is real and $\Delta E = E_{2} - E_1$ is the difference in the binding energy between the two states. The quantity $f(\boldsymbol{q})$ is the ground-state off-resonant form factor which is real. Eq.~\ref{eqn_psi1_ex}) and its coupled Schr\"odinger equation [Eq.~(\ref{eqn_psi0_nstate})] cannot be solved analytically. However, one can gain insights by substituting the monochromatic case solution for the amplitudes of the states involved in Rabi dynamics. In the monochromatic resonant case when there is no decay, the analytical solution for the chosen field [Eq.~(\ref{classicalvectorpotential})] is,
\begin{equation} \label{eqn_rabisolution_monogs}
C^{(0)}_{1}(t) = \frac{1}{\sqrt{3}} \exp(-i E_{1} t ) \cos(\frac{\Omega}{2}t)
\end{equation}
\begin{equation} \label{eqn_rabisolution_monoex}
C^{(0)}_{2}(t) = \frac{-i}{\sqrt{3}} \exp(-i E_{2}t) \sin(\frac{\Omega}{2}t)
\end{equation}
We note that for the case of an incident pulse, the true populations resemble the monochromatic solution modulated by the incident pulse envelope and a decay term. Substituting the monochromatic solution for the Rabi problem and dropping an overall phase factor, the source terms are of the form,
\begin{equation} \label{eqn_sourceterm_step2}
\begin{split}
\mathit{s_0} = & -\sqrt{\frac{2\pi}{3 V\omega_{k}} } \boldsymbol{\epsilon}^* \cdot \boldsymbol{\epsilon}_{in}  \bigg[ \mu ~\Delta E e^{i(\omega_{k} - \Delta E + \frac{i\Gamma_2}{2} ) t } \sin(\frac{\Omega}{2}t)  \\
 & ~~  \frac{i}{2}  A_0(t) e^{i(\omega_k - \omega_{in} + \frac{i\Gamma_1}{2}) t } f(\boldsymbol{q}) \cos(\frac{\Omega}{2}t) \bigg].
\end{split}
\end{equation}
It is evident from the above expression [Eq.~\ref{eqn_sourceterm_step2}] that under resonant conditions and at the centre of the energy spectrum defined in Eq.~\ref{energy_spectrum}, i.e. when $\omega_k = \Delta E$, the first term is purely real, and the second term is purely imaginary. Therefore, there is no interference for the final state irrespective of the incident field intensity or scattering angle. However for $\omega_k - \Delta E \neq 0$, constructive or destructive interference can occur depending on the sign, resulting in an asymmetry in the energy spectrum similar to the one observed in Fig.~\ref{Fig_ES_p25fs}. Since the energy spectrum of resonant fluorescence and the elastic scattering channels are otherwise symmetric, the lack of interference at $\omega_k = \Delta E$ explains why for this final state, the interference does not manifest in the DCS, which is calculated by integrating over the scattered photon energy. Additionally a non-zero detuning ($\delta = \omega_{in} - \Delta E$) in the incident x-ray field can impart opposite phases of $e^{i t \delta /2}$ and $e^{-i t \delta/2}$ on the amplitudes of the two states involved in the Rabi dynamics. This causes the asymmetry in the energy spectrum to depend on the choice of detuning of the incident field analogous to the effect observed in the photoelectron spectrum in Ref.~\cite{nandi2022_observationRabidynamics}.

The above discussion also offers insights into the case where the initial state for the Ne\textsuperscript{+} atom is chosen to be $\ket{1}$. In this case, the resonant fluorescence channel remains qualitatively the same but the yield is increased by a factor of $3$, as the entire population can participate in the Rabi dynamics. However, the interference between elastic scattering and resonant fluorescence does not significantly manifest in the DCS given Eq.~\ref{eqn_sourceterm_step2}. Nevertheless, the asymmetry can still appear in the energy spectrum depending on the scattering angle and outgoing polarization direction.

Similar to the expression in Eq.~(\ref{eqn_sourceterm_step2}), one can write the source terms for the final states $1s(2p_{+1})^{-1}$ and $1s(2p_{-1})^{-1}$.
\begin{equation} \label{eqn_sourceterm_step3}
\begin{split}
\mathit{s_{\pm}} =  &\sqrt{\frac{2\pi}{ 3 V\omega_{k}} } \bigg[ \boldsymbol{\epsilon}^* \cdot \bra{ 1_{\pm} }\sum\limits_{ b=1}^{N} \boldsymbol{P}_b \ket{ 2 }  \\ 
& ~ ~ \times  e^{i(\omega_{k} - \Delta E + \frac{i\Gamma_2}{2}) t }  (-i)  \sin(\frac{\Omega}{2}t) \\
& - \frac{i}{2}  A_0(t) e^{i(\omega_k - \omega_{in} + \frac{i\Gamma_1}{2}) t } ~\boldsymbol{\epsilon}^* \cdot \boldsymbol{\epsilon}_{in} ~ f(\boldsymbol{q})  \bigg] .
\end{split}
\end{equation}
In the above expression, the resonant fluorescence term is only non-zero if $\boldsymbol{\epsilon}$ has a non-zero component in x or y-direction. For $\theta_s = 90\degree$, for which both $\boldsymbol{k}$ and $\boldsymbol{\epsilon}$ are in the y-z plane, Eq.~(\ref{eqn_sourceterm_step3}) can be further simplified as
\begin{equation} \label{eqn_sourceterm_step4}
\begin{split}
\mathit{s_{\pm}} =  & - \sqrt{\frac{2\pi}{ 3 V\omega_{k}} }  \bigg[ \Delta E~\mu_{\pm1,y} ~ e^{i(\omega_{k} - \Delta E + \frac{i\Gamma_2}{2} ) t } \boldsymbol{\epsilon}^* \cdot \hat{y} \sin(\frac{\Omega}{2}t) \\
   & ~~ + \frac{i}{2}  A_0(t) e^{i(\omega_k - \omega_{in} + \frac{i\Gamma_1}{2}) t } ~\boldsymbol{\epsilon}^* \cdot \hat{z} 
~f( \boldsymbol{q})  \bigg].
\end{split}
\end{equation}
Here $\mu_{\pm1,y}$ denotes the transition dipole moments along y-direction for the corresponding states. These matrix elements are purely imaginary.
Therefore for these final states, if both terms are non-zero, the interference is always purely constructive or purely destructive at $\omega_k = \Delta E$. This leads to the interference pattern in the total DCS (coherent sum). The product of the two terms reach a maxima only when the azimuthal angle ($\phi_s$) is an odd multiple of $45\degree$. This results in the maximum and minimum of the interference term to occur at these azimuthal angles (Fig.~\ref{Channels_Interference_scheme}). A subtle point here is that for these final states ($\ket{1_{\pm}}$), the occurrence of constructive or destructive interference at the central peak is influenced by the incident field intensity. This is evident from the first term in Eq.~(\ref{eqn_sourceterm_step4}), which depends on $\sin(\frac{\Omega}{2}t)$ and can be either positive or negative.
Finally, we note that some of the properties discussed in this section can also be deduced in the weak-field limit from the Kramers-Heisenberg expression.

\bibliography{References.bib}

\begin{thebibliography}{45}%
\makeatletter
\providecommand \@ifxundefined [1]{%
 \@ifx{#1\undefined}
}%
\providecommand \@ifnum [1]{%
 \ifnum #1\expandafter \@firstoftwo
 \else \expandafter \@secondoftwo
 \fi
}%
\providecommand \@ifx [1]{%
 \ifx #1\expandafter \@firstoftwo
 \else \expandafter \@secondoftwo
 \fi
}%
\providecommand \natexlab [1]{#1}%
\providecommand \enquote  [1]{``#1''}%
\providecommand \bibnamefont  [1]{#1}%
\providecommand \bibfnamefont [1]{#1}%
\providecommand \citenamefont [1]{#1}%
\providecommand \href@noop [0]{\@secondoftwo}%
\providecommand \href [0]{\begingroup \@sanitize@url \@href}%
\providecommand \@href[1]{\@@startlink{#1}\@@href}%
\providecommand \@@href[1]{\endgroup#1\@@endlink}%
\providecommand \@sanitize@url [0]{\catcode `\\12\catcode `\$12\catcode `\&12\catcode `\#12\catcode `\^12\catcode `\_12\catcode `\%12\relax}%
\providecommand \@@startlink[1]{}%
\providecommand \@@endlink[0]{}%
\providecommand \url  [0]{\begingroup\@sanitize@url \@url }%
\providecommand \@url [1]{\endgroup\@href {#1}{\urlprefix }}%
\providecommand \urlprefix  [0]{URL }%
\providecommand \Eprint [0]{\href }%
\providecommand \doibase [0]{https://doi.org/}%
\providecommand \selectlanguage [0]{\@gobble}%
\providecommand \bibinfo  [0]{\@secondoftwo}%
\providecommand \bibfield  [0]{\@secondoftwo}%
\providecommand \translation [1]{[#1]}%
\providecommand \BibitemOpen [0]{}%
\providecommand \bibitemStop [0]{}%
\providecommand \bibitemNoStop [0]{.\EOS\space}%
\providecommand \EOS [0]{\spacefactor3000\relax}%
\providecommand \BibitemShut  [1]{\csname bibitem#1\endcsname}%
\let\auto@bib@innerbib\@empty
\bibitem [{\citenamefont {Kanter}\ \emph {et~al.}(2011)\citenamefont {Kanter}, \citenamefont {Kr\"assig}, \citenamefont {Li}, \citenamefont {March}, \citenamefont {Ho}, \citenamefont {Rohringer}, \citenamefont {Santra}, \citenamefont {Southworth}, \citenamefont {DiMauro}, \citenamefont {Doumy}, \citenamefont {Roedig}, \citenamefont {Berrah}, \citenamefont {Fang}, \citenamefont {Hoener}, \citenamefont {Bucksbaum}, \citenamefont {Ghimire}, \citenamefont {Reis}, \citenamefont {Bozek}, \citenamefont {Bostedt}, \citenamefont {Messerschmidt},\ and\ \citenamefont {Young}}]{Hiddenresonance_kanter}%
  \BibitemOpen
  \bibfield  {author} {\bibinfo {author} {\bibfnamefont {E.~P.}\ \bibnamefont {Kanter}}, \bibinfo {author} {\bibfnamefont {B.}~\bibnamefont {Kr\"assig}}, \bibinfo {author} {\bibfnamefont {Y.}~\bibnamefont {Li}}, \bibinfo {author} {\bibfnamefont {A.~M.}\ \bibnamefont {March}}, \bibinfo {author} {\bibfnamefont {P.}~\bibnamefont {Ho}}, \bibinfo {author} {\bibfnamefont {N.}~\bibnamefont {Rohringer}}, \bibinfo {author} {\bibfnamefont {R.}~\bibnamefont {Santra}}, \bibinfo {author} {\bibfnamefont {S.~H.}\ \bibnamefont {Southworth}}, \bibinfo {author} {\bibfnamefont {L.~F.}\ \bibnamefont {DiMauro}}, \bibinfo {author} {\bibfnamefont {G.}~\bibnamefont {Doumy}}, \bibinfo {author} {\bibfnamefont {C.~A.}\ \bibnamefont {Roedig}}, \bibinfo {author} {\bibfnamefont {N.}~\bibnamefont {Berrah}}, \bibinfo {author} {\bibfnamefont {L.}~\bibnamefont {Fang}}, \bibinfo {author} {\bibfnamefont {M.}~\bibnamefont {Hoener}}, \bibinfo {author} {\bibfnamefont {P.~H.}\ \bibnamefont {Bucksbaum}}, \bibinfo {author} {\bibfnamefont
  {S.}~\bibnamefont {Ghimire}}, \bibinfo {author} {\bibfnamefont {D.~A.}\ \bibnamefont {Reis}}, \bibinfo {author} {\bibfnamefont {J.~D.}\ \bibnamefont {Bozek}}, \bibinfo {author} {\bibfnamefont {C.}~\bibnamefont {Bostedt}}, \bibinfo {author} {\bibfnamefont {M.}~\bibnamefont {Messerschmidt}},\ and\ \bibinfo {author} {\bibfnamefont {L.}~\bibnamefont {Young}},\ }\bibfield  {title} {\bibinfo {title} {Unveiling and driving hidden resonances with high-fluence, high-intensity x-ray pulses},\ }\href {https://doi.org/10.1103/PhysRevLett.107.233001} {\bibfield  {journal} {\bibinfo  {journal} {Phys. Rev. Lett.}\ }\textbf {\bibinfo {volume} {107}},\ \bibinfo {pages} {233001} (\bibinfo {year} {2011})}\BibitemShut {NoStop}%
\bibitem [{\citenamefont {Rohringer}\ and\ \citenamefont {Santra}(2008)}]{Auger_Rabi_Santra}%
  \BibitemOpen
  \bibfield  {author} {\bibinfo {author} {\bibfnamefont {N.}~\bibnamefont {Rohringer}}\ and\ \bibinfo {author} {\bibfnamefont {R.}~\bibnamefont {Santra}},\ }\bibfield  {title} {\bibinfo {title} {Resonant auger effect at high x-ray intensity},\ }\href {https://doi.org/10.1103/PhysRevA.77.053404} {\bibfield  {journal} {\bibinfo  {journal} {Phys. Rev. A}\ }\textbf {\bibinfo {volume} {77}},\ \bibinfo {pages} {053404} (\bibinfo {year} {2008})}\BibitemShut {NoStop}%
\bibitem [{\citenamefont {Cavaletto}\ \emph {et~al.}(2012)\citenamefont {Cavaletto}, \citenamefont {Buth}, \citenamefont {Harman}, \citenamefont {Kanter}, \citenamefont {Southworth}, \citenamefont {Young},\ and\ \citenamefont {Keitel}}]{Cavaletto_ResFluor_PRA}%
  \BibitemOpen
  \bibfield  {author} {\bibinfo {author} {\bibfnamefont {S.~M.}\ \bibnamefont {Cavaletto}}, \bibinfo {author} {\bibfnamefont {C.}~\bibnamefont {Buth}}, \bibinfo {author} {\bibfnamefont {Z.}~\bibnamefont {Harman}}, \bibinfo {author} {\bibfnamefont {E.~P.}\ \bibnamefont {Kanter}}, \bibinfo {author} {\bibfnamefont {S.~H.}\ \bibnamefont {Southworth}}, \bibinfo {author} {\bibfnamefont {L.}~\bibnamefont {Young}},\ and\ \bibinfo {author} {\bibfnamefont {C.~H.}\ \bibnamefont {Keitel}},\ }\bibfield  {title} {\bibinfo {title} {Resonance fluorescence in ultrafast and intense x-ray free-electron-laser pulses},\ }\href {https://doi.org/10.1103/PhysRevA.86.033402} {\bibfield  {journal} {\bibinfo  {journal} {Phys. Rev. A}\ }\textbf {\bibinfo {volume} {86}},\ \bibinfo {pages} {033402} (\bibinfo {year} {2012})}\BibitemShut {NoStop}%
\bibitem [{\citenamefont {Li}\ \emph {et~al.}(2020)\citenamefont {Li}, \citenamefont {Labeye}, \citenamefont {Ho}, \citenamefont {Gaarde},\ and\ \citenamefont {Young}}]{Kai_phay_Neplus}%
  \BibitemOpen
  \bibfield  {author} {\bibinfo {author} {\bibfnamefont {K.}~\bibnamefont {Li}}, \bibinfo {author} {\bibfnamefont {M.}~\bibnamefont {Labeye}}, \bibinfo {author} {\bibfnamefont {P.~J.}\ \bibnamefont {Ho}}, \bibinfo {author} {\bibfnamefont {M.~B.}\ \bibnamefont {Gaarde}},\ and\ \bibinfo {author} {\bibfnamefont {L.}~\bibnamefont {Young}},\ }\bibfield  {title} {\bibinfo {title} {Resonant propagation of x rays from the linear to the nonlinear regime},\ }\href {https://doi.org/10.1103/PhysRevA.102.053113} {\bibfield  {journal} {\bibinfo  {journal} {Phys. Rev. A}\ }\textbf {\bibinfo {volume} {102}},\ \bibinfo {pages} {053113} (\bibinfo {year} {2020})}\BibitemShut {NoStop}%
\bibitem [{\citenamefont {Marinelli}\ \emph {et~al.}(2017)\citenamefont {Marinelli}, \citenamefont {MacArthur}, \citenamefont {Emma}, \citenamefont {Guetg}, \citenamefont {Field}, \citenamefont {Kharakh}, \citenamefont {Lutman}, \citenamefont {Ding},\ and\ \citenamefont {Huang}}]{Marinelli-APL-2017}%
  \BibitemOpen
  \bibfield  {author} {\bibinfo {author} {\bibfnamefont {A.}~\bibnamefont {Marinelli}}, \bibinfo {author} {\bibfnamefont {J.}~\bibnamefont {MacArthur}}, \bibinfo {author} {\bibfnamefont {P.}~\bibnamefont {Emma}}, \bibinfo {author} {\bibfnamefont {M.}~\bibnamefont {Guetg}}, \bibinfo {author} {\bibfnamefont {C.}~\bibnamefont {Field}}, \bibinfo {author} {\bibfnamefont {D.}~\bibnamefont {Kharakh}}, \bibinfo {author} {\bibfnamefont {A.~A.}\ \bibnamefont {Lutman}}, \bibinfo {author} {\bibfnamefont {Y.}~\bibnamefont {Ding}},\ and\ \bibinfo {author} {\bibfnamefont {Z.}~\bibnamefont {Huang}},\ }\bibfield  {title} {\bibinfo {title} {{Experimental demonstration of a single-spike hard-X-ray free-electron laser starting from noise}},\ }\href {https://doi.org/10.1063/1.4990716} {\bibfield  {journal} {\bibinfo  {journal} {Applied Physics Letters}\ }\textbf {\bibinfo {volume} {111}},\ \bibinfo {pages} {151101} (\bibinfo {year} {2017})},\ \Eprint
  {https://arxiv.org/abs/https://pubs.aip.org/aip/apl/article-pdf/doi/10.1063/1.4990716/13155677/151101\_1\_online.pdf} {https://pubs.aip.org/aip/apl/article-pdf/doi/10.1063/1.4990716/13155677/151101\_1\_online.pdf} \BibitemShut {NoStop}%
\bibitem [{\citenamefont {Huang}\ \emph {et~al.}(2017)\citenamefont {Huang}, \citenamefont {Ding}, \citenamefont {Feng}, \citenamefont {Hemsing}, \citenamefont {Huang}, \citenamefont {Krzywinski}, \citenamefont {Lutman}, \citenamefont {Marinelli}, \citenamefont {Maxwell},\ and\ \citenamefont {Zhu}}]{Huang-PRL-2017}%
  \BibitemOpen
  \bibfield  {author} {\bibinfo {author} {\bibfnamefont {S.}~\bibnamefont {Huang}}, \bibinfo {author} {\bibfnamefont {Y.}~\bibnamefont {Ding}}, \bibinfo {author} {\bibfnamefont {Y.}~\bibnamefont {Feng}}, \bibinfo {author} {\bibfnamefont {E.}~\bibnamefont {Hemsing}}, \bibinfo {author} {\bibfnamefont {Z.}~\bibnamefont {Huang}}, \bibinfo {author} {\bibfnamefont {J.}~\bibnamefont {Krzywinski}}, \bibinfo {author} {\bibfnamefont {A.~A.}\ \bibnamefont {Lutman}}, \bibinfo {author} {\bibfnamefont {A.}~\bibnamefont {Marinelli}}, \bibinfo {author} {\bibfnamefont {T.~J.}\ \bibnamefont {Maxwell}},\ and\ \bibinfo {author} {\bibfnamefont {D.}~\bibnamefont {Zhu}},\ }\bibfield  {title} {\bibinfo {title} {Generating single-spike hard x-ray pulses with nonlinear bunch compression in free-electron lasers},\ }\href {https://doi.org/10.1103/PhysRevLett.119.154801} {\bibfield  {journal} {\bibinfo  {journal} {Phys. Rev. Lett.}\ }\textbf {\bibinfo {volume} {119}},\ \bibinfo {pages} {154801} (\bibinfo {year} {2017})}\BibitemShut
  {NoStop}%
\bibitem [{\citenamefont {Duris}\ \emph {et~al.}(2020)\citenamefont {Duris}, \citenamefont {Li}, \citenamefont {Driver}, \citenamefont {Champenois}, \citenamefont {MacArthur}, \citenamefont {Lutman}, \citenamefont {Zhang}, \citenamefont {Rosenberger}, \citenamefont {Aldrich}, \citenamefont {Coffee} \emph {et~al.}}]{attoXLEAP2020}%
  \BibitemOpen
  \bibfield  {author} {\bibinfo {author} {\bibfnamefont {J.}~\bibnamefont {Duris}}, \bibinfo {author} {\bibfnamefont {S.}~\bibnamefont {Li}}, \bibinfo {author} {\bibfnamefont {T.}~\bibnamefont {Driver}}, \bibinfo {author} {\bibfnamefont {E.~G.}\ \bibnamefont {Champenois}}, \bibinfo {author} {\bibfnamefont {J.~P.}\ \bibnamefont {MacArthur}}, \bibinfo {author} {\bibfnamefont {A.~A.}\ \bibnamefont {Lutman}}, \bibinfo {author} {\bibfnamefont {Z.}~\bibnamefont {Zhang}}, \bibinfo {author} {\bibfnamefont {P.}~\bibnamefont {Rosenberger}}, \bibinfo {author} {\bibfnamefont {J.~W.}\ \bibnamefont {Aldrich}}, \bibinfo {author} {\bibfnamefont {R.}~\bibnamefont {Coffee}}, \emph {et~al.},\ }\bibfield  {title} {\bibinfo {title} {Tunable isolated attosecond x-ray pulses with gigawatt peak power from a free-electron laser},\ }\href@noop {} {\bibfield  {journal} {\bibinfo  {journal} {Nature Photonics}\ }\textbf {\bibinfo {volume} {14}},\ \bibinfo {pages} {30} (\bibinfo {year} {2020})}\BibitemShut {NoStop}%
\bibitem [{\citenamefont {Malyzhenkov}\ \emph {et~al.}(2020)\citenamefont {Malyzhenkov}, \citenamefont {Arbelo}, \citenamefont {Craievich}, \citenamefont {Dijkstal}, \citenamefont {Ferrari}, \citenamefont {Reiche}, \citenamefont {Schietinger}, \citenamefont {Jurani\ifmmode~\acute{c}\else \'{c}\fi{}},\ and\ \citenamefont {Prat}}]{Malyzhenkov-2020-PRR}%
  \BibitemOpen
  \bibfield  {author} {\bibinfo {author} {\bibfnamefont {A.}~\bibnamefont {Malyzhenkov}}, \bibinfo {author} {\bibfnamefont {Y.~P.}\ \bibnamefont {Arbelo}}, \bibinfo {author} {\bibfnamefont {P.}~\bibnamefont {Craievich}}, \bibinfo {author} {\bibfnamefont {P.}~\bibnamefont {Dijkstal}}, \bibinfo {author} {\bibfnamefont {E.}~\bibnamefont {Ferrari}}, \bibinfo {author} {\bibfnamefont {S.}~\bibnamefont {Reiche}}, \bibinfo {author} {\bibfnamefont {T.}~\bibnamefont {Schietinger}}, \bibinfo {author} {\bibfnamefont {P.}~\bibnamefont {Jurani\ifmmode~\acute{c}\else \'{c}\fi{}}},\ and\ \bibinfo {author} {\bibfnamefont {E.}~\bibnamefont {Prat}},\ }\bibfield  {title} {\bibinfo {title} {Single- and two-color attosecond hard x-ray free-electron laser pulses with nonlinear compression},\ }\href {https://doi.org/10.1103/PhysRevResearch.2.042018} {\bibfield  {journal} {\bibinfo  {journal} {Phys. Rev. Res.}\ }\textbf {\bibinfo {volume} {2}},\ \bibinfo {pages} {042018} (\bibinfo {year} {2020})}\BibitemShut {NoStop}%
\bibitem [{\citenamefont {Trebushinin}\ \emph {et~al.}(2023)\citenamefont {Trebushinin}, \citenamefont {Geloni}, \citenamefont {Serkez}, \citenamefont {Mercurio}, \citenamefont {Gerasimova}, \citenamefont {Maltezopoulos}, \citenamefont {Guetg},\ and\ \citenamefont {Schneidmiller}}]{Trebushinin-Photonics-2023}%
  \BibitemOpen
  \bibfield  {author} {\bibinfo {author} {\bibfnamefont {A.}~\bibnamefont {Trebushinin}}, \bibinfo {author} {\bibfnamefont {G.}~\bibnamefont {Geloni}}, \bibinfo {author} {\bibfnamefont {S.}~\bibnamefont {Serkez}}, \bibinfo {author} {\bibfnamefont {G.}~\bibnamefont {Mercurio}}, \bibinfo {author} {\bibfnamefont {N.}~\bibnamefont {Gerasimova}}, \bibinfo {author} {\bibfnamefont {T.}~\bibnamefont {Maltezopoulos}}, \bibinfo {author} {\bibfnamefont {M.}~\bibnamefont {Guetg}},\ and\ \bibinfo {author} {\bibfnamefont {E.}~\bibnamefont {Schneidmiller}},\ }\bibfield  {title} {\bibinfo {title} {Experimental demonstration of attoseconds-at-harmonics at the sase3 undulator of the european xfel},\ }\bibfield  {journal} {\bibinfo  {journal} {Photonics}\ }\textbf {\bibinfo {volume} {10}},\ \href {https://doi.org/10.3390/photonics10020131} {10.3390/photonics10020131} (\bibinfo {year} {2023})\BibitemShut {NoStop}%
\bibitem [{\citenamefont {Ho}\ \emph {et~al.}(2020{\natexlab{a}})\citenamefont {Ho}, \citenamefont {Daurer}, \citenamefont {Hantke}, \citenamefont {Bielecki}, \citenamefont {Al~Haddad}, \citenamefont {Bucher}, \citenamefont {Doumy}, \citenamefont {Ferguson}, \citenamefont {Fl{\"u}ckiger}, \citenamefont {Gorkhover}, \citenamefont {Iwan}, \citenamefont {Knight}, \citenamefont {Moeller}, \citenamefont {Osipov}, \citenamefont {Ray}, \citenamefont {Southworth}, \citenamefont {Svenda}, \citenamefont {Timneanu}, \citenamefont {Ulmer}, \citenamefont {Walter}, \citenamefont {Hajdu}, \citenamefont {Young}, \citenamefont {Maia},\ and\ \citenamefont {Bostedt}}]{Ho-2020-NatComm}%
  \BibitemOpen
  \bibfield  {author} {\bibinfo {author} {\bibfnamefont {P.~J.}\ \bibnamefont {Ho}}, \bibinfo {author} {\bibfnamefont {B.~J.}\ \bibnamefont {Daurer}}, \bibinfo {author} {\bibfnamefont {M.~F.}\ \bibnamefont {Hantke}}, \bibinfo {author} {\bibfnamefont {J.}~\bibnamefont {Bielecki}}, \bibinfo {author} {\bibfnamefont {A.}~\bibnamefont {Al~Haddad}}, \bibinfo {author} {\bibfnamefont {M.}~\bibnamefont {Bucher}}, \bibinfo {author} {\bibfnamefont {G.}~\bibnamefont {Doumy}}, \bibinfo {author} {\bibfnamefont {K.~R.}\ \bibnamefont {Ferguson}}, \bibinfo {author} {\bibfnamefont {L.}~\bibnamefont {Fl{\"u}ckiger}}, \bibinfo {author} {\bibfnamefont {T.}~\bibnamefont {Gorkhover}}, \bibinfo {author} {\bibfnamefont {B.}~\bibnamefont {Iwan}}, \bibinfo {author} {\bibfnamefont {C.}~\bibnamefont {Knight}}, \bibinfo {author} {\bibfnamefont {S.}~\bibnamefont {Moeller}}, \bibinfo {author} {\bibfnamefont {T.}~\bibnamefont {Osipov}}, \bibinfo {author} {\bibfnamefont {D.}~\bibnamefont {Ray}}, \bibinfo {author} {\bibfnamefont {S.~H.}\
  \bibnamefont {Southworth}}, \bibinfo {author} {\bibfnamefont {M.}~\bibnamefont {Svenda}}, \bibinfo {author} {\bibfnamefont {N.}~\bibnamefont {Timneanu}}, \bibinfo {author} {\bibfnamefont {A.}~\bibnamefont {Ulmer}}, \bibinfo {author} {\bibfnamefont {P.}~\bibnamefont {Walter}}, \bibinfo {author} {\bibfnamefont {J.}~\bibnamefont {Hajdu}}, \bibinfo {author} {\bibfnamefont {L.}~\bibnamefont {Young}}, \bibinfo {author} {\bibfnamefont {F.~R. N.~C.}\ \bibnamefont {Maia}},\ and\ \bibinfo {author} {\bibfnamefont {C.}~\bibnamefont {Bostedt}},\ }\bibfield  {title} {\bibinfo {title} {The role of transient resonances for ultra-fast imaging of single sucrose nanoclusters},\ }\href@noop {} {\bibfield  {journal} {\bibinfo  {journal} {Nature Communications}\ }\textbf {\bibinfo {volume} {11}},\ \bibinfo {pages} {167} (\bibinfo {year} {2020}{\natexlab{a}})}\BibitemShut {NoStop}%
\bibitem [{\citenamefont {Kuschel}\ \emph {et~al.}(2022)\citenamefont {Kuschel}, \citenamefont {Ho}, \citenamefont {Haddad}, \citenamefont {Zimmermann}, \citenamefont {Flueckiger}, \citenamefont {Ware}, \citenamefont {Duris}, \citenamefont {MacArthur}, \citenamefont {Lutman}, \citenamefont {Lin} \emph {et~al.}}]{Tais_2022_Xenon_preprint}%
  \BibitemOpen
  \bibfield  {author} {\bibinfo {author} {\bibfnamefont {S.}~\bibnamefont {Kuschel}}, \bibinfo {author} {\bibfnamefont {P.~J.}\ \bibnamefont {Ho}}, \bibinfo {author} {\bibfnamefont {A.~A.}\ \bibnamefont {Haddad}}, \bibinfo {author} {\bibfnamefont {F.}~\bibnamefont {Zimmermann}}, \bibinfo {author} {\bibfnamefont {L.}~\bibnamefont {Flueckiger}}, \bibinfo {author} {\bibfnamefont {M.~R.}\ \bibnamefont {Ware}}, \bibinfo {author} {\bibfnamefont {J.}~\bibnamefont {Duris}}, \bibinfo {author} {\bibfnamefont {J.~P.}\ \bibnamefont {MacArthur}}, \bibinfo {author} {\bibfnamefont {A.}~\bibnamefont {Lutman}}, \bibinfo {author} {\bibfnamefont {M.-F.}\ \bibnamefont {Lin}}, \emph {et~al.},\ }\bibfield  {title} {\bibinfo {title} {Enhanced ultrafast x-ray diffraction by transient resonances},\ }\href@noop {} {\bibfield  {journal} {\bibinfo  {journal} {arXiv preprint arXiv:2207.05472}\ } (\bibinfo {year} {2022})}\BibitemShut {NoStop}%
\bibitem [{\citenamefont {Ulmer}\ \emph {et~al.}(2023)\citenamefont {Ulmer}, \citenamefont {Kuschel}, \citenamefont {Langbehn}, \citenamefont {Hecht}, \citenamefont {Dold}, \citenamefont {R{\"o}nnebeck}, \citenamefont {Driver}, \citenamefont {Duris}, \citenamefont {Kamalov}, \citenamefont {Li} \emph {et~al.}}]{Tais_Ne_apstalk}%
  \BibitemOpen
  \bibfield  {author} {\bibinfo {author} {\bibfnamefont {A.}~\bibnamefont {Ulmer}}, \bibinfo {author} {\bibfnamefont {S.}~\bibnamefont {Kuschel}}, \bibinfo {author} {\bibfnamefont {B.}~\bibnamefont {Langbehn}}, \bibinfo {author} {\bibfnamefont {L.}~\bibnamefont {Hecht}}, \bibinfo {author} {\bibfnamefont {S.}~\bibnamefont {Dold}}, \bibinfo {author} {\bibfnamefont {L.}~\bibnamefont {R{\"o}nnebeck}}, \bibinfo {author} {\bibfnamefont {T.}~\bibnamefont {Driver}}, \bibinfo {author} {\bibfnamefont {J.}~\bibnamefont {Duris}}, \bibinfo {author} {\bibfnamefont {A.}~\bibnamefont {Kamalov}}, \bibinfo {author} {\bibfnamefont {X.}~\bibnamefont {Li}}, \emph {et~al.},\ }\bibfield  {title} {\bibinfo {title} {Exploring damage reduction and scattering cross section enhancement in attosecond x-ray imaging of neon near the k-edge},\ }\href@noop {} {\bibfield  {journal} {\bibinfo  {journal} {Bulletin of the American Physical Society}\ } (\bibinfo {year} {2023})}\BibitemShut {NoStop}%
\bibitem [{\citenamefont {Rudek}\ \emph {et~al.}(2012)\citenamefont {Rudek}, \citenamefont {Son}, \citenamefont {Foucar}, \citenamefont {Epp}, \citenamefont {Erk}, \citenamefont {Hartmann}, \citenamefont {Adolph}, \citenamefont {Andritschke}, \citenamefont {Aquila}, \citenamefont {Berrah}, \citenamefont {Bostedt}, \citenamefont {Bozek}, \citenamefont {Coppola}, \citenamefont {Filsinger}, \citenamefont {Gorke}, \citenamefont {Gorkhover}, \citenamefont {Graafsma}, \citenamefont {Gumprecht}, \citenamefont {Hartmann}, \citenamefont {Hauser}, \citenamefont {Herrmann}, \citenamefont {Hirsemann}, \citenamefont {Holl}, \citenamefont {Hoemke}, \citenamefont {Journel}, \citenamefont {Kaiser}, \citenamefont {Kimmel}, \citenamefont {Krasniqi}, \citenamefont {Kuehnel}, \citenamefont {Matysek}, \citenamefont {Messerschmidt}, \citenamefont {Miesner}, \citenamefont {Moeller}, \citenamefont {Moshammer}, \citenamefont {Nagaya}, \citenamefont {Nilsson}, \citenamefont {Potdevin}, \citenamefont {Pietschner}, \citenamefont
  {Reich}, \citenamefont {Rupp}, \citenamefont {Schaller}, \citenamefont {Schlichting}, \citenamefont {Schmidt}, \citenamefont {Schopper}, \citenamefont {Schorb}, \citenamefont {Schroeter}, \citenamefont {Schulz}, \citenamefont {Simon}, \citenamefont {Soltau}, \citenamefont {Strueder}, \citenamefont {Ueda}, \citenamefont {Weidenspointner}, \citenamefont {Santra}, \citenamefont {Ullrich}, \citenamefont {Rudenko},\ and\ \citenamefont {Rolles}}]{Rudek-2012-NatPho}%
  \BibitemOpen
  \bibfield  {author} {\bibinfo {author} {\bibfnamefont {B.}~\bibnamefont {Rudek}}, \bibinfo {author} {\bibfnamefont {S.-K.}\ \bibnamefont {Son}}, \bibinfo {author} {\bibfnamefont {L.}~\bibnamefont {Foucar}}, \bibinfo {author} {\bibfnamefont {S.~W.}\ \bibnamefont {Epp}}, \bibinfo {author} {\bibfnamefont {B.}~\bibnamefont {Erk}}, \bibinfo {author} {\bibfnamefont {R.}~\bibnamefont {Hartmann}}, \bibinfo {author} {\bibfnamefont {M.}~\bibnamefont {Adolph}}, \bibinfo {author} {\bibfnamefont {R.}~\bibnamefont {Andritschke}}, \bibinfo {author} {\bibfnamefont {A.}~\bibnamefont {Aquila}}, \bibinfo {author} {\bibfnamefont {N.}~\bibnamefont {Berrah}}, \bibinfo {author} {\bibfnamefont {C.}~\bibnamefont {Bostedt}}, \bibinfo {author} {\bibfnamefont {J.}~\bibnamefont {Bozek}}, \bibinfo {author} {\bibfnamefont {N.}~\bibnamefont {Coppola}}, \bibinfo {author} {\bibfnamefont {F.}~\bibnamefont {Filsinger}}, \bibinfo {author} {\bibfnamefont {H.}~\bibnamefont {Gorke}}, \bibinfo {author} {\bibfnamefont {T.}~\bibnamefont
  {Gorkhover}}, \bibinfo {author} {\bibfnamefont {H.}~\bibnamefont {Graafsma}}, \bibinfo {author} {\bibfnamefont {L.}~\bibnamefont {Gumprecht}}, \bibinfo {author} {\bibfnamefont {A.}~\bibnamefont {Hartmann}}, \bibinfo {author} {\bibfnamefont {G.}~\bibnamefont {Hauser}}, \bibinfo {author} {\bibfnamefont {S.}~\bibnamefont {Herrmann}}, \bibinfo {author} {\bibfnamefont {H.}~\bibnamefont {Hirsemann}}, \bibinfo {author} {\bibfnamefont {P.}~\bibnamefont {Holl}}, \bibinfo {author} {\bibfnamefont {A.}~\bibnamefont {Hoemke}}, \bibinfo {author} {\bibfnamefont {L.}~\bibnamefont {Journel}}, \bibinfo {author} {\bibfnamefont {C.}~\bibnamefont {Kaiser}}, \bibinfo {author} {\bibfnamefont {N.}~\bibnamefont {Kimmel}}, \bibinfo {author} {\bibfnamefont {F.}~\bibnamefont {Krasniqi}}, \bibinfo {author} {\bibfnamefont {K.-U.}\ \bibnamefont {Kuehnel}}, \bibinfo {author} {\bibfnamefont {M.}~\bibnamefont {Matysek}}, \bibinfo {author} {\bibfnamefont {M.}~\bibnamefont {Messerschmidt}}, \bibinfo {author} {\bibfnamefont {D.}~\bibnamefont
  {Miesner}}, \bibinfo {author} {\bibfnamefont {T.}~\bibnamefont {Moeller}}, \bibinfo {author} {\bibfnamefont {R.}~\bibnamefont {Moshammer}}, \bibinfo {author} {\bibfnamefont {K.}~\bibnamefont {Nagaya}}, \bibinfo {author} {\bibfnamefont {B.}~\bibnamefont {Nilsson}}, \bibinfo {author} {\bibfnamefont {G.}~\bibnamefont {Potdevin}}, \bibinfo {author} {\bibfnamefont {D.}~\bibnamefont {Pietschner}}, \bibinfo {author} {\bibfnamefont {C.}~\bibnamefont {Reich}}, \bibinfo {author} {\bibfnamefont {D.}~\bibnamefont {Rupp}}, \bibinfo {author} {\bibfnamefont {G.}~\bibnamefont {Schaller}}, \bibinfo {author} {\bibfnamefont {I.}~\bibnamefont {Schlichting}}, \bibinfo {author} {\bibfnamefont {C.}~\bibnamefont {Schmidt}}, \bibinfo {author} {\bibfnamefont {F.}~\bibnamefont {Schopper}}, \bibinfo {author} {\bibfnamefont {S.}~\bibnamefont {Schorb}}, \bibinfo {author} {\bibfnamefont {C.-D.}\ \bibnamefont {Schroeter}}, \bibinfo {author} {\bibfnamefont {J.}~\bibnamefont {Schulz}}, \bibinfo {author} {\bibfnamefont {M.}~\bibnamefont
  {Simon}}, \bibinfo {author} {\bibfnamefont {H.}~\bibnamefont {Soltau}}, \bibinfo {author} {\bibfnamefont {L.}~\bibnamefont {Strueder}}, \bibinfo {author} {\bibfnamefont {K.}~\bibnamefont {Ueda}}, \bibinfo {author} {\bibfnamefont {G.}~\bibnamefont {Weidenspointner}}, \bibinfo {author} {\bibfnamefont {R.}~\bibnamefont {Santra}}, \bibinfo {author} {\bibfnamefont {J.}~\bibnamefont {Ullrich}}, \bibinfo {author} {\bibfnamefont {A.}~\bibnamefont {Rudenko}},\ and\ \bibinfo {author} {\bibfnamefont {D.}~\bibnamefont {Rolles}},\ }\bibfield  {title} {\bibinfo {title} {Ultra-efficient ionization of heavy atoms by intense x-ray free-electron laser pulses},\ }\href {https://doi.org/10.1038/Nphoton.2012.261} {\bibfield  {journal} {\bibinfo  {journal} {Nat. Photon.}\ }\textbf {\bibinfo {volume} {6}},\ \bibinfo {pages} {858} (\bibinfo {year} {2012})}\BibitemShut {NoStop}%
\bibitem [{\citenamefont {Rudek}\ \emph {et~al.}(2013)\citenamefont {Rudek}, \citenamefont {Rolles}, \citenamefont {Son}, \citenamefont {Foucar}, \citenamefont {Erk}, \citenamefont {Epp}, \citenamefont {Boll}, \citenamefont {Anielski}, \citenamefont {Bostedt}, \citenamefont {Schorb}, \citenamefont {Coffee}, \citenamefont {Bozek}, \citenamefont {Trippel}, \citenamefont {Marchenko}, \citenamefont {Simon}, \citenamefont {Christensen}, \citenamefont {De}, \citenamefont {Wada}, \citenamefont {Ueda}, \citenamefont {Schlichting}, \citenamefont {Santra}, \citenamefont {Ullrich},\ and\ \citenamefont {Rudenko}}]{Rudek-2013-PRA}%
  \BibitemOpen
  \bibfield  {author} {\bibinfo {author} {\bibfnamefont {B.}~\bibnamefont {Rudek}}, \bibinfo {author} {\bibfnamefont {D.}~\bibnamefont {Rolles}}, \bibinfo {author} {\bibfnamefont {S.-K.}\ \bibnamefont {Son}}, \bibinfo {author} {\bibfnamefont {L.}~\bibnamefont {Foucar}}, \bibinfo {author} {\bibfnamefont {B.}~\bibnamefont {Erk}}, \bibinfo {author} {\bibfnamefont {S.}~\bibnamefont {Epp}}, \bibinfo {author} {\bibfnamefont {R.}~\bibnamefont {Boll}}, \bibinfo {author} {\bibfnamefont {D.}~\bibnamefont {Anielski}}, \bibinfo {author} {\bibfnamefont {C.}~\bibnamefont {Bostedt}}, \bibinfo {author} {\bibfnamefont {S.}~\bibnamefont {Schorb}}, \bibinfo {author} {\bibfnamefont {R.}~\bibnamefont {Coffee}}, \bibinfo {author} {\bibfnamefont {J.}~\bibnamefont {Bozek}}, \bibinfo {author} {\bibfnamefont {S.}~\bibnamefont {Trippel}}, \bibinfo {author} {\bibfnamefont {T.}~\bibnamefont {Marchenko}}, \bibinfo {author} {\bibfnamefont {M.}~\bibnamefont {Simon}}, \bibinfo {author} {\bibfnamefont {L.}~\bibnamefont {Christensen}},
  \bibinfo {author} {\bibfnamefont {S.}~\bibnamefont {De}}, \bibinfo {author} {\bibfnamefont {S.-i.}\ \bibnamefont {Wada}}, \bibinfo {author} {\bibfnamefont {K.}~\bibnamefont {Ueda}}, \bibinfo {author} {\bibfnamefont {I.}~\bibnamefont {Schlichting}}, \bibinfo {author} {\bibfnamefont {R.}~\bibnamefont {Santra}}, \bibinfo {author} {\bibfnamefont {J.}~\bibnamefont {Ullrich}},\ and\ \bibinfo {author} {\bibfnamefont {A.}~\bibnamefont {Rudenko}},\ }\bibfield  {title} {\bibinfo {title} {{Resonance-enhanced multiple ionization of krypton at an x-ray free-electron laser}},\ }\href {https://doi.org/{10.1103/PhysRevA.87.023413}} {\bibfield  {journal} {\bibinfo  {journal} {{Phys. Rev. A}}\ }\textbf {\bibinfo {volume} {{87}}},\ \bibinfo {pages} {023413} (\bibinfo {year} {{2013}})}\BibitemShut {NoStop}%
\bibitem [{\citenamefont {Ho}\ \emph {et~al.}(2014)\citenamefont {Ho}, \citenamefont {Bostedt}, \citenamefont {Schorb},\ and\ \citenamefont {Young}}]{Ho-2014-PRL}%
  \BibitemOpen
  \bibfield  {author} {\bibinfo {author} {\bibfnamefont {P.~J.}\ \bibnamefont {Ho}}, \bibinfo {author} {\bibfnamefont {C.}~\bibnamefont {Bostedt}}, \bibinfo {author} {\bibfnamefont {S.}~\bibnamefont {Schorb}},\ and\ \bibinfo {author} {\bibfnamefont {L.}~\bibnamefont {Young}},\ }\bibfield  {title} {\bibinfo {title} {{Theoretical tracking of resonance-enhanced multiple ionization pathways in x-ray free-electron laser pulses}},\ }\href {https://doi.org/{10.1103/PhysRevLett.113.253001}} {\bibfield  {journal} {\bibinfo  {journal} {{Phys. Rev. Lett.}}\ }\textbf {\bibinfo {volume} {{113}}},\ \bibinfo {pages} {{253001}} (\bibinfo {year} {{2014}})}\BibitemShut {NoStop}%
\bibitem [{\citenamefont {Ho}\ \emph {et~al.}(2015)\citenamefont {Ho}, \citenamefont {Kanter},\ and\ \citenamefont {Young}}]{Ho-2015-PRA}%
  \BibitemOpen
  \bibfield  {author} {\bibinfo {author} {\bibfnamefont {P.~J.}\ \bibnamefont {Ho}}, \bibinfo {author} {\bibfnamefont {E.~P.}\ \bibnamefont {Kanter}},\ and\ \bibinfo {author} {\bibfnamefont {L.}~\bibnamefont {Young}},\ }\bibfield  {title} {\bibinfo {title} {Resonance-mediated atomic ionization dynamics induced by ultraintense x-ray pulses},\ }\href {https://doi.org/10.1103/PhysRevA.92.063430} {\bibfield  {journal} {\bibinfo  {journal} {Phys. Rev. A}\ }\textbf {\bibinfo {volume} {92}},\ \bibinfo {pages} {063430} (\bibinfo {year} {2015})}\BibitemShut {NoStop}%
\bibitem [{\citenamefont {Ho}\ \emph {et~al.}(2023)\citenamefont {Ho}, \citenamefont {Ray}, \citenamefont {Lehmann}, \citenamefont {Fouda}, \citenamefont {Dunford}, \citenamefont {Kanter}, \citenamefont {Doumy}, \citenamefont {Young}, \citenamefont {Walko}, \citenamefont {Zheng} \emph {et~al.}}]{Ho-JCP-2023}%
  \BibitemOpen
  \bibfield  {author} {\bibinfo {author} {\bibfnamefont {P.~J.}\ \bibnamefont {Ho}}, \bibinfo {author} {\bibfnamefont {D.}~\bibnamefont {Ray}}, \bibinfo {author} {\bibfnamefont {C.~S.}\ \bibnamefont {Lehmann}}, \bibinfo {author} {\bibfnamefont {A.~E.}\ \bibnamefont {Fouda}}, \bibinfo {author} {\bibfnamefont {R.~W.}\ \bibnamefont {Dunford}}, \bibinfo {author} {\bibfnamefont {E.~P.}\ \bibnamefont {Kanter}}, \bibinfo {author} {\bibfnamefont {G.}~\bibnamefont {Doumy}}, \bibinfo {author} {\bibfnamefont {L.}~\bibnamefont {Young}}, \bibinfo {author} {\bibfnamefont {D.~A.}\ \bibnamefont {Walko}}, \bibinfo {author} {\bibfnamefont {X.}~\bibnamefont {Zheng}}, \emph {et~al.},\ }\bibfield  {title} {\bibinfo {title} {X-ray induced electron and ion fragmentation dynamics in ibr},\ }\href@noop {} {\bibfield  {journal} {\bibinfo  {journal} {The Journal of Chemical Physics}\ }\textbf {\bibinfo {volume} {158}} (\bibinfo {year} {2023})}\BibitemShut {NoStop}%
\bibitem [{\citenamefont {Rudenko}\ \emph {et~al.}(2017)\citenamefont {Rudenko}, \citenamefont {Inhester}, \citenamefont {Hanasaki}, \citenamefont {Li}, \citenamefont {Robatjazi}, \citenamefont {Erk}, \citenamefont {Boll}, \citenamefont {Toyota}, \citenamefont {Hao}, \citenamefont {Vendrell}, \citenamefont {Bomme}, \citenamefont {Savelyev}, \citenamefont {Rudek}, \citenamefont {Foucar}, \citenamefont {Southworth}, \citenamefont {Lehmann}, \citenamefont {Kraessig}, \citenamefont {Marchenko}, \citenamefont {Simon}, \citenamefont {Ueda}, \citenamefont {Ferguson}, \citenamefont {Bucher}, \citenamefont {Gorkhover}, \citenamefont {Carron}, \citenamefont {Alonso-Mori}, \citenamefont {Koglin}, \citenamefont {Correa}, \citenamefont {Williams}, \citenamefont {Boutet}, \citenamefont {Young}, \citenamefont {Bostedt}, \citenamefont {Son}, \citenamefont {Santra},\ and\ \citenamefont {Rolles}}]{Rudenko2017}%
  \BibitemOpen
  \bibfield  {author} {\bibinfo {author} {\bibfnamefont {A.}~\bibnamefont {Rudenko}}, \bibinfo {author} {\bibfnamefont {L.}~\bibnamefont {Inhester}}, \bibinfo {author} {\bibfnamefont {K.}~\bibnamefont {Hanasaki}}, \bibinfo {author} {\bibfnamefont {X.}~\bibnamefont {Li}}, \bibinfo {author} {\bibfnamefont {S.~J.}\ \bibnamefont {Robatjazi}}, \bibinfo {author} {\bibfnamefont {B.}~\bibnamefont {Erk}}, \bibinfo {author} {\bibfnamefont {R.}~\bibnamefont {Boll}}, \bibinfo {author} {\bibfnamefont {K.}~\bibnamefont {Toyota}}, \bibinfo {author} {\bibfnamefont {Y.}~\bibnamefont {Hao}}, \bibinfo {author} {\bibfnamefont {O.}~\bibnamefont {Vendrell}}, \bibinfo {author} {\bibfnamefont {C.}~\bibnamefont {Bomme}}, \bibinfo {author} {\bibfnamefont {E.}~\bibnamefont {Savelyev}}, \bibinfo {author} {\bibfnamefont {B.}~\bibnamefont {Rudek}}, \bibinfo {author} {\bibfnamefont {L.}~\bibnamefont {Foucar}}, \bibinfo {author} {\bibfnamefont {S.~H.}\ \bibnamefont {Southworth}}, \bibinfo {author} {\bibfnamefont {C.~S.}\ \bibnamefont
  {Lehmann}}, \bibinfo {author} {\bibfnamefont {B.}~\bibnamefont {Kraessig}}, \bibinfo {author} {\bibfnamefont {T.}~\bibnamefont {Marchenko}}, \bibinfo {author} {\bibfnamefont {M.}~\bibnamefont {Simon}}, \bibinfo {author} {\bibfnamefont {K.}~\bibnamefont {Ueda}}, \bibinfo {author} {\bibfnamefont {K.~R.}\ \bibnamefont {Ferguson}}, \bibinfo {author} {\bibfnamefont {M.}~\bibnamefont {Bucher}}, \bibinfo {author} {\bibfnamefont {T.}~\bibnamefont {Gorkhover}}, \bibinfo {author} {\bibfnamefont {S.}~\bibnamefont {Carron}}, \bibinfo {author} {\bibfnamefont {R.}~\bibnamefont {Alonso-Mori}}, \bibinfo {author} {\bibfnamefont {J.~E.}\ \bibnamefont {Koglin}}, \bibinfo {author} {\bibfnamefont {J.}~\bibnamefont {Correa}}, \bibinfo {author} {\bibfnamefont {G.~J.}\ \bibnamefont {Williams}}, \bibinfo {author} {\bibfnamefont {S.}~\bibnamefont {Boutet}}, \bibinfo {author} {\bibfnamefont {L.}~\bibnamefont {Young}}, \bibinfo {author} {\bibfnamefont {C.}~\bibnamefont {Bostedt}}, \bibinfo {author} {\bibfnamefont {S.-K.}\ \bibnamefont
  {Son}}, \bibinfo {author} {\bibfnamefont {R.}~\bibnamefont {Santra}},\ and\ \bibinfo {author} {\bibfnamefont {D.}~\bibnamefont {Rolles}},\ }\bibfield  {title} {\bibinfo {title} {Femtosecond response of polyatomic molecules to ultra-intense hard x-rays},\ }\href {https://doi.org/10.1038/nature22373} {\bibfield  {journal} {\bibinfo  {journal} {Nature}\ }\textbf {\bibinfo {volume} {546}},\ \bibinfo {pages} {129} (\bibinfo {year} {2017})}\BibitemShut {NoStop}%
\bibitem [{\citenamefont {Vinko}\ \emph {et~al.}(2012)\citenamefont {Vinko}, \citenamefont {Ciricosta}, \citenamefont {Cho}, \citenamefont {Engelhorn}, \citenamefont {Chung}, \citenamefont {Brown}, \citenamefont {Burian}, \citenamefont {Chalupsk{\'y}}, \citenamefont {Falcone}, \citenamefont {Graves}, \citenamefont {H{\'a}jkov{\'a}}, \citenamefont {Higginbotham}, \citenamefont {Juha}, \citenamefont {Krzywinski}, \citenamefont {Lee}, \citenamefont {Messerschmidt}, \citenamefont {Murphy}, \citenamefont {Ping}, \citenamefont {Scherz}, \citenamefont {Schlotter}, \citenamefont {Toleikis}, \citenamefont {Turner}, \citenamefont {Vysin}, \citenamefont {Wang}, \citenamefont {Wu}, \citenamefont {Zastrau}, \citenamefont {Zhu}, \citenamefont {Lee}, \citenamefont {Heimann}, \citenamefont {Nagler},\ and\ \citenamefont {Wark}}]{Vinko-Nature-2012}%
  \BibitemOpen
  \bibfield  {author} {\bibinfo {author} {\bibfnamefont {S.~M.}\ \bibnamefont {Vinko}}, \bibinfo {author} {\bibfnamefont {O.}~\bibnamefont {Ciricosta}}, \bibinfo {author} {\bibfnamefont {B.~I.}\ \bibnamefont {Cho}}, \bibinfo {author} {\bibfnamefont {K.}~\bibnamefont {Engelhorn}}, \bibinfo {author} {\bibfnamefont {H.-K.}\ \bibnamefont {Chung}}, \bibinfo {author} {\bibfnamefont {C.~R.~D.}\ \bibnamefont {Brown}}, \bibinfo {author} {\bibfnamefont {T.}~\bibnamefont {Burian}}, \bibinfo {author} {\bibfnamefont {J.}~\bibnamefont {Chalupsk{\'y}}}, \bibinfo {author} {\bibfnamefont {R.~W.}\ \bibnamefont {Falcone}}, \bibinfo {author} {\bibfnamefont {C.}~\bibnamefont {Graves}}, \bibinfo {author} {\bibfnamefont {V.}~\bibnamefont {H{\'a}jkov{\'a}}}, \bibinfo {author} {\bibfnamefont {A.}~\bibnamefont {Higginbotham}}, \bibinfo {author} {\bibfnamefont {L.}~\bibnamefont {Juha}}, \bibinfo {author} {\bibfnamefont {J.}~\bibnamefont {Krzywinski}}, \bibinfo {author} {\bibfnamefont {H.~J.}\ \bibnamefont {Lee}}, \bibinfo {author}
  {\bibfnamefont {M.}~\bibnamefont {Messerschmidt}}, \bibinfo {author} {\bibfnamefont {C.~D.}\ \bibnamefont {Murphy}}, \bibinfo {author} {\bibfnamefont {Y.}~\bibnamefont {Ping}}, \bibinfo {author} {\bibfnamefont {A.}~\bibnamefont {Scherz}}, \bibinfo {author} {\bibfnamefont {W.}~\bibnamefont {Schlotter}}, \bibinfo {author} {\bibfnamefont {S.}~\bibnamefont {Toleikis}}, \bibinfo {author} {\bibfnamefont {J.~J.}\ \bibnamefont {Turner}}, \bibinfo {author} {\bibfnamefont {L.}~\bibnamefont {Vysin}}, \bibinfo {author} {\bibfnamefont {T.}~\bibnamefont {Wang}}, \bibinfo {author} {\bibfnamefont {B.}~\bibnamefont {Wu}}, \bibinfo {author} {\bibfnamefont {U.}~\bibnamefont {Zastrau}}, \bibinfo {author} {\bibfnamefont {D.}~\bibnamefont {Zhu}}, \bibinfo {author} {\bibfnamefont {R.~W.}\ \bibnamefont {Lee}}, \bibinfo {author} {\bibfnamefont {P.~A.}\ \bibnamefont {Heimann}}, \bibinfo {author} {\bibfnamefont {B.}~\bibnamefont {Nagler}},\ and\ \bibinfo {author} {\bibfnamefont {J.~S.}\ \bibnamefont {Wark}},\ }\bibfield  {title}
  {\bibinfo {title} {Creation and diagnosis of a solid-density plasma with an x-ray free-electron laser},\ }\href {https://doi.org/10.1038/nature10746} {\bibfield  {journal} {\bibinfo  {journal} {Nature}\ }\textbf {\bibinfo {volume} {482}},\ \bibinfo {pages} {59} (\bibinfo {year} {2012})}\BibitemShut {NoStop}%
\bibitem [{\citenamefont {Venkatesh}\ and\ \citenamefont {Robicheaux}(2020)}]{NLCPRA_1}%
  \BibitemOpen
  \bibfield  {author} {\bibinfo {author} {\bibfnamefont {A.}~\bibnamefont {Venkatesh}}\ and\ \bibinfo {author} {\bibfnamefont {F.}~\bibnamefont {Robicheaux}},\ }\bibfield  {title} {\bibinfo {title} {Simulation of nonlinear {C}ompton scattering from bound electrons},\ }\href {https://doi.org/10.1103/PhysRevA.101.013409} {\bibfield  {journal} {\bibinfo  {journal} {Phys. Rev. A}\ }\textbf {\bibinfo {volume} {101}},\ \bibinfo {pages} {013409} (\bibinfo {year} {2020})}\BibitemShut {NoStop}%
\bibitem [{\citenamefont {Venkatesh}(2022)}]{PhDThesis2022}%
  \BibitemOpen
  \bibfield  {author} {\bibinfo {author} {\bibfnamefont {A.}~\bibnamefont {Venkatesh}},\ }\emph {\bibinfo {title} {Theoretical Methods for Non-Relativistic Quantum and Classical Scattering Processes}},\ \href@noop {} {Ph.D. thesis},\ \bibinfo  {school} {Purdue University} (\bibinfo {year} {2022})\BibitemShut {NoStop}%
\bibitem [{\citenamefont {Krebs}\ \emph {et~al.}(2019)\citenamefont {Krebs}, \citenamefont {Reis},\ and\ \citenamefont {Santra}}]{Krebs}%
  \BibitemOpen
  \bibfield  {author} {\bibinfo {author} {\bibfnamefont {D.}~\bibnamefont {Krebs}}, \bibinfo {author} {\bibfnamefont {D.~A.}\ \bibnamefont {Reis}},\ and\ \bibinfo {author} {\bibfnamefont {R.}~\bibnamefont {Santra}},\ }\bibfield  {title} {\bibinfo {title} {Time-dependent {QED} approach to x-ray nonlinear {C}ompton scattering},\ }\href {https://doi.org/10.1103/PhysRevA.99.022120} {\bibfield  {journal} {\bibinfo  {journal} {Phys. Rev. A}\ }\textbf {\bibinfo {volume} {99}},\ \bibinfo {pages} {022120} (\bibinfo {year} {2019})}\BibitemShut {NoStop}%
\bibitem [{\citenamefont {Loudon}(1983)}]{Loudon}%
  \BibitemOpen
  \bibfield  {author} {\bibinfo {author} {\bibfnamefont {R.}~\bibnamefont {Loudon}},\ }\href@noop {} {\emph {\bibinfo {title} {The Quantum Theory of Light}}}\ (\bibinfo  {publisher} {Oxford University Press, New York},\ \bibinfo {year} {1983})\BibitemShut {NoStop}%
\bibitem [{\citenamefont {Sakurai}(1967)}]{Sakurai_adv}%
  \BibitemOpen
  \bibfield  {author} {\bibinfo {author} {\bibfnamefont {J.}~\bibnamefont {Sakurai}},\ }\href@noop {} {\emph {\bibinfo {title} {Advanced Quantum Mechanics}}}\ (\bibinfo  {publisher} {Addison-Wesley Publishing Company},\ \bibinfo {year} {1967})\BibitemShut {NoStop}%
\bibitem [{\citenamefont {Fano}(1985)}]{fano1985propensity}%
  \BibitemOpen
  \bibfield  {author} {\bibinfo {author} {\bibfnamefont {U.}~\bibnamefont {Fano}},\ }\bibfield  {title} {\bibinfo {title} {Propensity rules: An analytical approach},\ }\href@noop {} {\bibfield  {journal} {\bibinfo  {journal} {Physical Review A}\ }\textbf {\bibinfo {volume} {32}},\ \bibinfo {pages} {617} (\bibinfo {year} {1985})}\BibitemShut {NoStop}%
\bibitem [{\citenamefont {Buffa}\ \emph {et~al.}(1988)\citenamefont {Buffa}, \citenamefont {Cavalieri}, \citenamefont {Fini},\ and\ \citenamefont {Matera}}]{buffa1988_amplitude_ResFluor}%
  \BibitemOpen
  \bibfield  {author} {\bibinfo {author} {\bibfnamefont {R.}~\bibnamefont {Buffa}}, \bibinfo {author} {\bibfnamefont {S.}~\bibnamefont {Cavalieri}}, \bibinfo {author} {\bibfnamefont {L.}~\bibnamefont {Fini}},\ and\ \bibinfo {author} {\bibfnamefont {M.}~\bibnamefont {Matera}},\ }\bibfield  {title} {\bibinfo {title} {Resonance fluorescence of a two-level atom driven by a short laser pulse: extension to the off-resonance excitation},\ }\href@noop {} {\bibfield  {journal} {\bibinfo  {journal} {Journal of Physics B: Atomic, Molecular and Optical Physics}\ }\textbf {\bibinfo {volume} {21}},\ \bibinfo {pages} {239} (\bibinfo {year} {1988})}\BibitemShut {NoStop}%
\bibitem [{\citenamefont {Robinson}\ and\ \citenamefont {Berman}(1984)}]{Robinson_Berman1984_buffa_citation}%
  \BibitemOpen
  \bibfield  {author} {\bibinfo {author} {\bibfnamefont {E.}~\bibnamefont {Robinson}}\ and\ \bibinfo {author} {\bibfnamefont {P.}~\bibnamefont {Berman}},\ }\bibfield  {title} {\bibinfo {title} {Comment on'the resonance fluorescence of a two-level system driven by a smooth pulse},\ }\href@noop {} {\bibfield  {journal} {\bibinfo  {journal} {Journal of Physics B: Atomic and Molecular Physics}\ }\textbf {\bibinfo {volume} {17}},\ \bibinfo {pages} {L847} (\bibinfo {year} {1984})}\BibitemShut {NoStop}%
\bibitem [{\citenamefont {Ho}\ and\ \citenamefont {Knight}(2017{\natexlab{a}})}]{HFS_Phay_2017}%
  \BibitemOpen
  \bibfield  {author} {\bibinfo {author} {\bibfnamefont {P.~J.}\ \bibnamefont {Ho}}\ and\ \bibinfo {author} {\bibfnamefont {C.}~\bibnamefont {Knight}},\ }\bibfield  {title} {\bibinfo {title} {Large-scale atomistic calculations of clusters in intense x-ray pulses},\ }\href@noop {} {\bibfield  {journal} {\bibinfo  {journal} {Journal of Physics B: Atomic, Molecular and Optical Physics}\ }\textbf {\bibinfo {volume} {50}},\ \bibinfo {pages} {104003} (\bibinfo {year} {2017}{\natexlab{a}})}\BibitemShut {NoStop}%
\bibitem [{\citenamefont {Press}\ \emph {et~al.}(1992)\citenamefont {Press}, \citenamefont {Teukolsky},\ and\ \citenamefont {Vetterling}}]{numericalrecipes}%
  \BibitemOpen
  \bibfield  {author} {\bibinfo {author} {\bibfnamefont {W.~H.}\ \bibnamefont {Press}}, \bibinfo {author} {\bibfnamefont {S.~A.}\ \bibnamefont {Teukolsky}},\ and\ \bibinfo {author} {\bibfnamefont {W.~T.}\ \bibnamefont {Vetterling}},\ }\href@noop {} {\emph {\bibinfo {title} {Numerical Recipes in {C}: The Art of Scientific Computing, Second Edition}}}\ (\bibinfo  {publisher} {Cambridge University Press},\ \bibinfo {year} {1992})\BibitemShut {NoStop}%
\bibitem [{Com(2024)}]{Companion_letter}%
  \BibitemOpen
  \href@noop {} {\bibinfo {title} {Companion letter}} (\bibinfo {year} {2024})\BibitemShut {NoStop}%
\bibitem [{Ele(2024)}]{Elettra_database}%
  \BibitemOpen
  \href {https://vuo.elettra.eu/services/elements/WebElements.html} {\bibinfo {title} {Elettra synchrotron database}},\ \bibinfo {howpublished} {\url{https://vuo.elettra.eu/services/elements/WebElements.html}} (\bibinfo {year} {2024}),\ \bibinfo {note} {accessed: 2024-03-27}\BibitemShut {NoStop}%
\bibitem [{\citenamefont {Eberly}(1998)}]{Pulsearea_defn_Eberly}%
  \BibitemOpen
  \bibfield  {author} {\bibinfo {author} {\bibfnamefont {J.}~\bibnamefont {Eberly}},\ }\bibfield  {title} {\bibinfo {title} {Area theorem rederived},\ }\href {https://doi.org/10.1364/OE.2.000173} {\bibfield  {journal} {\bibinfo  {journal} {Opt. Express}\ }\textbf {\bibinfo {volume} {2}},\ \bibinfo {pages} {173} (\bibinfo {year} {1998})}\BibitemShut {NoStop}%
\bibitem [{\citenamefont {Cavaletto}(2023)}]{Cavaletto_discussion}%
  \BibitemOpen
  \bibfield  {author} {\bibinfo {author} {\bibfnamefont {S.}~\bibnamefont {Cavaletto}},\ }\href@noop {} {}\bibinfo {howpublished} {Private communication} (\bibinfo {year} {2023})\BibitemShut {NoStop}%
\bibitem [{\citenamefont {Rzazewski}\ and\ \citenamefont {Florjanczyk}(1984)}]{rzazewski1984_multipeakresfluorescence}%
  \BibitemOpen
  \bibfield  {author} {\bibinfo {author} {\bibfnamefont {K.}~\bibnamefont {Rzazewski}}\ and\ \bibinfo {author} {\bibfnamefont {M.}~\bibnamefont {Florjanczyk}},\ }\bibfield  {title} {\bibinfo {title} {The resonance fluorescence of a two-level system driven by a smooth pulse},\ }\href@noop {} {\bibfield  {journal} {\bibinfo  {journal} {Journal of Physics B: Atomic and Molecular Physics}\ }\textbf {\bibinfo {volume} {17}},\ \bibinfo {pages} {L509} (\bibinfo {year} {1984})}\BibitemShut {NoStop}%
\bibitem [{\citenamefont {Robinson}(1986)}]{robinson1986_temporaldiffraction}%
  \BibitemOpen
  \bibfield  {author} {\bibinfo {author} {\bibfnamefont {E.}~\bibnamefont {Robinson}},\ }\bibfield  {title} {\bibinfo {title} {Temporal'diffraction'and eigenvalue interpretation of the resonance fluorescence spectrum of two-level systems driven by short pulses},\ }\href@noop {} {\bibfield  {journal} {\bibinfo  {journal} {Journal of Physics B: Atomic and Molecular Physics}\ }\textbf {\bibinfo {volume} {19}},\ \bibinfo {pages} {L657} (\bibinfo {year} {1986})}\BibitemShut {NoStop}%
\bibitem [{\citenamefont {Lewenstein}\ \emph {et~al.}(1986)\citenamefont {Lewenstein}, \citenamefont {Zakrzewski},\ and\ \citenamefont {Rzazewski}}]{lewenstein1986theory_santraref}%
  \BibitemOpen
  \bibfield  {author} {\bibinfo {author} {\bibfnamefont {M.}~\bibnamefont {Lewenstein}}, \bibinfo {author} {\bibfnamefont {J.}~\bibnamefont {Zakrzewski}},\ and\ \bibinfo {author} {\bibfnamefont {K.}~\bibnamefont {Rzazewski}},\ }\bibfield  {title} {\bibinfo {title} {Theory of fluorescence spectra induced by short laser pulses},\ }\href@noop {} {\bibfield  {journal} {\bibinfo  {journal} {JOSA B}\ }\textbf {\bibinfo {volume} {3}},\ \bibinfo {pages} {22} (\bibinfo {year} {1986})}\BibitemShut {NoStop}%
\bibitem [{LCL(2024)}]{LCLS_specswebsite}%
  \BibitemOpen
  \href {https://lcls.slac.stanford.edu/sites/default/files/2023-07/LCLS-Parameters-Run-22.pdf} {\bibinfo {title} {Projected {R}un 22 {LCLS} {FEL} parameters}},\ \bibinfo {howpublished} {\url{https://lcls.slac.stanford.edu/sites/default/files/2023-07/LCLS-Parameters-Run-22.pdf}} (\bibinfo {year} {2024}),\ \bibinfo {note} {accessed: 2024-03-27}\BibitemShut {NoStop}%
\bibitem [{Eux(2024)}]{Euxfel_specswebsite}%
  \BibitemOpen
  \href {https://www.xfel.eu/facility/instruments/mid/instrument_specifications/index_eng.html} {\bibinfo {title} {Instrument specifications}},\ \bibinfo {howpublished} {\url{https://www.xfel.eu/facility/instruments/mid/instrument_specifications/index_eng.html}} (\bibinfo {year} {2024}),\ \bibinfo {note} {accessed: 2024-03-25}\BibitemShut {NoStop}%
\bibitem [{\citenamefont {Barty}\ \emph {et~al.}(2012)\citenamefont {Barty}, \citenamefont {Caleman}, \citenamefont {Aquila}, \citenamefont {Timneanu}, \citenamefont {Lomb}, \citenamefont {White}, \citenamefont {Andreasson}, \citenamefont {Arnlund}, \citenamefont {Bajt}, \citenamefont {Barends}, \citenamefont {Barthelmess}, \citenamefont {Bogan}, \citenamefont {Bostedt}, \citenamefont {Bozek}, \citenamefont {Coffee}, \citenamefont {Coppola}, \citenamefont {Davidsson}, \citenamefont {DePonte}, \citenamefont {Doak}, \citenamefont {Ekeberg}, \citenamefont {Elser}, \citenamefont {Epp}, \citenamefont {Erk}, \citenamefont {Fleckenstein}, \citenamefont {Foucar}, \citenamefont {Fromme}, \citenamefont {Graafsma}, \citenamefont {Gumprecht}, \citenamefont {Hajdu}, \citenamefont {Hampton}, \citenamefont {Hartmann}, \citenamefont {Hartmann}, \citenamefont {Hauser}, \citenamefont {Hirsemann}, \citenamefont {Holl}, \citenamefont {Hunter}, \citenamefont {Johansson}, \citenamefont {Kassemeyer}, \citenamefont {Kimmel},
  \citenamefont {Kirian}, \citenamefont {Liang}, \citenamefont {Maia}, \citenamefont {Malmerberg}, \citenamefont {Marchesini}, \citenamefont {Martin}, \citenamefont {Nass}, \citenamefont {Neutze}, \citenamefont {Reich}, \citenamefont {Rolles}, \citenamefont {Rudek}, \citenamefont {Rudenko}, \citenamefont {Scott}, \citenamefont {Schlichting}, \citenamefont {Schulz}, \citenamefont {Seibert}, \citenamefont {Shoeman}, \citenamefont {Sierra}, \citenamefont {Soltau}, \citenamefont {Spence}, \citenamefont {Stellato}, \citenamefont {Stern}, \citenamefont {Str{\"u}der}, \citenamefont {Ullrich}, \citenamefont {Wang}, \citenamefont {Weidenspointner}, \citenamefont {Weierstall}, \citenamefont {Wunderer},\ and\ \citenamefont {Chapman}}]{Barty-NatPho-2012}%
  \BibitemOpen
  \bibfield  {author} {\bibinfo {author} {\bibfnamefont {A.}~\bibnamefont {Barty}}, \bibinfo {author} {\bibfnamefont {C.}~\bibnamefont {Caleman}}, \bibinfo {author} {\bibfnamefont {A.}~\bibnamefont {Aquila}}, \bibinfo {author} {\bibfnamefont {N.}~\bibnamefont {Timneanu}}, \bibinfo {author} {\bibfnamefont {L.}~\bibnamefont {Lomb}}, \bibinfo {author} {\bibfnamefont {T.~A.}\ \bibnamefont {White}}, \bibinfo {author} {\bibfnamefont {J.}~\bibnamefont {Andreasson}}, \bibinfo {author} {\bibfnamefont {D.}~\bibnamefont {Arnlund}}, \bibinfo {author} {\bibfnamefont {S.}~\bibnamefont {Bajt}}, \bibinfo {author} {\bibfnamefont {T.~R.~M.}\ \bibnamefont {Barends}}, \bibinfo {author} {\bibfnamefont {M.}~\bibnamefont {Barthelmess}}, \bibinfo {author} {\bibfnamefont {M.~J.}\ \bibnamefont {Bogan}}, \bibinfo {author} {\bibfnamefont {C.}~\bibnamefont {Bostedt}}, \bibinfo {author} {\bibfnamefont {J.~D.}\ \bibnamefont {Bozek}}, \bibinfo {author} {\bibfnamefont {R.}~\bibnamefont {Coffee}}, \bibinfo {author} {\bibfnamefont
  {N.}~\bibnamefont {Coppola}}, \bibinfo {author} {\bibfnamefont {J.}~\bibnamefont {Davidsson}}, \bibinfo {author} {\bibfnamefont {D.~P.}\ \bibnamefont {DePonte}}, \bibinfo {author} {\bibfnamefont {R.~B.}\ \bibnamefont {Doak}}, \bibinfo {author} {\bibfnamefont {T.}~\bibnamefont {Ekeberg}}, \bibinfo {author} {\bibfnamefont {V.}~\bibnamefont {Elser}}, \bibinfo {author} {\bibfnamefont {S.~W.}\ \bibnamefont {Epp}}, \bibinfo {author} {\bibfnamefont {B.}~\bibnamefont {Erk}}, \bibinfo {author} {\bibfnamefont {H.}~\bibnamefont {Fleckenstein}}, \bibinfo {author} {\bibfnamefont {L.}~\bibnamefont {Foucar}}, \bibinfo {author} {\bibfnamefont {P.}~\bibnamefont {Fromme}}, \bibinfo {author} {\bibfnamefont {H.}~\bibnamefont {Graafsma}}, \bibinfo {author} {\bibfnamefont {L.}~\bibnamefont {Gumprecht}}, \bibinfo {author} {\bibfnamefont {J.}~\bibnamefont {Hajdu}}, \bibinfo {author} {\bibfnamefont {C.~Y.}\ \bibnamefont {Hampton}}, \bibinfo {author} {\bibfnamefont {R.}~\bibnamefont {Hartmann}}, \bibinfo {author} {\bibfnamefont
  {A.}~\bibnamefont {Hartmann}}, \bibinfo {author} {\bibfnamefont {G.}~\bibnamefont {Hauser}}, \bibinfo {author} {\bibfnamefont {H.}~\bibnamefont {Hirsemann}}, \bibinfo {author} {\bibfnamefont {P.}~\bibnamefont {Holl}}, \bibinfo {author} {\bibfnamefont {M.~S.}\ \bibnamefont {Hunter}}, \bibinfo {author} {\bibfnamefont {L.}~\bibnamefont {Johansson}}, \bibinfo {author} {\bibfnamefont {S.}~\bibnamefont {Kassemeyer}}, \bibinfo {author} {\bibfnamefont {N.}~\bibnamefont {Kimmel}}, \bibinfo {author} {\bibfnamefont {R.~A.}\ \bibnamefont {Kirian}}, \bibinfo {author} {\bibfnamefont {M.}~\bibnamefont {Liang}}, \bibinfo {author} {\bibfnamefont {F.~R. N.~C.}\ \bibnamefont {Maia}}, \bibinfo {author} {\bibfnamefont {E.}~\bibnamefont {Malmerberg}}, \bibinfo {author} {\bibfnamefont {S.}~\bibnamefont {Marchesini}}, \bibinfo {author} {\bibfnamefont {A.~V.}\ \bibnamefont {Martin}}, \bibinfo {author} {\bibfnamefont {K.}~\bibnamefont {Nass}}, \bibinfo {author} {\bibfnamefont {R.}~\bibnamefont {Neutze}}, \bibinfo {author}
  {\bibfnamefont {C.}~\bibnamefont {Reich}}, \bibinfo {author} {\bibfnamefont {D.}~\bibnamefont {Rolles}}, \bibinfo {author} {\bibfnamefont {B.}~\bibnamefont {Rudek}}, \bibinfo {author} {\bibfnamefont {A.}~\bibnamefont {Rudenko}}, \bibinfo {author} {\bibfnamefont {H.}~\bibnamefont {Scott}}, \bibinfo {author} {\bibfnamefont {I.}~\bibnamefont {Schlichting}}, \bibinfo {author} {\bibfnamefont {J.}~\bibnamefont {Schulz}}, \bibinfo {author} {\bibfnamefont {M.~M.}\ \bibnamefont {Seibert}}, \bibinfo {author} {\bibfnamefont {R.~L.}\ \bibnamefont {Shoeman}}, \bibinfo {author} {\bibfnamefont {R.~G.}\ \bibnamefont {Sierra}}, \bibinfo {author} {\bibfnamefont {H.}~\bibnamefont {Soltau}}, \bibinfo {author} {\bibfnamefont {J.~C.~H.}\ \bibnamefont {Spence}}, \bibinfo {author} {\bibfnamefont {F.}~\bibnamefont {Stellato}}, \bibinfo {author} {\bibfnamefont {S.}~\bibnamefont {Stern}}, \bibinfo {author} {\bibfnamefont {L.}~\bibnamefont {Str{\"u}der}}, \bibinfo {author} {\bibfnamefont {J.}~\bibnamefont {Ullrich}}, \bibinfo {author}
  {\bibfnamefont {X.}~\bibnamefont {Wang}}, \bibinfo {author} {\bibfnamefont {G.}~\bibnamefont {Weidenspointner}}, \bibinfo {author} {\bibfnamefont {U.}~\bibnamefont {Weierstall}}, \bibinfo {author} {\bibfnamefont {C.~B.}\ \bibnamefont {Wunderer}},\ and\ \bibinfo {author} {\bibfnamefont {H.~N.}\ \bibnamefont {Chapman}},\ }\bibfield  {title} {\bibinfo {title} {Self-terminating diffraction gates femtosecond x-ray nanocrystallography measurements},\ }\href {https://doi.org/10.1038/nphoton.2011.297} {\bibfield  {journal} {\bibinfo  {journal} {Nature Photonics}\ }\textbf {\bibinfo {volume} {6}},\ \bibinfo {pages} {35} (\bibinfo {year} {2012})}\BibitemShut {NoStop}%
\bibitem [{\citenamefont {Ho}\ \emph {et~al.}(2016)\citenamefont {Ho}, \citenamefont {Knight}, \citenamefont {Tegze}, \citenamefont {Faigel}, \citenamefont {Bostedt},\ and\ \citenamefont {Young}}]{Ho-2016-PRA}%
  \BibitemOpen
  \bibfield  {author} {\bibinfo {author} {\bibfnamefont {P.~J.}\ \bibnamefont {Ho}}, \bibinfo {author} {\bibfnamefont {C.}~\bibnamefont {Knight}}, \bibinfo {author} {\bibfnamefont {M.}~\bibnamefont {Tegze}}, \bibinfo {author} {\bibfnamefont {G.}~\bibnamefont {Faigel}}, \bibinfo {author} {\bibfnamefont {C.}~\bibnamefont {Bostedt}},\ and\ \bibinfo {author} {\bibfnamefont {L.}~\bibnamefont {Young}},\ }\bibfield  {title} {\bibinfo {title} {Atomistic three-dimensional coherent x-ray imaging of nonbiological systems},\ }\href {https://doi.org/10.1103/PhysRevA.94.063823} {\bibfield  {journal} {\bibinfo  {journal} {Phys. Rev. A}\ }\textbf {\bibinfo {volume} {94}},\ \bibinfo {pages} {063823} (\bibinfo {year} {2016})}\BibitemShut {NoStop}%
\bibitem [{\citenamefont {Ho}\ \emph {et~al.}(2021)\citenamefont {Ho}, \citenamefont {Knight},\ and\ \citenamefont {Young}}]{Ho-2021-SD}%
  \BibitemOpen
  \bibfield  {author} {\bibinfo {author} {\bibfnamefont {P.~J.}\ \bibnamefont {Ho}}, \bibinfo {author} {\bibfnamefont {C.}~\bibnamefont {Knight}},\ and\ \bibinfo {author} {\bibfnamefont {L.}~\bibnamefont {Young}},\ }\bibfield  {title} {\bibinfo {title} {{Fluorescence intensity correlation imaging with high spatial resolution and elemental contrast using intense x-ray pulses}},\ }\href {https://doi.org/10.1063/4.0000105} {\bibfield  {journal} {\bibinfo  {journal} {Structural Dynamics}\ }\textbf {\bibinfo {volume} {8}},\ \bibinfo {pages} {044101} (\bibinfo {year} {2021})},\ \Eprint {https://arxiv.org/abs/https://pubs.aip.org/aca/sdy/article-pdf/doi/10.1063/4.0000105/13835747/044101\_1\_online.pdf} {https://pubs.aip.org/aca/sdy/article-pdf/doi/10.1063/4.0000105/13835747/044101\_1\_online.pdf} \BibitemShut {NoStop}%
\bibitem [{\citenamefont {Carniato}\ \emph {et~al.}(2012)\citenamefont {Carniato}, \citenamefont {Selles}, \citenamefont {Journel}, \citenamefont {Guillemin}, \citenamefont {Stolte}, \citenamefont {El~Khoury}, \citenamefont {Marin}, \citenamefont {Gel'mukhanov}, \citenamefont {Lindle},\ and\ \citenamefont {Simon}}]{Thomson_resonant_interference2012}%
  \BibitemOpen
  \bibfield  {author} {\bibinfo {author} {\bibfnamefont {S.}~\bibnamefont {Carniato}}, \bibinfo {author} {\bibfnamefont {P.}~\bibnamefont {Selles}}, \bibinfo {author} {\bibfnamefont {L.}~\bibnamefont {Journel}}, \bibinfo {author} {\bibfnamefont {R.}~\bibnamefont {Guillemin}}, \bibinfo {author} {\bibfnamefont {W.~C.}\ \bibnamefont {Stolte}}, \bibinfo {author} {\bibfnamefont {L.}~\bibnamefont {El~Khoury}}, \bibinfo {author} {\bibfnamefont {T.}~\bibnamefont {Marin}}, \bibinfo {author} {\bibfnamefont {F.}~\bibnamefont {Gel'mukhanov}}, \bibinfo {author} {\bibfnamefont {D.~W.}\ \bibnamefont {Lindle}},\ and\ \bibinfo {author} {\bibfnamefont {M.}~\bibnamefont {Simon}},\ }\bibfield  {title} {\bibinfo {title} {Thomson-resonant interference effects in elastic x-ray scattering near the cl k edge of hcl},\ }\href@noop {} {\bibfield  {journal} {\bibinfo  {journal} {The Journal of Chemical Physics}\ }\textbf {\bibinfo {volume} {137}} (\bibinfo {year} {2012})}\BibitemShut {NoStop}%
\bibitem [{\citenamefont {Nandi}\ \emph {et~al.}(2022)\citenamefont {Nandi}, \citenamefont {Olofsson}, \citenamefont {Bertolino}, \citenamefont {Carlstr{\"o}m}, \citenamefont {Zapata}, \citenamefont {Busto}, \citenamefont {Callegari}, \citenamefont {Di~Fraia}, \citenamefont {Eng-Johnsson}, \citenamefont {Feifel} \emph {et~al.}}]{nandi2022_observationRabidynamics}%
  \BibitemOpen
  \bibfield  {author} {\bibinfo {author} {\bibfnamefont {S.}~\bibnamefont {Nandi}}, \bibinfo {author} {\bibfnamefont {E.}~\bibnamefont {Olofsson}}, \bibinfo {author} {\bibfnamefont {M.}~\bibnamefont {Bertolino}}, \bibinfo {author} {\bibfnamefont {S.}~\bibnamefont {Carlstr{\"o}m}}, \bibinfo {author} {\bibfnamefont {F.}~\bibnamefont {Zapata}}, \bibinfo {author} {\bibfnamefont {D.}~\bibnamefont {Busto}}, \bibinfo {author} {\bibfnamefont {C.}~\bibnamefont {Callegari}}, \bibinfo {author} {\bibfnamefont {M.}~\bibnamefont {Di~Fraia}}, \bibinfo {author} {\bibfnamefont {P.}~\bibnamefont {Eng-Johnsson}}, \bibinfo {author} {\bibfnamefont {R.}~\bibnamefont {Feifel}}, \emph {et~al.},\ }\bibfield  {title} {\bibinfo {title} {Observation of rabi dynamics with a short-wavelength free-electron laser},\ }\href@noop {} {\bibfield  {journal} {\bibinfo  {journal} {Nature}\ }\textbf {\bibinfo {volume} {608}},\ \bibinfo {pages} {488} (\bibinfo {year} {2022})}\BibitemShut {NoStop}%
\bibitem [{\citenamefont {Ho}\ \emph {et~al.}(2020{\natexlab{b}})\citenamefont {Ho}, \citenamefont {Daurer}, \citenamefont {Hantke}, \citenamefont {Bielecki}, \citenamefont {Al~Haddad}, \citenamefont {Bucher}, \citenamefont {Doumy}, \citenamefont {Ferguson}, \citenamefont {Fl{\"u}ckiger}, \citenamefont {Gorkhover} \emph {et~al.}}]{imagingsucrose_phay}%
  \BibitemOpen
  \bibfield  {author} {\bibinfo {author} {\bibfnamefont {P.~J.}\ \bibnamefont {Ho}}, \bibinfo {author} {\bibfnamefont {B.~J.}\ \bibnamefont {Daurer}}, \bibinfo {author} {\bibfnamefont {M.~F.}\ \bibnamefont {Hantke}}, \bibinfo {author} {\bibfnamefont {J.}~\bibnamefont {Bielecki}}, \bibinfo {author} {\bibfnamefont {A.}~\bibnamefont {Al~Haddad}}, \bibinfo {author} {\bibfnamefont {M.}~\bibnamefont {Bucher}}, \bibinfo {author} {\bibfnamefont {G.}~\bibnamefont {Doumy}}, \bibinfo {author} {\bibfnamefont {K.~R.}\ \bibnamefont {Ferguson}}, \bibinfo {author} {\bibfnamefont {L.}~\bibnamefont {Fl{\"u}ckiger}}, \bibinfo {author} {\bibfnamefont {T.}~\bibnamefont {Gorkhover}}, \emph {et~al.},\ }\bibfield  {title} {\bibinfo {title} {The role of transient resonances for ultra-fast imaging of single sucrose nanoclusters},\ }\href@noop {} {\bibfield  {journal} {\bibinfo  {journal} {Nature communications}\ }\textbf {\bibinfo {volume} {11}},\ \bibinfo {pages} {167} (\bibinfo {year} {2020}{\natexlab{b}})}\BibitemShut {NoStop}%
\bibitem [{\citenamefont {Ho}\ and\ \citenamefont {Knight}(2017{\natexlab{b}})}]{largescale_rateequation_montecarlo_phay}%
  \BibitemOpen
  \bibfield  {author} {\bibinfo {author} {\bibfnamefont {P.~J.}\ \bibnamefont {Ho}}\ and\ \bibinfo {author} {\bibfnamefont {C.}~\bibnamefont {Knight}},\ }\bibfield  {title} {\bibinfo {title} {Large-scale atomistic calculations of clusters in intense x-ray pulses},\ }\href@noop {} {\bibfield  {journal} {\bibinfo  {journal} {Journal of Physics B: Atomic, Molecular and Optical Physics}\ }\textbf {\bibinfo {volume} {50}},\ \bibinfo {pages} {104003} (\bibinfo {year} {2017}{\natexlab{b}})}\BibitemShut {NoStop}%
\end{thebibliography}%

\end{document}